\documentclass[12pt]{article}\usepackage{nand-tree}
\input{MyQcircuit.sty}
\begin{document}

\title{
\begin{figure}[h]
\begin{center}
\setlength{\unitlength}{6pt}
\begin{picture}(16,16)
\thicklines
\multiput(0,0)(4,0){5}{\line(0,1){16}}
\multiput(0,0)(0,4){5}{\line(1,0){16}}
\thinlines
\multiput(0,0)(1,0){17}{\line(0,1){16}}
\multiput(0,0)(0,1){17}{\line(1,0){16}}
\put(1,13){$\bullet$}
\put(1,12){$\bullet$}
\put(2,11){$\bullet$}
\put(2,10){$\bullet$}
\put(3,9){$\bullet$}
\put(3,8){$\bullet$}
\put(4,7){$\bullet$}
\put(4,6){$\bullet$}
\put(5,5){$\bullet$}
\put(5,4){$\bullet$}
\put(6,3){$\bullet$}
\put(6,2){$\bullet$}
\put(7,1){$\bullet$}
\put(7,0){$\bullet$}
\put(2,14){$\bullet$}
\put(3,14){$\bullet$}
\put(4,13){$\bullet$}
\put(5,13){$\bullet$}
\put(6,12){$\bullet$}
\put(7,12){$\bullet$}
\put(8,11){$\bullet$}
\put(9,11){$\bullet$}
\put(10,10){$\bullet$}
\put(11,10){$\bullet$}
\put(12,9){$\bullet$}
\put(13,9){$\bullet$}
\put(14,8){$\bullet$}
\put(15,8){$\bullet$}
\end{picture}\nonumber
\end{center}\end{figure}
How to Compile Some\\
NAND Formula Evaluators}

\author{Robert R. Tucci\\
        P.O. Box 226\\
        Bedford,  MA   01730\\
        tucci@ar-tiste.com}

\date{ \today}

\maketitle

\vskip .01 cm
\section*{Abstract}

We say a unitary operator
acting on a set of qubits has been
compiled if
it has been expressed
as a SEO (sequence of elementary
operations, like CNOTs and single-qubit operations).
SEO's are often represented as quantum circuits.
arXiv:quant-ph/0702144
by Farhi-Goldstone-Gutmann
has inspired a recent flurry of papers,
that
propose quantum algorithms for
evaluating NAND formulas
via quantum walks over tree graphs.
These algorithms
use
two types
of unitary evolution:
oracle and non-oracle.
Non-oracle evolutions
are independent of the
NAND formula input,
whereas oracle evolutions depend
on this input.
In this paper we
compile (i.e., give
explicit SEOs and their
associated quantum circuits for)
the oracle and non-oracle evolution operators
used in some of these NAND formula
evaluators. We consider
here only the case of
balanced binary NAND trees.
Our compilation methods are based
on the CSD (Cosine Sine Decomposition),
a matrix decomposition from Linear Algebra.
The CS decomposition has been
used very successfully in the past
to compile
unstructured unitary matrices exactly.

\section{Introduction and Motivation}
(The picture on the front page
is an incidence matrix for a
binary tree. Empty squares
indicate zero entries
and
bullets indicate
non-zero entries.)

Recently,
Farhi-Goldstone-Gutmann
proposed in FGG07
\cite{FGG07}
a quantum algorithm
for evaluating certain boolean formulas
$\phi:\{0,1\}^\nlvs\rarrow \{0,1\}$
on $\nlvs$ variables.
FGG07 considers only
those $\phi()$
representable as balanced
binary
 NAND trees. These are
perfectly bifurcating trees,
with a NAND at every non-leaf node,
with $\nlvs$ leaves and $\log_2(\nlvs)$ levels,
where each leaf represents a different input
variable of $\phi()$.
The FGG07 algorithm
evaluates these
balanced binary NAND formulas
in time $\calo(\sqrt{\nlvs})$.
The FGG07 algorithm uses
a continuous time quantum walk.
The input to the NAND formula
is entered into the quantum
walk by means of an
``oracle Hamiltonian",
which is applied constantly and
continuously during the quantum walk.
One describes this situation as
a {\bf continuous querying of the oracle}.

A few days after FGG07 appeared
at ArXiv,
Cleve et al
published Cle07 \cite{Cleve-quick}
 at the same library.
This short two page paper points out that
one can translate the FGG07 algorithm
to a counterpart algorithm.
The counterpart
algorithm
enters the input
into the quantum walk
during
a finite number of
interruptions or queries.
This alternative method
of entering the input is
described as a {\bf discrete querying
of the oracle}.
Cle07 shows that
the counterpart
algorithm
will yield a solution to a
balanced binary NAND formula after
$\calo((\nlvs)^{\frac{1}{2}+\epsilon})$
queries, for any $\epsilon>0$.

Next, Chi07 \cite{caltech}, by
Childs et al, appeared on ArXiv.
Chi07
proposes an FGG07-inspired,
discrete-queries algorithm
for evaluating arbitrary (not necessarily balanced
or binary) NAND formulas. The Chi07 algorithm,
like the Cle07 one, requires
$\calo((\nlvs)^{\frac{1}{2}+\epsilon})$
queries to evaluate a NAND formula.
It uses a quantum walk on a
tree graph, with
a tail of nodes attached to the root node
of the tree.
In contrast, the
FGG07 algorithm
attaches
a runway to the tree. The runway is
a finite line of nodes
attached at its middle node to the root
node of the tree. The Chi07 algorithm
uses quantum phase estimation
to
deduce the value of the NAND formula.
In contrast,
FGG07 uses a scattering
experiment to deduce this value.

Next,
Amb07\cite{Amb}, by Ambainis, appeared
on ArXiv.
Amb07 describes an
FGG07-inspired, discrete-queries algorithm
that can evaluate balanced binary NAND formulas
with
$\calo((\nlvs)^{\frac{1}{2}})$
queries.
Thus, the Amb07 algorithm improves
over
Cle07 and Chi07
by removing the $\epsilon$.
Amb07 also gives an algorithm that
allows one to evaluate
arbitrary NAND formulas
with
$\calo((\nlvs)^{
\frac{1}{2}+
\calo(\frac{1}{\sqrt{\log_2\nlvs}})
})$ queries.

We say a unitary operator
acting on a set of qubits has been
compiled if
it has been expressed
as a SEO (sequence of elementary
operations, like CNOTs and single
qubit operations).
SEO's are often represented as quantum circuits.

This paper, like Cle07, Chi07 and Amb07,
considers FSS07-inspired algorithms with
discrete querying.
Such algorithms
use
two types
of unitary evolution:
oracle and non-oracle.
Non-oracle evolutions
are independent of the
NAND formula input,
whereas oracle evolutions depend
on this input.
The goal of this  paper is
to compile these
oracle and non-oracle
evolutions.
We present explicit SEO's
and their associated
circuits, for these evolutions.
Such circuits will be required
for most physical implementations of
the algorithms proposed in
Cle07, Chi07 and Amb07, but are
not given by those papers. Such
missing circuits constitute a large
gap in our practical knowledge
that this paper is intended
to fill.

Some papers (for example,
Refs.\cite{Cleve-two-trees,
Cleve-using-Suz,Childs-thesis,Hines})
have previously addressed the issue of
compiling
quantum walks over trees and other graphs.
Some of these papers use oracles to encode
the graph topology. Our
approach is different.
We do not use any oracles
to encode the graph topology.

This paper considers
only balanced binary NAND trees,
although I think some of the methods
of this paper can be applied to more
general trees.

Our compilation methods are based
on the CSD (Cosine Sine
Decomposition)\cite{Golub},
a matrix decomposition from Linear Algebra.
This decomposition is
explained in Section \ref{sec-csd}.
For now, suffice it to say
that the CS decomposition has
been used very successfully in the past
to compile
unstructured unitary matrices
exactly.

\section{Notation}

In this section, we will
define some notation that is
used throughout this paper.
For additional information about our
notation, we recommend that
the reader
consult Ref.\cite{Paulinesia}.
Ref.\cite{Paulinesia} is
a review article, written
by the author of this paper, which
uses the same notation as this paper.

We will often
use the symbol $\nb$ for the number ($\geq 1$) of qubits and
$\ns = 2^\nb$ for the number of states with $\nb$ qubits.
The quantum computing literature
often uses $n$ for $\nb$ and $N$
for $\ns$, but we will avoid this
notation. We prefer to use $n$
for the number operator, defined below.

Let $Bool =\{0, 1\}$. As usual, let $\ZZ, \RR, \CC$ represent the set
of integers (negative  and non-negative),
real numbers, and
complex numbers, respectively.
For integers $a$, $b$
 such that $a\leq b$, let
$Z_{a,b}=\{a, a+1,
\ldots b-1, b\}$.
For $\Gamma$ equal to $\ZZ$ or $\RR$, let
$\Gamma^{>0}$ and  $\Gamma^{\geq 0}$
represent the set of
positive
and
non-negative $\Gamma$
numbers, respectively.
For any positive integer $k$
and any set $S$, let
$S^k$ denote
the Cartesian product of
$k$ copies of $S$; i.e.,  the set of
all $k$-tuples
of elements of $S$.

We will use $\theta(S)$
to represent the ``truth function";
$\theta(S)$ equals 1 if statement $S$ is true
and 0 if $S$ is false.
For example, the Kronecker delta
function is defined by
$\delta^y_x=\delta(x,y) = \theta(x=y)$.

Let
$\overline{0}=1$
and $\overline{1}=0$.
If $\veca = a_{\nb-1} \ldots a_2 a_1 a_0$,
where $a_\mu\in Bool$, then
$dec(\veca) = \sum^{\nb-1}_{\mu=0} 2^\mu a_\mu=a$.
Conversely, $\veca=bin(a)$.

We define the single-qubit states $\ket{0}$ and $\ket{1}$ by

\beq
\ket{0} =
\left[
\begin{array}{c}
1 \\ 0
\end{array}
\right]
\;\;,\;\;
\ket{1} =
\left[
\begin{array}{c}
0 \\ 1
\end{array}
\right]
\;.
\eeq
If $\veca \in Bool^{\nb}$, we define the
$\nb$-qubit state $\ket{\veca}$ as the following tensor product

\beq
\ket{\veca} = \ket{a_{\nb -1}}
\otimes \ldots \ket{a_1} \otimes \ket{a_0}
\;.
\eeq
For example,

\beq
\ket{01} =
\left[
\begin{array}{c}
1 \\ 0
\end{array}
\right]
\otimes
\left[
\begin{array}{c}
0 \\ 1
\end{array}
\right]
=
\left[
\begin{array}{c}
0 \\ 1 \\ 0 \\0
\end{array}
\right]
\;.
\eeq

When we write a matrix, and
leave some of its entries
blank, those blank entries
should be interpreted as zeros.

$I_k$ and $0_k$ will represent the
$k\times k$ unit and zero matrices, respectively.
For any matrix $A\in\CC^{p\times q}$,
$A^*$ will stand for its complex
conjugate, $A^T$ for its transpose, and
$A^\dagger$ for its Hermitian conjugate.
For any two same-sized square matrices $A$ and $B$,
 we define the o-dot product $\odot$  by
 $A \odot B = A B A^{\dagger}$.

For any matrix $A$ and positive integer $k$,
let

\beq
A^{\otimes k} =
\underbrace{A\otimes \cdots
\otimes A \otimes A}_{k \mbox{\tiny\;\;copies of } A}
\;,
\eeq

\beq
A^{\oplus k} =
\underbrace{A\oplus \cdots
\oplus A \oplus A}_{k \mbox{\tiny\;\;copies of } A}
\;.
\eeq

Suppose $\beta\in Z_{0,\nb-1}$ and
$M$ is any $2\times 2$ matrix. We define
$M(\beta)$ by

\beq
M(\beta) =
I_2 \otimes
\cdots \otimes
I_2 \otimes
M \otimes
I_2 \otimes
\cdots \otimes
I_2
\;,
\label{eq-m-beta-def}
\eeq
where the matrix $M$ on the right
hand side is located
at qubit position $\beta$ in the tensor product
of $\nb$ $2\times 2$ matrices.
The numbers that label qubit positions in the
tensor product increase from
right to left ($\leftarrow$),
and the rightmost qubit is taken
to be at position 0.

The Pauli matrices
are

\beq
\sigx=
\left(
\begin{array}{cc}
0&1\\
1&0
\end{array}
\right)
\;,
\;\;
\sigy=
\left(
\begin{array}{cc}
0&-i\\
i&0
\end{array}
\right)
\;,
\;\;
\sigz=
\left(
\begin{array}{cc}
1&0\\
0&-1
\end{array}
\right)
\;.
\eeq
Let $\vec{\sigma} = (\sigx, \sigy , \sigz)$.
For any $\veca\in\RR^3$,
let $\sigma_{\veca}=\vec{\sigma}\cdot\veca$.

The one-qubit Hadamard
matrix $H$ is defined as:

\beq
H= \frac{1}{\sqrt{2}}
\left[
\begin{array}{cc}
1&1\\
1&-1
\end{array}
\right]
\;.
\eeq
The $\nb$-qubit Hadamard matrix is defined
as $H^{\otimes \nb}$.

The number operator $n$ for
a single qubit is defined by

\beq
n =
\left[
\begin{array}{cc}
0 & 0 \\
0 & 1
\end{array}
\right]
=
\frac{ 1 - \sigz}{2}
\;.
\eeq
Note that

\beq
n \ket{0} = 0\ket{0} = 0
\;\;,\;\;
n \ket{1} = 1\ket{1}
\;.
\eeq
We will often use $\nbar$ as shorthand for

\beq
\nbar =
1-n =
\left[
\begin{array}{cc}
1 & 0 \\
0 & 0
\end{array}
\right]
=
\frac{ 1 + \sigz}{2}
\;.
\eeq
Define $P_0$ and $P_1$ by

\beq
P_0 = \nbar =
\left[
\begin{array}{cc}
1 & 0 \\
0 & 0
\end{array}
\right]=
\ket{0}\bra{0}
\;\;,\;\;
P_1 = n =
\left[
\begin{array}{cc}
0 & 0 \\
0 & 1
\end{array}
\right]
=
\ket{1}\bra{1}
\;.
\eeq
Two other related $2\times 2$ matrices are

\beq
\sigm =
\left[
\begin{array}{cc}
0 & 0 \\
1 & 0
\end{array}
\right]=
\ket{1}\bra{0}
\;\;,\;\;
\sigp=
\left[
\begin{array}{cc}
0 & 1 \\
0 & 0
\end{array}
\right]
=
\ket{0}\bra{1}
\;.
\eeq
$P_0$ and $P_1$ are orthogonal projection
operators and they add to one:

\beq
P_a P_b = \delta(a, b) P_b
\;\;\;\;\; {\rm for} \;\; a,b\in Bool
\;,
\eeq

\beq
P_0 +  P_1 = I_2
\;.
\eeq

For $\veca \in Bool^\nb$, let

\beq
P_{\veca} = P_{a_{\nb-1}} \otimes \cdots
\otimes P_{a_2} \otimes P_{a_1} \otimes P_{a_0}
\;.
\eeq
For example,
with 2 qubits we have

\beq
P_{00} = P_0 \otimes P_0 = diag(1, 0, 0, 0)
\;,
\eeq

\beq
P_{01} = P_0 \otimes P_1 = diag(0, 1, 0, 0)
\;,
\eeq

\beq
P_{10} = P_1 \otimes P_0 = diag(0, 0, 1, 0)
\;,
\eeq

\beq
P_{11} = P_1 \otimes P_1 = diag(0, 0, 0, 1)
\;.
\eeq
Note that

\beq
P_\veca P_\vecb = \delta(\veca, \vecb) P_\vecb
\;\;\;\;\; {\rm for} \;\; \veca,\vecb\in Bool^\nb
\;,
\eeq

\beq
\sum_{\veca\in Bool^\nb }
P_{\veca} =
I_2 \otimes I_2 \otimes \cdots \otimes I_2 = I_{2^\nb}
\;.
\eeq

If $\veca\in Bool^\nb$, and $\beta\in Z_{0,\nb-1}$,
then we will denote
$\sum_{a_\beta\in Bool}P_{\veca}$
by replacing the symbol $a_\beta$
in $P_{\veca}$ by a dot. For example,
$\sum_{a_1\in Bool}P_{a_2,a_1,a_0}=
P_{a_1,\cdot,a_0}$

Next we explain our circuit diagram notation.
We label single qubits (or qubit
positions) by a Greek letter or by an
integer. When we use integers,
the topmost qubit wire is 0, the next one
down is 1, then
2, etc.
{\it Note that in our circuit diagrams,
time flows from the right to the left
of the diagram.} Careful:
Many workers in Quantum
Computing draw their diagrams
so that time flows from
left to right. We eschew their
convention because
it forces one to reverse
the order of the operators
every time one wishes to convert
between a circuit
diagram
and its algebraic equivalent
in Dirac notation.

$h.c.$ will stand for
Hermitian conjugate. For example,
$A + h.c.$ will denote $A + A^\dagger$,
and
$\left[\begin{array}{cc} & h.c.
\\A&\end{array}\right]$ will denote
$\left[\begin{array}{cc} & A^\dagger
\\A&\end{array}\right]$.
This notation is useful when
$A$ is a long expression that
we do not wish to repeat.

SVD will stand for Singular Value Decomposition.
An SVD of a matrix $A$
consists of  matrices $U,V,\Delta$
such that
$A=U \Delta V^\dagger$,
where the matrices $U,V$ are unitary
and $\Delta$ is a non-negative diagonal
matrix. We will say that this SVD is
``one-sided" if one ``side"
equals the identity matrix;
i.e., if $U=1$ or $V=1$.

When we use $\prod_{i=1,3,2}Q_{i}$
for non-commuting operators $Q_i$,
we mean $Q_1 Q_3, Q_2$, not $Q_1 Q_2 Q_3$
or any other permutation. Thus,
the multiplication must be performed
in the order indicated.

\section{Review of Some Prerequisites}
\subsection{Circulant Matrices}\label{sec-circulant}

In this section, we will review
some properties of
circulant matrices\cite{circulant}.
These properties will be
used later on, to compile
the evolution
operator for a random walk on a
closed loop.

A {\bf circulant matrix}
 is any
matrix of the form

\beq
C =
\left[
\begin{array}{cccccc}
c_0 & c_1 & c_2 & \cdots & c_{n-2} & c_{n-1}\\
c_{n-1} & c_0 & c_1 & \cdots & c_{n-3} & c_{n-2}\\
c_{n-2} & c_{n-1} & c_0 & \cdots & c_{n-4} & c_{n-3}\\
\vdots&\vdots&\vdots& &\vdots&\vdots\\
c_2 & c_3 & c_4 & \cdots & c_{0} & c_{1}\\
c_1 & c_2 & c_3 & \cdots & c_{n-1} & c_{0}
\end{array}
\right]
\;,
\eeq
for $n=1,2,3,\ldots$ and
$c_j\in\CC$.
Note that each row of $C$ is a
cyclic shift of the
row above it.

If we denote the eigenvalues
and eigenvectors of $C$ by

\beq
Cv = \lambda v
\;,
\eeq
then it is easy to show
(see Ref.\cite{circulant})
that the $n$ eigenvalues and
corresponding
eigenvectors of $C$ are given by

\beq
\lam_m = \sum_{k=0}^{n-1} c_k (\omega^m)^k
\;\;,\;\;
v^{(m)} =
\left[
\begin{array}{c}
1\\
\omega^m\\
(\omega^m)^2\\
\vdots\\
(\omega^m)^{n-1}
\end{array}
\right]
\;,\;\;{\rm where}\;\;
\omega = e^{-i \frac{2\pi}{n}}
\;,
\eeq
for $m\in Z_{0,n-1}$.
Define matrices $U$ and $D$ by

\beq
U= (v^{(0)},v^{(1)},v^{(2)},\cdots,v^{(n-1)})
\;,\;\;
D= diag(\lam_0,\lam_1,\lam_2,\cdots,\lam_{n-1})
\;.
\eeq
Then

\beq
CU= UD
\;,\;\;
C= U D U^\dagger
\;.
\eeq
The matrix $U$ is in
fact the discrete Fourier transform (DFT) matrix.

In this paper, we will
consider the case
 that $c_1=c_{n-1}=\coco$
and all other $c_k$ are zero.
In this case,

\beq
\lam_m = \coco(\omega^m + (\omega^m)^{n-1})=
\coco(\omega^m + \omega^{-m})=
\coco 2 \cos(\frac{2\pi m}{n})
\;.
\eeq

\subsection{Trotter Rescaling,
Lie and Suzuki Approximants}

In this section, we will review the
Trotter rescaling of the Lie and
Suzuki approximants of $e^{A+B}$.
More information
about this topic may be
found in Ref.\cite{HatSuz}.

Suppose $A,B\in\CC^{n\times n}$
and $t\in\RR$.
Define
\beq
L_1(t) = e^{tA}e^{tB}
\;.
\eeq
In this paper, we will refer to
$L_1(t)$ as the
{\bf Lie first-order approximant}
of $e^{t(A+B)}$.
We call it a first-order
approximant because, for small $t$,
according to the Baker-Campbell-Hausdorff
expansion\cite{Zachos, Reinsch},

\beq
L_1(t)=e^{t(A+B) +
\frac{t^2}{2}[A,B] + \calo(t^3)}
= e^{t(A+B)} + \calo(t^2)
\;.
\eeq

\begin{figure}[h]
    \begin{center}
    \epsfig{file=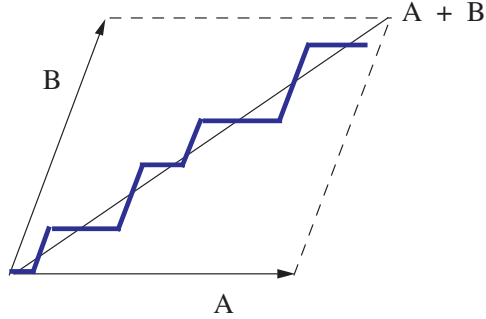, height=1.75in}
    \caption{Lie algebra ``Physicist's picture" of Trotter
    rescaling. The system moves from 0 to $A + B$,
by moving in small increments in the $A$ and $B$
directions.
    }
    \label{fig-trotter}
    \end{center}
\end{figure}

But what if $t$ isn't small?
Even when $t$
is not small, one can still use
the Lie approximant to approximate
$e^{t(A + B)}$. Indeed,
if $N$ is a very large integer, then

\beqa
L_1^N(\frac{t}{N})&=&
\left(
e^{\frac{t}{N}A}
e^{\frac{t}{N}B}
\right)^N
\\
&=&
\left(
e^{\frac{t}{N}(A+B)
+ \frac{t^2}{2N^2}[A,B] + \calo(\frac{t^3}{N^3})}
\right)^N
\\
&=&
e^{t(A+B)
+ \frac{t^2}{2N}[A,B] + \calo(\frac{t^3}{N^2})}
\\
&=&
e^{t(A+B)}
+ \calo(\frac{t^2}{N})
\;.
\eeqa
Henceforth, will refer
to this nice trick as
a {\bf Trotter rescaling} of an
approximant (in this case, the Lie
approximant). See Fig.\ref{fig-trotter}.

Next define
\beq
S_2(t) = e^{t\frac{A}{2}}e^{tB}e^{t\frac{A}{2}}
\;.
\eeq
We will refer to $S_2(t)$ as
the {\bf Suzuki second-order approximant}
.
One can show\cite{Zachos} that
for small $t$:

\beq
S_2(t)=e^{t(A+B) +
\frac{t^3}{6}[\frac{A}{2}+B,[B,A]] + \calo(t^5)}
= e^{t(A+B)} + \calo(t^3)
\;.
\eeq
Suzuki also defined
higher order approximants
based on $S_2(t)$.
For $k=1,2,3,\ldots$ , define
the {\bf Suzuki $(2k+2)$th-order approximant}
$S_{2k+2}(t)$ by

\beq
S_{2k+2}(t) =
S_{2k}^2(a_{2k} t)
S_{2k}((1-4a_{2k})t))
S_{2k}^2(a_{2k} t)
\;,
\eeq
for some $a_{2k}\in \RR$.
It is possible to show that for
$k=1,2,3,
\dots$ and small $t$:

\beq
S_{2k+2}(t)=
e^{t(A+B)}
+ \calo(t^{2k+3})
\;\;{\rm ,\;\;if}\;\;
a_{2k} = \frac{1}
{4 - 4^{\frac{1}{2k+1}}}
\;.
\eeq
$(a_{2k})_{k=1,2,3,\ldots}$
is a monotone decreasing sequence
with $a_2 = 0.4145...$ and $\lim_{k\rarrow \infty} a_{2k} = 1/3$.

As with the Lie approximant,
is possible to do a Trotter
rescaling of the Suzuki approximants.
One finds that for $k=1,2,3\dots$,
large $N$ and fixed $t$:

\beq
S_{2k}^N(\frac{t}{N})=
e^{t(A+B)}
+ \calo(\frac{t^{2k+1}}{N^{2k}})
\;.
\eeq

Henceforth, what we
have called $N$ so far in this section
will be renamed $N_T$, to distinguish it
from all the other $N$'s used in this paper.
The $T$  in $N_T$ stands
for Trotter, since it represents the number
of Trotter time slices (trots?).

If we call $N_{exp}$ the number
of factors of the type $e^{\tau A}$
or $e^{\tau B}$ for some $\tau\in \RR$, then we can compare
$N_{exp}$ and the error for the Trotterized
Lie and Suzuki approximants\footnote{
If $N_{exp}(S_{2k})$
is the number of exponentials in
$S_{2k}$,
then $N_{exp}(S_2)=3$,
$N_{exp}(S_4)=11$, $N_{exp}(S_6)=51$.
In general,
$N_{exp}(S_{2k+2})= N_{exp}(S_{2k})5-4$
for $k=1,2,\ldots$.
Solving the difference equation
$f(k+1)=f(k)5 -4$ with $f(1)=3$
yields $f(k)=2(5)^{k-1}+1$.
Finally, note that
$N_{exp}^{T-S}=f(k)N_T-(N_T-1)$.}:

\begin{center}
\begin{tabular}{l||c|c|}
& $N_{exp}$ & error\\ \hline\hline
$L_1^{N_T}(\frac{t}{N_T})$& $2N_T$ & $\calo(\frac{t^2}{N_T})$\\ \hline
$S_{2k}^{N_T}(\frac{t}{N_T})$& $2(5)^{k-1}N_T + 1$ & $\calo(\frac{t^{2k+1}}{N_T^{2k}})$
\\ \hline
\end{tabular}
\end{center}

Let $N_{exp}^{T-S}$ stand for the
number of exponentials in
the Trotterized Suzuki
approximant.
Define
\beq
\epsilon=\frac{t^{2k+1}}{N_T^{2k}}
\;.
\label{eq-eps-def}
\eeq
Eq.(\ref{eq-eps-def}) for $\epsilon$
can be inverted to get
$N_T= \frac{t^{1+\frac{1}{2k}}}{\epsilon^{\frac{1}{2k}}}$.
This expression for $N_T$ as a function of $\epsilon$
can be substituted into
the value for $N_{exp}^{T-S}$
given in the above table, to get:

\beq
N_{exp}^{T-S}=
2(5)^{k-1}
\frac{t^{1+\frac{1}{2k}}}{\epsilon^{\frac{1}{2k}}}
+1
\;.
\eeq
The previous expression for
$N_{exp}^{T-S}$ has a minimum for small $\epsilon$.
For example, suppose $\epsilon=10^{-12}$.
For $k=1$, $N_{exp}^{T-S}\sim\epsilon^{\frac{-1}{2k}}=10^{6}$.
For $k=2$, $N_{exp}^{T-S}\sim\epsilon^{\frac{-1}{2k}}=10^{3}$
As $k\rarrow \infty$, $N_{exp}^{T-S}\rarrow
2(5)^{k-1}t\rarrow \infty$. It is
possible to use calculus to obtain
the value of $k$ that minimizes
$N_{exp}^{T-S}$. One finds\footnote{Ref.\cite{Cleve-using-Suz}
does a similar minimization for
a more general Suzuki approximant
that approximates $e^{i(A_1 + A_2+\ldots+A_m)}$
where $m\geq 2$. We will only use $m=2$ in this paper.}

\beq
k \approx \sqrt{\frac{\ln(\frac{t}{\epsilon})}{2\ln(5)}}
\;,\;\;
N_{exp}^{T-S}\approx
\frac{2t}{5}e^{ [2\ln(5)\ln(\frac{t}{\epsilon})]^\frac{1}{2}}
\;.
\label{eq-best-nexp}
\eeq

\subsection{Multiply Controlled NOTs}
In this section, we will review
how to compile
multiply controlled $U(2)$ operators.
More information about this topic may be
found in Ref.\cite{Bar}.

Suppose $U\in U(2)$.
If $\tau$ and $\kappa$ are two different
qubit positions, gate
$U(\tau)^{n(\kappa)}$ (or $U(\tau)^{\nbar(\kappa)}$)
is called a {\bf controlled $U$}
with target $\tau$ and control $\kappa$.
When $U=\sigx$, this reduces to a
CNOT (controlled NOT).
 If
 $\tau$,$\kappa_1$ and $\kappa_0$
are 3 different qubit positions,
$\sigx(\tau)^{n(\kappa_1)n(\kappa_0)}$
is called a {\bf Toffoli gate} with target
$\tau$ and controls $\kappa_1, \kappa_0$.
Suppose $N_K\geq 2$ is an integer and
$\vec{b}\in Bool^{N_K}$.
Suppose
$\tau, \kappa_{N_K-1},\kappa_{N_K-2},
\ldots,\kappa_1,\kappa_0$ are distinct qubits
and $\vec{\kappa}=
(\kappa_{N_K-1},\kappa_{N_K-2},
\ldots,\kappa_1,\kappa_0)$.
Gate
$U(\tau)^{P_{\vec{b}}
(\vec{\kappa})}$
is called a {\bf multiply
controlled $U$}
with target $\tau$ and $N_K$
controls
$\vec{\kappa}$.
When $U=\sigx$, this reduces to an
{\bf MCNOT (multiply controlled
NOT)}.

Our goal in this section is to show
a reasonably efficient way of compiling
any multiply controlled $U(2)$ operator;
i.e., to express it as a SEO consisting
of CNOTs and single-qubit operations.

For any $U\in U(2)$, a
multiply controlled $U$
can be
expressed in terms of  MCNOTs as follows:

\beq
\begin{array}{c}
\Qcircuit @C=1em @R=.25em @!R{
&\dotgate&\qw
\\
&\dotgate&\qw
\\
&\dotgate&\qw
\\
&\dotgate&\qw
\\
&\qw&\rstick{\ket{0}}\qw
\\
&\gate{U}\qwx[-5]&\qw
}
\end{array}
\;\;\;\;\;\;\;=
\begin{array}{c}
\Qcircuit @C=1em @R=.25em @!R{
&\dotgate&\qw&\dotgate&\qw
\\
&\dotgate&\qw&\dotgate&\qw
\\
&\dotgate&\qw&\dotgate&\qw
\\
&\dotgate&\qw&\dotgate&\qw
\\
&\timesgate\qwx[-4]&\dotgate&\timesgate\qwx[-4]&\rstick{\ket{0}}\qw
\\
&\qw&\gate{U}\qwx[-1]&\qw&\qw
}
\end{array}
\;.
\label{eq-contr-u}
\eeq
This requires
one ancilla qubit in state $\ket{0}$.
The ancilla is reusable.

Remember that our goal is to show how
to compile any multiply controlled $U(2)$ operator.
Now all we need to show is how to
compile MCNOTs.
MCNOTs
can be expressed in terms of
Toffoli gates as follows:

\beq
\begin{array}{c}
\Qcircuit @C=1em @R=1.25em @!R{
&\dotgate&\qw
\\
&\dotgate&\qw
\\
&\dotgate&\qw
\\
&\dotgate&\qw
\\
&\dotgate&\qw
\\
&\qw&\rstick{\ket{0}}\qw
\\
&\qw&\rstick{\ket{0}}\qw
\\
&\qw&\rstick{\ket{0}}\qw
\\
&\timesgate\qwx[-8]&\qw
}
\end{array}
\;\;\;\;\;\;=\;\;\;\;\;\;
\begin{array}{c}
\Qcircuit @C=1em @R=1.25em @!R{
\lstick{\p{0}}&\dotgate&\qw&\qw&\qw&\qw&\qw&\dotgate&\qw
\\
\lstick{\p{1}}&\dotgate&\qw&\qw&\qw&\qw&\qw&\dotgate&\qw
\\
\lstick{\p{2}}&\qw&\dotgate&\qw&\qw&\qw&\dotgate&\qw&\qw
\\
\lstick{\p{3}}&\qw&\qw&\dotgate&\qw&\dotgate&\qw&\qw&\qw
\\
\lstick{\p{4}}&\qw&\qw&\qw&\dotgate&\qw&\qw&\qw&\qw
\\
\lstick{\p{5}}&\timesgate\qwx[-5]&\dotgate&\qw&\qw&\qw&\dotgate&\timesgate\qwx[-5]&\rstick{\ket{0}}\qw
\\
\lstick{\p{6}}&\qw&\timesgate\qwx[-4]&\dotgate&\qw&\dotgate&\timesgate\qwx[-4]&\qw&\rstick{\ket{0}}\qw
\\
\lstick{\p{7}}&\qw&\qw&\timesgate\qwx[-4]&\dotgate&\timesgate\qwx[-4]&\qw&\qw&\rstick{\ket{0}}\qw
\\
\lstick{\p{8}}&\qw&\qw&\qw&\timesgate\qwx[-4]&\qw&\qw&\qw&\qw
}
\end{array}
\;.
\eeq
For $N_K\geq 3$ controls, this requires
$N_K-2$ ancilla qubits in state $\ket{0}$.
The ancillas are reusable.

So now all we need to show is how to
compile a Toffoli gate.
The following identity,
true for any $V\in U(2)$,
is useful in this regard:

\beq
\begin{array}{c}
\Qcircuit @C=1em @R=1em @!R{
&\dotgate&\qw
\\
&\dotgate&\qw
\\
&\gate{V}\qwx[-2]&\qw
}
\end{array}
=
\begin{array}{c}
\Qcircuit @C=1em @R=1em @!R{
&\qw&\dotgate&\qw&\dotgate&\dotgate&\qw
\\
&\dotgate&\timesgate\qwx[-1]&\dotgate&\timesgate\qwx[-1]&\qw&\qw
\\
&\gate{V^{\frac{1}{2}}}\qwx[-1]&\qw&\gate{V^{-\frac{1}{2}}}\qwx[-1]&\qw&\gate{V^{\frac{1}{2}}}\qwx[-2]&\qw
}
\end{array}
\;.
\eeq
When $V=\sigx$, this becomes:

\beq
\begin{array}{c}
\Qcircuit @C=1em @R=1em @!R{
&\dotgate&\qw
\\
&\dotgate&\qw
\\
&\gate{\sigx}\qwx[-2]&\qw
}
\end{array}
=
\begin{array}{c}
\Qcircuit @C=1em @R=1em @!R{
&\qw&\dotgate&\qw&\dotgate&\dotgate&\qw
\\
&\dotgate&\timesgate\qwx[-1]&\dotgate&\timesgate\qwx[-1]&\qw&\qw
\\
&\gate{\sigx^{\frac{1}{2}}}\qwx[-1]&\qw&\gate{\sigx^{\frac{1}{2}}}\qwx[-1]&\qw&\gate{\sigx^{\frac{1}{2}}}\qwx[-2]&\qw
}
\end{array}
\;.
\eeq

Thus, we can compile a Toffoli gate
if we can compile a singly
controlled $(\sigx)^{\frac{1}{2}}$.
A singly controlled $U(2)$ operator
also arose in Eq.(\ref{eq-contr-u}). So
we should show how to compile a general
singly controlled $U(2)$ operator.
This
can be done using the following quaternion
trick\cite{Tuc-dressed-nots}.
Observe that
an arbitrary $U\in U(2)$
need not have unit determinant. For
example, $\sigx$ and
$(\sigx)^{\frac{1}{2}}$ don't.
So we need to express
$U$ as a product of an
$SU(2)$ matrix and a phase:
$U= e^{i\phi} e^{i\theta\sigma_{\hat{n}}}$,
where $\hat{n}$ is a unit vector in $\RR^3$,
and $\phi, \theta\in \RR$.
Then use the fact that any
special unitary matrix
$e^{i\theta\sigma_{\hat{n}}}$
can always
be expressed as $\sigma_{\hat{a}}\sigma_{\hat{b}}$,
where
$\hat{a},\hat{b}$ are unit vectors in $\RR^3$.
Indeed,
$e^{i\theta\sigma_{\hat{n}}}=
\cos\theta + i\hat{n}\sin\theta$, and
$\sigma_{\hat{a}}\sigma_{\hat{b}}=
\hat{a}\cdot\hat{b} + i\hat{a}\times\hat{b}$,
so set
$\cos\theta=\hat{a}\cdot\hat{b}$
and $\hat{n}\sin\theta=\hat{a}\times\hat{b}$.
Thus

\beq
U=
e^{i\phi}\sigma_{\hat{a}}\sigma_{\hat{b}}
\;.
\eeq
Let $R$ and
$S$ be $SU(2)$ matrices that rotate
$\hat{a}$ and $\hat{b}$ into the x axis:

\beq
\sigma_{\hat{a}} = R\sigx R^\dagger
\;,\;\;
\sigma_{\hat{b}} = S\sigx S^\dagger
\;.
\eeq
Then

\beqa
\begin{array}{c}
\Qcircuit @C=1em @R=1em @!R{
&\dotgate&\rstick{\alpha}\qw
\\
&\gate{U}\qwx[-1]&\rstick{\beta}\qw
}
\end{array}
\;\;\;&=&
U(\beta)^{n(\alpha)}
\\
&=&
e^{i\phi n(\alpha)}
\sigma_{\hat{a}}(\beta)^{n(\alpha)}
\sigma_{\hat{b}}(\beta)^{n(\alpha)}
\\
&=&
\begin{array}{c}
\Qcircuit @C=1em @R=1em @!R{
&\gate{e^{i\phi n}}&\qw&\dotgate&\qw&\dotgate&\qw&\qw
\\
&\qw&\gate{R}&\timesgate\qwx[-1]&\gate{R^\dagger S}&\timesgate\qwx[-1]&\gate{S^\dagger}&\qw
}
\end{array}
\;.
\eeqa

Thus, we have achieved our goal
of showing how to compile any
multiply controlled $U(2)$ operator.

Let $G$ be any multiply controlled
$U(2)$ gate
with $N_K$
controls, where $N_K\geq 1$.
A consequence of the results of this section
is that
$G$
can always be expressed as a
SEO with $\calo(N_K)$ CNOTs.

\subsection{CS Decomposition}\label{sec-csd}

In this section, we will
define the CSD (Cosine Sine Decomposition)\cite{Golub}
and review how it has been applied
to quantum computing prior to this
paper.

Suppose that $U$ is an $N\times N$
unitary matrix, where $N$ is an even number.
The CSD Theorem
states\footnote{Actually, this
is only a special case of the CSD
Theorem---the case which is
most relevant to quantum computing.
The general version of the
CSD Theorem does not restrict
the dimension of $U$ to be even,
or even restrict the blocks
into which $U$ is
partitioned to be of
equal size.}
that one can always
express $U$ in the form

\begin{subequations}
\label{eq-csd-long}
\beq
U =
\left[
\begin{array}{cc}
L_0&0\\
0&L_1
\end{array}
\right]
D
\left[
\begin{array}{cc}
R_0&0\\
0&R_1
\end{array}
\right]
\;,
\eeq
where the left and right matrices
$L_0,L_1,R_0,R_1$ are
$\frac{N}{2}\times \frac{N}{2}$
unitary matrices, and

\beq
D=
\left[
\begin{array}{cc}
D_{00}&D_{01}\\
D_{10}&D_{11}
\end{array}
\right]
\;,
\eeq

\beq
D_{00}=D_{11}=diag(C_1,C_2, \dots,C_{\frac{N}{2}})
\;,
\eeq

\beq
D_{01}=diag(S_1,S_2, \dots,S_{\frac{N}{2}})
\;,\;\;D_{10}=-D_{01}
\;.
\eeq
\end{subequations}
For all $i\in Z_{1,\frac{N}{2}}$,
\;\;$C_i=\cos\theta_i$
and
$S_i=\sin\theta_i$
for some angle $\theta_i$.
Eqs.(\ref{eq-csd-long}) can be expressed more
succinctly as

\beq
U=(L_0\oplus L_1)
e^{i\sigy\otimes \Theta}
(R_0\oplus R_1)
\;,
\eeq
 where
$\Theta=diag(\theta_1,
\theta_2, \dots, \theta_{\frac{N}{2}})$.

\begin{figure}[h]
    \begin{center}
    \epsfig{file=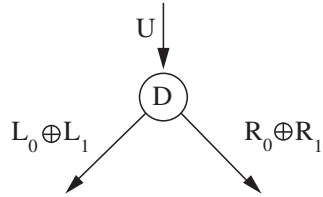, height=1.0in}
    \caption{Diagrammatic representation of
    CS decomposition, Eq.(\ref{eq-csd-long}).
    }
    \label{fig-csd-once}
    \end{center}
\end{figure}

Fig.\ref{fig-csd-once} is a diagrammatic
representation of the CSD. Note that:
(1)Matrix $U$ is assigned to the incoming arrow.
(2)Matrix $D$ is assigned to the node.
(3)Matrices $R_0\oplus R_1$ and
$L_0\oplus L_1$ are each assigned
to an outgoing arrow.

\begin{figure}[h]
    \begin{center}
    \epsfig{file=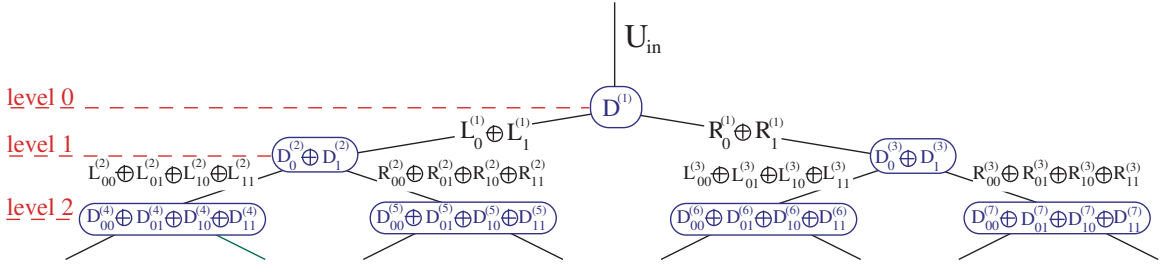, height=1.35in}
    \caption{A CSD binary tree.
It arises from the recursive application
of the CSD.}
    \label{fig-csd-tree}
    \end{center}
\end{figure}

The CS decomposition was first used
for quantum compiling
in Tuc99\cite{Tuc99}.
In the Tuc99 compiling algorithm,
the CSD is used recursively.
A nice way of picturing this
recursive use of the CSD
is to represent each CSD application
by a node, as in Fig\ref{fig-csd-once}.
The recursion connects these nodes
so as to form a
binary tree, as shown in
Fig.\ref{fig-csd-tree}.
In Fig.\ref{fig-csd-tree},
we start with an initial
unitary matrix $U_{in}$
entering the root node, which we define as level 0.
Without loss of generality, we can assume that
the dimension of $U_{in}$ is $2^\nb$ for some $\nb\geq 1$.
(If initially $U_{in}$'s dimension is not a power of 2,
 we replace it by a direct sum
$U_{in}\oplus I_r$ whose dimension is a power of two.)
We apply the CSD method to $U_{in}$.
This yields for level 0 a
D matrix $D_0^{(1)}$,
two unitary matrices
$L_0^{(1)}$ and $L_1^{(1)}$
on the left side and two unitary
matrices $R_0^{(1)}$ and $R_1^{(1)}$
on the right side.
Then we apply the CSD method to
each of the 4 matrices
$L_0^{(1)}, L_1^{(1)}, R_0^{(1)}$ and $R_1^{(1)}$
 that were produced
in the previous step. Then we apply the
CSD method to each of the 16 $R$ and $L$ matrices that were produced
in the previous step. And so on.
The nodes of level $\nb$
don't have $R, L$ arrows coming out
of them
since the $D$ matrices for those nodes
are all $1\times 1$.

Call a central matrix either (1) a single D matrix, or
(2) a direct sum $D_1 \oplus D_2 \oplus \cdots  \oplus D_r$ of D matrices,
or (3) a diagonal unitary matrix. From
Fig.\ref{fig-csd-tree}, it is clear that
the initial matrix $U_{in}$ can be expressed as a product of
central matrices,
with each node of the tree providing
one of the central matrices in the product.
We can use this factorization
of $U_{in}$
into central
matrices to compile
$U_{in}$, if we can find a
method for
decomposing any central matrix into a SEO.
Tuc99 gives such a method.

Let's say a few more words about
central matrices.
For simplicity, consider
matrices of size $2^\nb$
with $\nb=3$. Let
$R_y(\phi)=e^{i\phi\sigy}$.
An $8\times 8$ $D$ matrix
is of the form

\begin{subequations}
\label{eq-central-mats-3bits}
\beq
D=
\sum_{a,b\in Bool}
R_y(\phi_{ab})\otimes P_a\otimes P_b
\;,
\label{eq-d-3bits}
\eeq
for some $\phi_{ab}\in\RR$.
An $8\times 8$ matrix
which is a  sum of two $D$ matrices
 is of the form

\beq
D_0\oplus D_1=
\sum_{a,b\in Bool}
P_a\otimes R_y(\phi'_{ab})\otimes P_b
\;,
\label{eq-dd-3bits}
\eeq
for some $\phi'_{ab}\in\RR$.
An $8\times 8$ matrix
which is a  sum of
four $D$ matrices
 is of the form

\beq
D_{00}\oplus D_{01}\oplus D_{10}\oplus D_{11}
=
\sum_{a,b\in Bool}
P_a\otimes  P_b \otimes R_y(\phi''_{ab})
\;,
\label{eq-dddd-3bits}
\eeq
\end{subequations}
for some $\phi''_{ab}\in\RR$.
Thus, the $D$,
$D_0\oplus D_1$
and
$D_{00}\oplus D_{01}\oplus D_{10}\oplus D_{11}$
matrices
given by Eqs.(\ref{eq-central-mats-3bits})
only
differ by a qubit permutation
(apart from the fact that they have
different angles).
Note that $D$ of Eq.(\ref{eq-d-3bits}) can also
be written in the form

\begin{subequations}
\label{eq-d-mat-3-bits}
\begin{eqnarray}
D&=&
\exp\left(i\sum_{a,b\in Bool}\phi_{a,b}
\sigy\otimes P_a\otimes P_b\right)
\\&=&
\exp\left(i\sum_{a,b\in Bool}\phi_{a,b}
\sigy(2)P_a(1) P_b(0)\right)
\;.
\end{eqnarray}
\end{subequations}
For $\nb\geq 1$, Eq.(\ref{eq-d-mat-3-bits})
generalizes
to

\begin{subequations}
\label{eq-d-nb-bits}
\begin{eqnarray}
D&=&
\exp\left(i\sigy\otimes
\sum_{\vecb\in Bool^{\nb-1}}
\phi_\vecb P_\vecb\right)
\\&=&
\exp\left(i\sigy(\nb-1)
\sum_{\vecb\in Bool^{\nb-1}}
\phi_\vecb P_\vecb(\nb-2,\ldots,2,1,0)\right)
\;.
\end{eqnarray}
\end{subequations}
In general, a central matrix
acting on $\nb$ qubits
is
a matrix of the form
exhibited by $D$ in Eq.(\ref{eq-d-nb-bits}),
or a qubit permutation
thereof, or a diagonal unitary matrix.

In my papers that followed Tuc99,
I've begun calling a $D$ matrix
of the form Eq.(\ref{eq-d-nb-bits})
a ``multiplexor".\footnote{``multiplexor"
means ``multi-fold" in Latin.
A special type of electronic
device is also called a multiplexor
or multiplexer.}
When I want to be more
precise, I call it an
$R_y(2)$-multiplexor with target qubit $\nb-1$
and control qubits
$\nb-2,\dots,2,1,0$. The $R_y(2)$
term refers to the fact that
the set of operations acting
on the target qubit are $2\times 2$
qubit rotations $R_y(\phi)=e^{i\phi\sigy}$
for some $\phi\in\RR$. More
generally, one can speak
of $U(2)$-multiplexors. Henceforth in this paper,
I'll continue using this multiplexor
nomenclature, even though it's
not used in Tuc99.

\begin{figure}[h]
    \begin{center}
    \epsfig{file=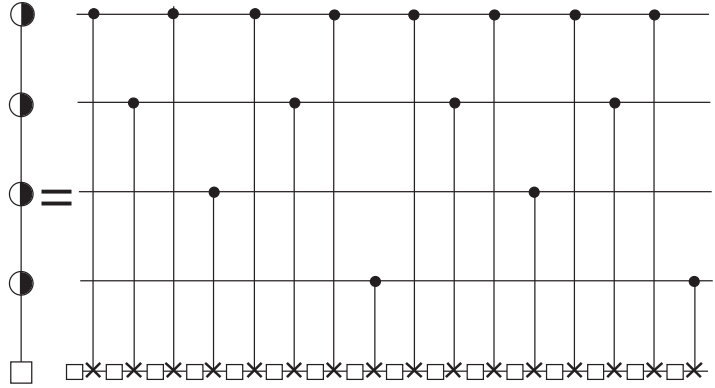, height=2.0in}
    \caption{
    A possible decomposition
    of an $R_y(2)$-multiplexor
    with 4 controls.
    }
    \label{fig-4controls}
    \end{center}
\end{figure}

Tuc99 gives identities for
decomposing an arbitrary
$R_y(2)$-multiplexor
and a diagonal unitary
matrix into a SEO with $2^\nb$ CNOTs.
Fig.\ref{fig-4controls} shows an example
of the SEO decomposition
found in Tuc99 for an $R_y(2)$-multiplexor. In Fig.\ref{fig-4controls},
0,1,2,3 are the control qubits,
and 4 is the target qubit. The empty
square vertices represent $R_y(2)$
gates. The symbol to the left of the equal sign,
the one with
the ``half-moon" vertices, was invented
by the authors of Ref.\cite{Hels04b}
 to represent
an $R_y(2)$-multiplexor.

The Tuc99 algorithm
is implemented
in a computer program called Qubiter,
available at Ref.\cite{qbtr}.

Later papers\cite{Mich04,Hels04b,Tuc-qbtr-mod}
have improved the Tuc99 algorithm.
Instead of growing the CSD tree
all the way down to where the
leaf nodes contain
diagonal unitary matrices, they stop
one level earlier,
when the leaf
nodes contain
$U(2)$-multiplexors.
Then they apply a nice technique,
due to Vidal and Dawson\cite{VD},
 for  expressing
arbitrary $U(2)$ matrices
using just 3 CNOTs.
They are able to express
an arbitrary
structureless $U(2^\nb)$ matrix
using $\frac{1}{2}4^\nb + \calo(2^\nb)$
CNOTs (and also some qubit rotations, of course).
It's easy to show\cite{bound}
that the number
of CNOTs needed to express a
structureless $U(2^\nb)$ matrix
is greater or equal to
$\frac{1}{4}(4^\nb-3\nb-1)$.
Roughly speaking,
in these algorithms, a
unitary matrix is decomposed into
$2^\nb$ $U(2)$-multiplexors, and
each of these multiplexors is expressed
as a SEO with $2^\nb$ CNOTs.

Note that quantum compiling papers
usually measure circuit complexity
by counting the total number of CNOTs.
Others count the total
number of elementary gates.
One could also count just the total number of
control vertices in the circuit diagram.
These 3 measures are linearly related so
they have the same big O behavior
when expressed as a function of the
number of qubits.
It is expected that the quantum
computers of the future will be able to
perform
single-qubit operations
much faster than
CNOTs, so it makes sense to
ignore the single-qubit operations and
count only the CNOTs.

The Tuc99 algorithm and its variants
expect a
structureless unitary matrix as input. They
do not
assume any  symmetries in the structure
of the input matrix being decomposed.
But the CSD does have
some free degrees
of freedom (for example, certain phase
choices) which are chosen
arbitrarily.
The number of these degrees of freedom
increases as the symmetry of the input
matrix increases.
If the input matrix does have a
symmetric structure, it is
sometimes possible to
chose those free degrees of freedom in
a way that reduces the length of
the output SEO. For example,
Refs.\cite{Tuc-fft,Yumi}
show that the
Coppersmith decomposition\cite{Copper}
of the discrete Fourier transform
matrix
 can
be obtained
using the Tuc99 algorithm and
some judicious choices for the
free degrees of freedom.

\section{Some Identities for\\
Exponentials of Partitioned Matrices}

In this section, we present some
identities
satisfied by $\exp(iM)$,
where
$M$ is a Hermitian matrix
partitioned into four equal sized blocks.

Suppose $M_{00}$, $M_{01}$, $M_{10}$
and $M_{11}$ are square matrices of the
same dimension, and

\beq
M=\left[
\begin{array}{cc}
M_{00} & M_{01}\\
M_{10} & M_{11}
\end{array}
\right]
\;.
\label{eq-m-4blocks}
\eeq
We will say $M$ is {\bf block diagonal} if
$M_{10}=M_{01}=0$ and {\bf block anti-diagonal} if
$M_{00}=M_{11}=0$. We are interested
in the case were $M$ is Hermitian.

If the $M$ of Eq.(\ref{eq-m-4blocks})
is block diagonal, then
$\exp(iM) = \exp(iM_{00})\oplus\exp(iM_{11})$.
But suppose $M$ is block anti-diagonal.
What can we say about $\exp(iM)$ then?

\begin{Lemma}\label{lem-anti-d-svd}
Suppose $F\in \CC^{n\times n}$
and
$F = V \Delta U^\dagger$
is an SVD,
so $V,U$ are unitary matrices
and $\Delta$ is a non-negative diagonal
matrix. Then

\beq
\exp(i
\left[
\begin{array}{cc}
0 & F^\dagger \\
F & 0
\end{array}
\right]
)
=
\left[
\begin{array}{cc}
U & 0 \\
0 & V
\end{array}
\right]
\left[
\begin{array}{cc}
\cos(\Delta) & i\sin(\Delta) \\
i\sin(\Delta) & \cos(\Delta)
\end{array}
\right]
\left[
\begin{array}{cc}
U^\dagger & 0 \\
0 & V^\dagger
\end{array}
\right]
\;.
\label{eq-svd-exp-anti-block}
\eeq
\end{Lemma}
\proof
Let LHS and RHS stand for the left
and right hand sides of
Eq.(\ref{eq-svd-exp-anti-block}). Then

\beqa
LHS
&=&
\exp(i
\left[
\begin{array}{cc}
U & 0 \\
0 & V
\end{array}
\right]
\left[
\begin{array}{cc}
0 & \Delta \\
\Delta & 0
\end{array}
\right]
\left[
\begin{array}{cc}
U^\dagger & 0 \\
0 & V^\dagger
\end{array}
\right]
)
\\
&=&
\left[
\begin{array}{cc}
U & 0 \\
0 & V
\end{array}
\right]
\exp(i
\left[
\begin{array}{cc}
0 & \Delta \\
\Delta & 0
\end{array}
\right]
)\left[
\begin{array}{cc}
U^\dagger & 0 \\
0 & V^\dagger
\end{array}
\right]\\
&=& RHS
\;.
\eeqa

LHS  can still be exponentiated
without having to find the SVD of $F$.
See Appendix \ref{app-exp-anti-block}
if interested.
\qed

Suppose the $M$ of Eq.(\ref{eq-m-4blocks})
only has $M_{11}=0$, so it isn't fully block
diagonal.
What can we say about $\exp(iM)$ then?

\begin{Lemma}\label{lem-time-ordered}
Suppose $A,B\in \CC^{n\times n}$
and $A$ is a Hermitian matrix. Then

\beq
\exp(i
\left[
\begin{array}{cc}
A & B^\dagger \\
B & 0
\end{array}
\right]
)=
\exp(i
\left[
\begin{array}{cc}
A & 0 \\
0 & 0
\end{array}
\right]
)
\calt\left\{
\exp(i
\int_0^1 dt
\left[
\begin{array}{cc}
0 & h.c. \\
Be^{itA}& 0
\end{array}
\right]
)
\right\}
\;,
\eeq
where $\calt\{\}$ indicates a time-ordered exponential.
\end{Lemma}
{\bf first proof:}
Define the following three matrices:

\beq
Q =
\left[
\begin{array}{cc}
A & B^\dagger \\
B & 0
\end{array}
\right]
\;,\;\;
Q_A =
\left[
\begin{array}{cc}
A & 0\\
0 & 0
\end{array}
\right]
\;,\;\;
Q_B =
\left[
\begin{array}{cc}
0 & B^\dagger \\
B & 0
\end{array}
\right]
\;.
\label{eq-q-qa-qb}
\eeq
For any $t\in \RR$,
consider the $t$-dependent
matrix

\beq
x(t) = e^{-itQ_A} e^{itQ}
\;.
\label{eq-xt-def}
\eeq
The derivative of $x(t)$ is

\beqa
x'(t)&=&
e^{-itQ_A}
(-iQ_A + i Q)
e^{itQ}
\\
&=&(i e^{-itQ_A}\;Q_B\;e^{itQ_A})x(t)
\;.
\eeqa
Hadamard's identity\cite{Zachos}
tells us that

\beq
e^{-itQ_A}\;Q_B\;e^{itQ_A}=
e^{-it\ad{Q_A}}Q_B
\;.
\eeq
It is easy to check that for
$n=1,2,\ldots$,

\beq
\ad{Q_A}^n  Q_B =
\left[
\begin{array}{cc}
0 & A^n B^\dagger \\
B (-A)^n& 0
\end{array}
\right]
\;.
\eeq
Therefore,

\beqa
i e^{-itQ_A}\;Q_B\;e^{itQ_A}
&=& i \sum_{n=0}^{\infty}
\frac{(-it)^n}{n!}
\left[
\begin{array}{cc}
0 & A^n B^\dagger \\
B (-A)^n& 0
\end{array}
\right]\\
&=&
i
\left[
\begin{array}{cc}
0 & e^{-itA}B^\dagger\\
B e^{itA}& 0
\end{array}
\right]
\;.
\eeqa
Thus, $x(t)$ defined
by Eq.(\ref{eq-xt-def}) satisfies the differential
equation:

\beq
x'(t) =
i
\left[
\begin{array}{cc}
0 & e^{-itA}B^\dagger\\
B e^{itA}& 0
\end{array}
\right]
x(t)
\;.
\label{eq-xt-dif-eq-sp}
\eeq
This differential equation
for $x(t)$ is of the form

\beq
x'(t) = M(t) x(t)
\;,
\label{eq-x-m-diff-eq}
\eeq
where $M(t)$ is a
$t$-dependent matrix.
If we express the derivative
on the left hand side of
Eq.(\ref{eq-x-m-diff-eq})
in terms of infinitesimals,
we find

\beq
x(t+dt)\approx [1 + M(t)dt]x(t)
\approx e^{M(t)dt} x(t)
\;.
\label{eq-x-t-recur}
\eeq
Iterating Eq.(\ref{eq-x-t-recur}) gives:

\beq
x(t) = \calt\left\{
e^{\int_0^t dt' M(t')}\right\}
x(0)
\;.
\label{eq-sol-xt-dif-eq}
\eeq
Eq.(\ref{eq-sol-xt-dif-eq})
is the solution of any differential
equation of the form
Eq.(\ref{eq-x-m-diff-eq}).
We can use Eq.(\ref{eq-sol-xt-dif-eq})
to solve Eq.(\ref{eq-xt-dif-eq-sp})
with the initial condition
$x(0)=1$ (from
Eq.(\ref{eq-xt-def})).

{\bf second proof:}
Here is an alternative proof that uses
infinitesimals instead of
differential equations.
For any large integer $N$, let
$\Delta t= \frac{1}{N}$ and
$t_k= k\Delta t$ for $k\in Z_{0, N}$.
Thus, $t_0=0$ and
$t_N=1$. Then
\beqa
e^{-i Q_A} e^{i Q}
&\approx&
e^{-i Q_A}
(e^{i\Delta t \; Q_B}
e^{i\Delta t \; Q_A})^{N}
\\
&\approx&
\prod_{j=N, \ldots, 2, 1}
\left\{
e^{-i t_j Q_A}
e^{i\Delta t Q_B}
e^{i t_j Q_A}
\right\}
\\
&\approx&
\prod_{j=N, \ldots, 2, 1}
\exp(
e^{-i t_j Q_A}\;
i\Delta t\; Q_B\;
e^{i t_j Q_A}
)
\\
&\approx&
\prod_{j=N, \ldots, 2, 1}
\exp(i\Delta t
\left[
\begin{array}{cc}
0 & e^{-it_jA}B^\dagger\\
B e^{it_jA}
\end{array}
\right])
\\
&=&
\calt \left\{
\exp(i\int_0^1 dt
\left[
\begin{array}{cc}
0 & e^{-itA}B^\dagger\\
B e^{itA}
\end{array}
\right]
)
\right\}
\;.
\eeqa
\qed

In the previous Lemma, the
magnitude of the matrices $A$ and $B$ was
arbitrary. If we assume that $A$ and $B$
are $\calo(\coco)$, where $\coco$ is small,
then one can say more about the
exponential of an $M$ such that $M_{11}=0$.

\begin{Lemma}
Suppose $A,B\in \CC^{n\times n}$,
$A$ is a Hermitian matrix,
and $A,B$ are both $\calo(\coco)$,
where $\coco$ is small. Then

\beq
\exp(i
\left[
\begin{array}{cc}
A & B^\dagger \\
B & 0
\end{array}
\right]
)=
\exp(i
\left[
\begin{array}{cc}
A & 0 \\
0 & 0
\end{array}
\right]
)
\exp(
\left[
\begin{array}{cc}
T_1 + T_2 & i\bbar^\dagger\\
i\bbar & -i \frac{BAB^\dagger}{6}
\end{array}
\right]
)
+ \calo(\coco^5)
\;,
\label{eq-approx-with-t1-t2}
\eeq
where

\beq
T_1 = \frac{i}{12}(B^\dagger B A + A B^\dagger B)
\;,
\eeq

\beq
T_2 = (\frac{1}{24})
(-B^\dagger B A^2 + A^2 B^\dagger B)
\;,
\eeq

\beq
\bbar = B \sinc (\frac{A}{2}) e^{i \frac{A}{2}}
\;.
\eeq
Also, to lower order,

\beq
\exp(i
\left[
\begin{array}{cc}
A & B^\dagger \\
B & 0
\end{array}
\right]
)=
\exp(i
\left[
\begin{array}{cc}
A & 0 \\
0 & 0
\end{array}
\right]
)
\exp(i
\left[
\begin{array}{cc}
0 & \bbar^\dagger\\
\bbar & 0
\end{array}
\right]
)
+ \calo(\coco^3)
\;.
\label{eq-simplest-recursion}
\eeq
\end{Lemma}
{\bf proof:}
(The arXiv source code for this
paper includes an Octave/Matlab
subroutine called {\tt qtree\_cbh.m}
that checks this Lemma.)
If $x$ and $y$ are both of order
$\coco$ and $\coco$ is small, then the
Campbell-Baker-Hausdorff  Expansion
is \cite{Zachos, Reinsch}

\beq
e^x e^y = \exp\left\{
x+y +\frac{1}{2}[x,y]
+\frac{1}{12}
([x,.]^2 y +
[y,.]^2 x)
+\frac{1}{24}
[y,[x,[y,x]]]
+\calo(\coco^5)\right\}
\;.
\label{eq-cbh}
\eeq
If $Q_A$ and $Q_B$ are defined
by Eq.(\ref{eq-q-qa-qb}),
then this Lemma follows
from Eq.(\ref{eq-cbh}) with

\beq
x = - i Q_A, \;\;
y = i (Q_A + Q_B)
\;.
\eeq
The definition of $\bbar$
was inspired from Lemma \ref{lem-time-ordered}
and
the observation
that

\beq
\int_0^1 dt \;e^{itA} =
\frac{e^{iA}-1}{iA} =
\sinc(\frac{A}{2})
e^{i \frac{A}{2}}
\;,
\eeq
where $\sinc(x)=\sin(x)/x$
as usual.
\qed

\section{General Strategy for \\
Compiling Any Quantum Walk}
\label{sec-gen-comp}

In this section, we describe
our general strategy
for compiling the
FGG07 algorithm (or, more precisely,
 the
FGG07-inspired algorithm with
discrete queries that was first proposed
in Cle07).
The compilation strategy
described in this section is
also useful for compiling other
quantum walks.

The standard definition of
the evolution operator
in Quantum Mechanics is
$U= e^{-itH}$, where
$t$ is time and $H$
is a Hamiltonian. Throughout
this paper, we will set
$t = -1$ so $U = e^{iH}$.
If $H$ is proportional
to a coupling constant $\coco$,
reference to time can be
 restored easily by
replacing the symbol $\coco$ by
$-t\coco$, and the symbol $H$ by $-tH$.

\begin{figure}[h]
    \begin{center}
    \epsfig{file=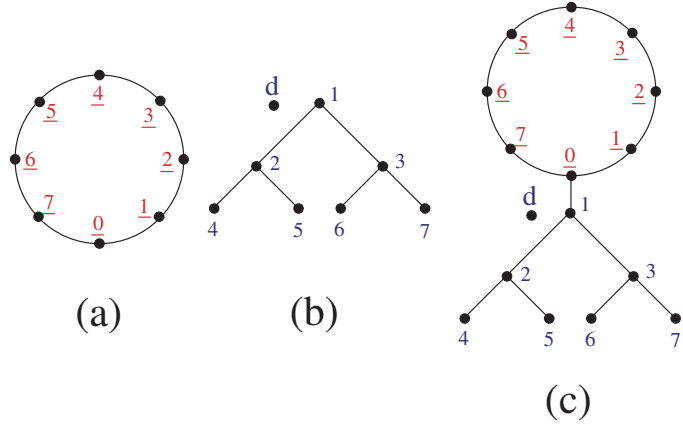, height=2.25in}
    \caption{(a)loop graph (b)tree graph (c)tree and loop
    graphs glued together
    }
    \label{fig-tree-loop}
    \end{center}
\end{figure}

Fig.\ref{fig-tree-loop}(a)
shows an example of a ``loop"
graph. It has 8 nodes (nodes=states),
labelled $0, 1, \ldots, 7$.
We wish to consider
the following Hamiltonian
for transitions along the edges
 of
this loop graph:

\beq
H_{lp} = (\coco)
\begin{array}{c|c|c|c|c|c|c|c|c|}
&\p{0}&\p{1}&\p{2}&\p{3}&\p{4}&\p{5}&\p{6}&\p{7} \\ \hline
\p{0}&&1&&&&&&1 \\ \hline
\p{1}&1&&1&&&&& \\ \hline
\p{2}&&1&&1&&&& \\ \hline
\p{3}&&&1&&1&&& \\ \hline
\p{4}&&&&1&&1&& \\ \hline
\p{5}&&&&&1&&1& \\ \hline
\p{6}&&&&&&1&&1 \\ \hline
\p{7}&1&&&&&&1& \\ \hline
\end{array}
\;,
\label{eq-h-loop}
\eeq
where $\coco\in \RR$.

A binary tree with $\Lam+1$ levels
has $1+2+2^2 +\ldots 2^{\Lam}=
2^{\Lam+1}-1$ nodes.
To reach $2^{\Lam+1}$ nodes, we
add an extra ``dead"  or ``dud" node,
labelled with the letter ``$d$".
This $d$ node is not connected
to any other node in the graph.
If we include this
dud node, then the number of leaves $\nlvs$
is exactly half the number of nodes:
$\nlvs = \frac{1}{2}2^{\Lam+1}=2^\Lam$.

Fig.\ref{fig-tree-loop}(b) shows an example
of a binary-tree
graph with $\Lam=2$. It has 8 nodes,
labelled $d, 1,2, \ldots, 7$.
We wish to consider
the following Hamiltonian
for transitions along the
edges of this binary-tree graph:

\beq
H_{tr} =
\begin{array}{c|c|c|c|c|c|c|c|c|}
&\p{d}&\p{1}&\p{2}&\p{3}&\p{4}&\p{5}&\p{6}&\p{7} \\ \hline
\p{d}&\;\;\;\;  & & & & & & &  \\ \hline
\p{1}& & &\coco_{12}&\coco_{13}& & & &  \\ \hline
\p{2}& &\coco_{12}& & &\coco_{24}&\coco_{25}& &  \\ \hline
\p{3}& &\coco_{13}& & & & &\coco_{36}&\coco_{37}  \\ \hline
\p{4}& & &\coco_{24}& & & & &  \\ \hline
\p{5}& & &\coco_{25}& & & & &  \\ \hline
\p{6}& & & &\coco_{36}& & & &  \\ \hline
\p{7}& & & &\coco_{37}& & & &  \\ \hline
\end{array}
\;,
\label{eq-h-tr}
\eeq
where $\coco_{p,q}\in \RR$ for all $p,q$.

Fig.\ref{fig-tree-loop}(c)
 shows an example of the graph
used in FGG07.
It contains a binary tree
attached to
a loop which serves as a
``runway"
for a wave packet.
For the algorithm of
FGG07 to work, the runway must
be much longer than $\sqrt{\nlvs}$
as $\nlvs\rarrow \infty$.\footnote{
FGG07 ``requests" a runway
of length $\nlvs$. However,
only a stretch of length $c\sqrt{\nlvs}$,
for some large constant $c$,
is ever used(traversed)
by the narrow wavepacket during
the duration of the experiment.
Chi07 and Amb07 request a
tail of length $c\sqrt{\nlvs}$.}
In Fig.\ref{fig-tree-loop}(c), the tree
has 8 nodes (including the dud one),
and the runway has
8 nodes.
Let's call the tree nodes
$tr_{j}$,where $j\in Z_{0,7}$,
and the runway nodes
$lp_{j}$, where $j\in Z_{0,7}$.
We wish to consider
the following Hamiltonian
for transitions along the edges of
 this tree and
runway graph:

\beq
H =
\left[
\begin{array}{c|c|c|c|c|c|c}
\multicolumn{4}{c|}{}&h^\dagger_{gl}&\;\;\;\;\;&\\ \cline{5-7}
\multicolumn{4}{c|}{}&&&\\ \cline{5-7}
\multicolumn{4}{c|}{}&&&\\ \cline{5-7}
\multicolumn{4}{c|}{\rb{4ex}{$H_{lp}$}}&&&\\ \hline
h_{gl}&&&&\multicolumn{2}{c|}{}&\\ \cline{1-4}\cline{7-7}
&&&&\multicolumn{2}{c|}{\rb{1ex}{$H_{tr}$}}&h_{in}^\dagger\\ \hline
\;\;\;\;\;&\;\;\;\;\;&\;\;\;\;\;&\;\;\;\;\;&&h_{in}&
\end{array}
\right]
\;.
\label{eq-h-fagogu}
\eeq
If $\coco$ is
the coupling constant that
appears in the loop Hamiltonian $H_{lp}$,
then we will assume, as
FGG07 does, that
the coupling constants
$\{\coco_{p,q}\}_{\forall p,q}$
that appear in the tree Hamiltonian $H_{tr}$
are all equal to $\coco$.
Besides $H_{tr}$ and $H_{lp}$,
which have already been discussed,
the Hamiltonian of Eq.(\ref{eq-h-fagogu})
includes blocks with
$h_{gl}$, $h_{in}$,
and their Hermitian conjugates.
The tree-loop ``glue" Hamiltonian
$H_{gl}$ is defined by

\beq
H_{gl} = h_{gl} + h_{gl}^\dagger=
\coco
(\ket{lp_0}\bra{tr_1} + h.c.)
\;.
\eeq
It corresponds to the graph
edge that connects the tree and
the runway.
The input (oracle)
Hamiltonian $H_{in}$
 is defined by

\beq
H_{in} = h_{in} + h_{in}^\dagger=
\coco
\sum_k
x_k (\ket{tr_k}\bra{in_k} + h.c.)
\;,
\eeq
where $k$ runs over all
tree leaves, and $x_k \in Bool$
are the NAND formula inputs.

The compilation strategy of
this paper is to split the
full Hamiltonian of
Eq.(\ref{eq-h-fagogu})
into two parts, which we
call the
{\bf bulk Hamiltonian} $H_{bulk}$
and the {\bf boundary corrections Hamiltonian}
$H_{corr}$:

\beq
H = H_{bulk} + H_{corr}
\;.
\eeq
We will set

\beq
H_{bulk} = H_{lp} + H_{tr}
\;\;\;,\;\;\;
H_{corr} = H_{gl} + H_{in}
\;.
\eeq
Note that $[H_{lp},H_{tr}]=0$
and $[H_{gl},H_{in}]=0$.
In subsequent sections
we will show how to compile
the evolution operators
$\exp(iH_{lp})$ and
$\exp(iH_{tr})$,
and therefore $\exp(iH_{bulk})$.
We will also show how to compile
$\exp(iH_{gl})$ and
$\exp(iH_{in})$,
and therefore $\exp(iH_{corr})$.
Exact compilation of
$\exp(iH_{bulk}+iH_{corr})$,
although possible
via an algorithm such as that of Tuc99,
is numerically very laborious.
Our strategy is to avoid such
numerical calculation by combining
the compilations of
$\exp(iH_{bulk})$
and
$\exp(iH_{corr})$
via the Trotterized Suzuki
approximation.

Actually, FGG07 assumes
a runway that is a straight line, but
replacing
a straight line
 by a  large enough closed loop
does not change the
algorithm significantly. Purists
can always convert the loop Hamiltonian
into a line Hamiltonian by adding
an extra loop-cutting Hamiltonian

\beq
H_{lp-cut} =
-\coco(\ket{lp_{3}}\bra{lp_4} + h.c.)
\;
\eeq
to the corrections Hamiltonian.
It is also possible to convert a loop
into a ``tail" like the one used in
Chi07 and Amb07, by
adding a loop-cutting Hamiltonian
that cuts the loop at the edge
that connects nodes $lp_{0}$ and $lp_{7}$.

\section{Compiling Loop Graphs}

In this section, we will show how
to compile (exactly) the evolution operator
$e^{iH_{lp}}$ for a loop graph.

Eq.(\ref{eq-h-loop}) is an example
of a loop Hamiltonian
$H_{lp}$. Since $H_{lp}$ is a circulant
matrix, $e^{iH_{lp}} = U e^{-iD} U^\dagger$,
where $U$ is a discrete Fourier
transform matrix and $D$ is a diagonal real
matrix. The $U$ and $U^\dagger$
can be immediately compiled
via the Coppersmith decomposition\cite{Copper},
using $\calo(\nb^2)$ CNOTs,
where $\ns=2^\nb$ is the number of
states in the loop. $e^{-iD}$
is a diagonal unitary matrix. Tuc99
shows how to compile an $\ns\times\ns$
 diagonal unitary matrix using $\calo(\ns)$
CNOTs.

Suppose $\theta_\vecb\in\RR$
for all $\vecb\in Bool^\nb$.
Suppose $\vecbet$ are $\nb$
distinct qubit positions and $\alpha$
is the position
of an additional ancilla qubit.
Note that

\beqa
\exp\left(
i\sum_\vecb \theta_\vecb P_\vecb(\vecbet)
\right)
&=&
\exp\left(
i\sigz(\alpha)\sum_\vecb \theta_\vecb P_\vecb(\vecbet)
\right)
\ket{0}_\alpha
\\&=&
e^{-i\frac{\pi}{4}\sigx(\alpha)}
\exp\left(i\sigy(\alpha)\sum_\vecb
\theta_\vecb P_\vecb(\vecbet)
\right)
e^{i\frac{\pi}{4}\sigx(\alpha)}
\ket{0}_\alpha
\;.
\eeqa
We used $\sigz\ket{0}=\ket{0}$ and
$e^{-i\frac{\pi}{4}\sigx}
\sigy
e^{i\frac{\pi}{4}\sigx}
=\sigz$. Thus, if one
can
compile any
$R_y(2)$-multiplexor with $\nb$
controls, then
one can
compile
any $2^\nb\times 2^\nb$  diagonal
unitary matrix.

\section{Compiling Glue and Cuts}

In this section, we will show how to compile
the evolution operators
$e^{iH_{cut}}$ for a loop cut and
$e^{iH_{gl}}$ for the tree-loop glue.

Consider the loop cut first.
For example, suppose we want to cut
the loop graph of
Fig.\ref{fig-tree-loop}(a) at the edge
connecting the states $\ket{0}$
and $\ket{7}$.

Define
\beq
U=[\sigx(2)\sigx(1)]^{n(0)}
\;.
\eeq
Then

\beqa
|7><0| +h.c.&=&
\ket{111}\bra{000} + h.c.\\
&=&
U(\ket{001}\bra{000} + h.c.)U^\dagger
\\
&=&
U\sigx(0)^{\nbar(2)\nbar(1)}U^\dagger
\label{eq-sigx-recognize}
\;.
\eeqa
To get Eq.(\ref{eq-sigx-recognize}),
we used the fact that
$\sigx(\beta) = (\ket{0}\bra{1}+h.c.)_\beta$
for any qubit position $\beta$.
Finally, note that

\beq
e^{iH_{cut}}=
e^{-i\coco(\ket{7}\bra{0} + h.c.)}
=
U
[e^{-i\coco \sigx(0)}]^{\nbar(2)\nbar(1)}
U^\dagger
\;.
\eeq

In general, suppose
we want to cut the edge
that connects states
 $\ket{j}$ and
$\ket{k}$, where $j$ and $k$
are the decimal names of the states.
Let $\vec{j}=bin(j)$ and
let $j_\beta$ be the component of
$\vec{j}$ at bit position $\beta$.
Define $\vec{k}$ and $k_\beta$ analogously.
$j\neq k$ so $\vec{j}$ and
$\vec{k}$ must differ
at one bit position, at least. Let
$\beta_o$ be one such bit position.
Assume $j_{\beta_o}=1$ and $k_{\beta_o}=0$.
Define
\beq
U=[\prod_{\beta: j_\beta\neq k_\beta,
\beta\neq \beta_o}\sigx(\beta)]^{n(\beta_o)}
\;.
\eeq
Then

\beq
|j><k| +h.c.
=
U\sigx(\beta_o)^{
\prod_{\beta: \beta\neq\beta_o}\{P_{k_\beta}(\beta)\}
}U^\dagger
\;.
\eeq
Therefore,

\beq
e^{iH_{cut}}=
e^{-i\coco(\ket{j}\bra{k} + h.c.)}
=
U
[e^{-i\coco \sigx(\beta_o)}]^{
\prod_{\beta: \beta\neq\beta_o}P_{k_\beta}(\beta)
}
U^\dagger
\;.
\eeq

Compiling $e^{iH_{gl}}$
is identical to compiling $e^{iH_{cut}}$.
The only difference is that the coupling
constant $\coco$ is replaced by
$-\coco$.

\section{Compiling Line Graphs}

In this section, we will show
how to compile (approximately)
the evolution operator $e^{iH_{line}}$
of a line graph (open loop).
Previously, we showed how to compile
the evolution operators
$e^{iH_{lp}}$
and $e^{iH_{cut}}$
for a loop graph and a cut.
A possible compilation
of $e^{iH_{line}}$
can be achieved by combining
$e^{iH_{lp}}$
and $e^{iH_{cut}}$
via Trotterized Suzuki. Alternatively,
one can compile
$e^{iH_{line}}$
directly, as
will be shown in this section.

Eq.(\ref{eq-h-loop}) is an example
of a line graph Hamiltonian $H_{line}$,
 provided
we set to zero the entries $(0,7)$
and $(7,0)$.
In Eq.(\ref{eq-h-loop}),
we've label states by the decimal
numbers 0 to 7, and we have
assumed that transitions between
these states are such that
their label $x\in Z_{0,7}$
can only vary by $\Delta x = \pm 1$.
As in spectroscopy, let's call
this constraint on $\Delta x$
a selection rule. Although
compiling $e^{iH_{line}}$
with an $H_{line}$ that
satisfies $\Delta x = \pm 1$
is possible, a cleaner,
simpler compilation\cite{Hines}
 can be achieved
if $H_{line}$ satisfies
a different selection rule; namely,
the constraint that states can only change
to other states iff the initial
and final states are adjacent
in a Gray ordering.
In a Gray ordering, states
are labelled by a binary number,
and adjacent states have labels that differ
only at one bit position.

\begin{figure}[h]
    \begin{center}
    \epsfig{file=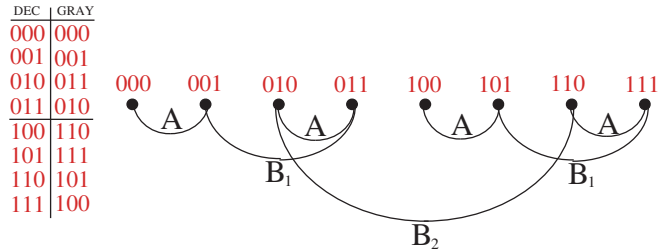, height=1.35in}
    \caption{
    Line graph for eight states.
    States are connected iff their
    binary representations differ
    only at one bit position; i.e.,
    connected states are adjacent in a Gray ordering.
    }
    \label{fig-gray-line}
    \end{center}
\end{figure}

Fig.\ref{fig-gray-line} shows
eight states, ordered linearly
in a decimal ordering from left to right.
Edges connecting the states indicate
possible transitions between the states.
These transitions can only occur between
states that are adjacent in a Gray ordering.
The Hamiltonian $H_{line}$ that describes transitions
along the edges of the graph of
Fig.\ref{fig-gray-line} is the following. (Note
that the states along the rows and columns of
$H_{line}$ are ordered in a decimal ordering.)

\begin{eqnarray}
H_{line} &=& (\coco)
\begin{array}{c|c|c|c|c|c|c|c|c|}
&\p{000}&\p{001}&\p{010}&\p{011}&\p{100}&\p{101}&\p{110}&\p{111} \\ \hline
\p{000}&0&1& & & & & & \\ \hline
\p{001}&1&0& &1& & & & \\ \hline
\p{010}& & &0&1& & &1& \\ \hline
\p{011}& &1&1&0& & & & \\ \hline
\p{100}& & & & &0&1& & \\ \hline
\p{101}& & & & &1&0& &1\\ \hline
\p{110}& & &1& & & &0&1\\ \hline
\p{111}& & & & & &1&1&0\\ \hline
\end{array}\\
&=&
\coco\left[
\begin{array}{c|c|c|c}
\sigx& P_1 & & \\ \hline
P_1&\sigx& &P_0\\ \hline
 & &\sigx&P_1\\ \hline
 &P_0&P_1& \sigx
\end{array}
\right]\\
&=&
\coco(\left[
\begin{array}{c|c|c|c}
\p{\sigx}& & & \\ \hline
&\p{\sigx}& &\\ \hline
 & &\p{\sigx}&\\ \hline
 &&& \p{\sigx}
\end{array}
\right]
+
\left[
\begin{array}{c|c|c|c}
& \p{P_1} & & \\ \hline
\p{P_1}&& &\\ \hline
 & &&\p{P_1}\\ \hline
 &&\p{P_1}&
\end{array}
\right]
+
\left[
\begin{array}{c|c|c|c}
\;&\;&\;&\;\\ \hline
&& &\p{P_0}\\ \hline
 & &&\\ \hline
 &\p{P_0}&&
\end{array}
\right])
\\
&=&
\coco \{I\otimes I \otimes \sigx
+
I\otimes \sigx \otimes P_1
+
\sigx\otimes P_1 \otimes P_0 \}
\\
&=&
\coco \{
\sigx(0) +
\sigx(1)n(0) +
\sigx(2)n(1)\nbar(0)
\}
\;.
\label{eq-h-line-3bit}
\end{eqnarray}
where $\coco\in \RR$.
Define

\beq
\begin{array}{l}
A = \coco\sigx(0)\\
B_1 = \coco\sigx(1)n(0)\\
B_2 = \coco\sigx(2)n(1)\nbar(0)
\end{array}
\;.
\eeq
Note that the $B_j$ commute
because they live in different states
(see Fig.\ref{fig-gray-line}).
Thus,

\begin{eqnarray}
e^{iH_{line}} &=&
e^{i(A + B_1 + B_2)} =
e^{iA}e^{iB_1}e^{iB_2}  + \calo(\coco^2)\\
&=&
\begin{array}{c}
\Qcircuit @C=1em @R=.25em @!R{
&\gate{R_x}&\dotgate&\ogate&\qw
\\
&\qw &\gate{R_x}\qwx&\dotgate\qwx&\qw
\\
&\qw&\qw&\gate{R_x}\qwx &\qw
}
\end{array}
+ \calo(\coco^2)
\;
\end{eqnarray}
if we use a first-order Lie approximant,
and

\begin{eqnarray}
e^{iH_{line}} &=&
e^{i(A + B_1 + B_2)} =
e^{i\frac{A}{2}}e^{iB_1}e^{iB_2}
e^{i\frac{A}{2}}  + \calo(\coco^3)\\
&=&
\begin{array}{c}
\Qcircuit @C=1em @R=.25em @!R{
&\gate{R_x}&\dotgate&\ogate&\gate{R_x}&\qw
\\
&\qw &\gate{R_x}\qwx&\dotgate\qwx&\qw&\qw
\\
&\qw&\qw&\gate{R_x}\qwx &\qw&\qw
}
\end{array}
+ \calo(\coco^3)
\;
\end{eqnarray}
if we use a second-order Suzuki approximant.

With four bits, one gets

\begin{eqnarray}
H_{line} &=&
\coco\left[
\begin{array}{c|c|c|c|c|c|c|c}
\sigx& P_1 & & &&&&\\ \hline
P_1& \sigx & &P_0&&&&\\ \hline
&  &\sigx&P_1&&&P_0&\\ \hline
 &P_0&P_1&\sigx&&&&\\ \hline

 &&&&\sigx& P_1 & &\\ \hline
 &&&&P_1& \sigx & &P_0\\ \hline
&  &P_0&&&&\sigx&P_1\\ \hline
 &&&&&P_0&P_1&\sigx\\ \hline
\end{array}
\right]\\
&=&
\coco \{
\sigx(0) +
\sigx(1)n(0) +
\sigx(2)n(1)\nbar(0)+
\sigx(3)n(2)\nbar(1)\nbar(0)
\}
\;.
\label{eq-h-line-4bit}
\end{eqnarray}
where $\coco\in \RR$.
Define

\beq
\begin{array}{l}
A = \coco\sigx(0)\\
B_1 = \coco\sigx(1)n(0)\\
B_2 = \coco\sigx(2)n(1)\nbar(0)\\
B_3 = \coco\sigx(3)n(2)\nbar(1)\nbar(0)
\end{array}
\;.
\eeq
Then

\begin{eqnarray}
e^{iH_{line}} &=&
e^{i(A + B_1 + B_2 + B_3)} =
e^{iA}e^{iB_1}e^{iB_2}e^{iB_3}  + \calo(\coco^2)\\
&=&
\begin{array}{c}
\Qcircuit @C=1em @R=.25em @!R{
&\gate{R_x}&\dotgate&\ogate&\ogate&\qw
\\
&\qw &\gate{R_x}\qwx&\dotgate\qwx&\ogate\qwx&\qw
\\
&\qw&\qw&\gate{R_x}\qwx &\dotgate\qwx&\qw
\\&\qw
&\qw&\qw&\gate{R_x}\qwx&\qw
}
\end{array}
+ \calo(\coco^2)
\;
\end{eqnarray}
if we use a first-order Lie approximant,
and similarly if we use a
second-order Suzuki approximant.

Generalization to arbitrary number
of bits is obvious.

\section{Generalities about Compiling Tree Graphs}
\label{sec-comp-tree-gen}

In this section, we
will introduce  several
key ideas that are useful
in compiling the evolution
operator $\exp(iH_{tr})$
for a general binary tree graph.

For $\lam=1,2,3,\ldots$,
define the $\twotolam\times\twotolam$
matrix $A_\twotolam$
to be, up to a constant factor,
the Hamiltonian $H_{tr}$ for a binary tree
with $2^\lam$ nodes.
By studying Eq.(\ref{eq-h-tr}),
one realizes that
the family of matrices
$\{A_\twotolam\}_{\forall \lam}$
can be specified in terms of
a family of matrices
$\{B_\twotolam\}_{\forall \lam}$
as follows. Let
$A_2=0$, and

\beq
A_{2^{\lam+1}}=
\left[
\begin{array}{cc}
A_\twotolam & B^\dagger_\twotolam \\
B_\twotolam & 0
\end{array}
\right]
\;,
\label{eq-recursion-a-mat}
\eeq
for $\lam=1,2,3,\cdots$.
(The subscripts of $A_\twotolam$ and $B_\twotolam$
indicate the dimension of these matrices.)
For example,
according to Eq.(\ref{eq-recursion-a-mat}),
$A_{16}$ has the form:

\beq
A_{16} =
\left[
\begin{array}{c|c|c|c|cccc}
0_2 & B_2^\dagger
&\multicolumn{2}{c|}{}
&\multicolumn{4}{c}{}
\\ \cline{1-2}
B_2&0_2
&\multicolumn{2}{c|}{\rb{1.5ex}{$B_4^\dagger$}}
&\multicolumn{4}{c}{}
\\ \cline{1-4}
\multicolumn{2}{c|}{}
&\multicolumn{2}{c|}{}
&\multicolumn{4}{c}{}
\\
\multicolumn{2}{c|}{\rb{1.5ex}{$B_4$}}
&\multicolumn{2}{c|}{\rb{1.5ex}{$0_4$}}
&\multicolumn{4}{c}{\rb{4ex}{$B_8^\dagger$}}
\\ \hline
\multicolumn{4}{c|}{}
&\multicolumn{4}{c}{}
\\
\multicolumn{4}{c|}{}
&\multicolumn{4}{c}{}
\\
\multicolumn{4}{c|}{}
&\multicolumn{4}{c}{}
\\
\multicolumn{4}{c|}{\rb{4ex}{$B_8$}}
&\multicolumn{4}{c}{\rb{4ex}{$0_8$}}
\end{array}
\right]
\;.
\label{eq-a16-def}
\eeq

\subsection{CSD-ready Evolutions}

By applying
Eq.(\ref{eq-simplest-recursion})
repeatedly, we can factor $e^{i A_{16}}$
with $A_{16}$ given by Eq.(\ref{eq-a16-def})
as follows:

\beq
e^{i A_{16}}=
\Gamma_1 \Gamma_2
\Gamma_3 + \calo(\coco^3)
\;,
\label{eq-gamma-bslash-3}
\eeq
where

\begin{subequations}
\beq
\Gamma_3 =
\exp(i
\left[
\begin{array}{cc}
0 & \bbar^\dagger_8 \\
\bbar_8 & 0
\end{array}
\right]
)
\;,
\eeq

\beq
\Gamma_2 =
\exp(i
\left[
\begin{array}{cc|c}
0 & \bbar^\dagger_4 & \\
\bbar_4 & 0 & \\ \hline
& & 0_{8}
\end{array}
\right]
)
\;,
\eeq
and

\beq
\Gamma_1 =
\exp(i
\left[
\begin{array}{cc|c}
0 & \bbar^\dagger_2 & \\
\bbar_2 & 0 & \\ \hline
& & 0_{12}
\end{array}
\right]
)
\;.
\eeq
\end{subequations}
$\bbar_\twotolam$ is defined
for $\lam \in Z_{1,3}$ by

\beq
\bbar_\twotolam=
B_\twotolam
\sinc(\frac{A_\twotolam}{2})
e^{i\frac{A_\twotolam}{2}}
\;.
\eeq

The above example
generalizes as follows.
For $\Lam=1,2,3,\ldots$,
and
$\lam\in Z_{1, \Lam}$,
we have\footnote{
An important special case is when
the $d_\lam$ (defined by
Eq.(\ref{eq-const-delta-assump}))
are all equal to $\sqrt{2}\coco$.
In this case, a slight
redefinition of $\Gamma^\lam$ in
Eq.(\ref{eq-gamma-defs})
makes Eq.(\ref{eq-strati})
good to order $\calo(\coco^4)$
instead of $\calo(\coco^3)$.
See Appendix \ref{app-all-d-same}.}

\beq
e^{i A_{2^{\Lam+1}}}=
\Gamma_1
\Gamma_2
\dots\Gamma_\Lam
+ \calo(\coco^3)
\;,
\label{eq-strati}
\eeq

\beq
\Gamma_\lam =
\exp(i
\left[
\begin{array}{cc|c}
0 & \bbar^\dagger_\twotolam & \\
\bbar_\twotolam & 0 & \\ \hline
& & 0_{2^{\Lam+1}-2^{\lam+1}}
\end{array}
\right]
)
\;,
\label{eq-gamma-defs}
\eeq
and

\beq
\bbar_\twotolam=
B_\twotolam
\sinc(\frac{A_\twotolam}{2})
e^{i\frac{A_\twotolam}{2}}
\;.
\label{eq-tb-def}
\eeq

Since these $\Gamma_\lam$
are ready for an application of the
CSD, we will henceforth refer to them
 as {\bf CSD-ready evolutions (or factors)}
 of $\exp(iA_{2^{\Lam+1}})$, and
to
Eq.(\ref{eq-strati}) as a
factorization of $\exp(iA_{2^{\Lam+1}})$
into CSD-ready factors.

\subsection{SVD of $B_{\twotolam}$}
\label{sec-svd-b}

Next we will calculate an SVD for
each of the matrices
$B_{\twotolam}$ that appear in
Eq.(\ref{eq-recursion-a-mat}).
These SVD's will be useful to us in future
sections.

We will use $E_4$ to denote
the $4\times 4$ matrix that
exchanges (swaps) 2 qubits:

\beq
E_4 =
\begin{array}{c|c|c|c|c|c|c|c|c|}
&\p{00}&\p{01}&\p{10}&\p{11} \\ \hline
\p{00}&1& & & \\ \hline
\p{01}& & &1& \\ \hline
\p{10}& &1& & \\ \hline
\p{11}& & & &1\\ \hline
\end{array}
\;.
\eeq

Given $a_j,b_j\in\RR$ for some
index $j$, define

\beq
\rho_j = \sqrt{a^2_j + b^2_j}
\;,\;\;
U_j =
\frac{1}{\rho_j}
\left[
\begin{array}{cc}
b_j & a_j \\
-a_j & b_j
\end{array}
\right]
\;.
\label{eq-eta-u-defs}
\eeq

The matrix $B_2$
that appears in
Eq.(\ref{eq-recursion-a-mat})
is of the form

\beq
B_2 =
\left[
\begin{array}{cc}
0 & a \\
0 & b
\end{array}
\right]
\;.
\eeq
These values of $a$ and
$b$ define a matrix
 $U$ and a scalar $\rho$ via
Eq.(\ref{eq-eta-u-defs})
(with the $j$ subscript absent).
The SVD of $B_2$ is one-sided and
can be found in closed form. It's
given by:

\beq
B_2 = F_2
\left[
\begin{array}{cc}
0 & 0 \\
0 & \Delta_1
\end{array}
\right]
\;,
\eeq
where

\begin{subequations}
\label{eq-b2-f-del}
\beq
F_2 = U
\;,
\eeq
and

\beq
\Delta_1 = \rho
\;.
\eeq
\end{subequations}
When expressed in  operator rather than matrix
notation, Eqs.(\ref{eq-b2-f-del}) become

\begin{subequations}
\beq
F_2(0) = U(0)
\;,
\eeq
and

\beq
\Delta_1 = \rho
\;.
\eeq
\end{subequations}

The matrix $B_4$
that appears in
Eq.(\ref{eq-recursion-a-mat})
is of the form

\beq
B_4 =
\left[
\begin{array}{cc|cc}
\;\; & \;\;  & a_0 & 0 \\
  &   & b_0 & 0 \\ \hline
  &   & 0   & a_1 \\
  &   & 0   & b_1
\end{array}
\right]
\;.
\eeq
For $j=0,1$, these values of $a_j$ and
$b_j$ define  a matrix
 $U_j$ and a scalar $\rho_j$ via
Eq.(\ref{eq-eta-u-defs}).
The SVD of $B_4$ is one-sided and
can be found in closed form. It's
given by:

\beqa
B_4 &=&
\left[
\begin{array}{cc}
U_0 &  \\
 & U_1
\end{array}
\right]
\left[
\begin{array}{cc|cc}
\;\;  & \;\;  & 0 & 0 \\
  &   & \rho_0 & 0 \\ \hline
  &   & 0   & 0 \\
  &   &  0  &\rho_1
\end{array}
\right]
\\
&=&
F_4
\left[
\begin{array}{cc}
0_2 &  \\
& \Delta_2
\end{array}
\right]
\;,
\eeqa
where

\begin{subequations}
\label{eq-b4-f-del}
\beq
F_4 =
(U_0\oplus U_1)
E_4
\;,
\eeq
and

\beq
\Delta_2 = diag(\rho_0, \rho_1)
\;.
\eeq
\end{subequations}
When expressed in  operator rather than matrix
notation, Eqs.(\ref{eq-b4-f-del}) become

\begin{subequations}
\beq
F_4(1,0) =
U_0^{\nbar(1)}(0)
U_1^{n(1)}(0)
E_4(1,0)
\;,
\eeq
and

\beq
\Delta_2(0) = \sum_{b\in Bool} \rho_b P_b(0)
\;.
\eeq
\end{subequations}

The matrix $B_8$
that appears in
Eq.(\ref{eq-recursion-a-mat})
is of the form

\beq
B_8 =
\left[
\begin{array}{cc|cc|cc|cc}
\;\;\; & \;\;\;  &\;\;\; & \;\;\;
& a_{00} & 0 & 0 & 0 \\
&&&
& b_{00} & 0 & 0 & 0 \\ \hline
&&&
& 0 & a_{01} & 0 & 0 \\
&&&
& 0 & b_{01} & 0 & 0 \\ \hline
&&&
& 0 & 0 & a_{10} & 0 \\
&&&
& 0 & 0 & b_{10} & 0 \\ \hline
&&&
& 0 & 0 & 0 & a_{11} \\
&&&
& 0 & 0 & 0 & b_{11} \\
\end{array}
\right]
\;.
\eeq
For $j\in\{00,01,10,11\}$,
these values of $a_j$ and
$b_j$ define  a matrix
 $U_j$ and a scalar $\rho_j$ via
Eq.(\ref{eq-eta-u-defs}).
The SVD of $B_8$ is one-sided and
can be found in closed form. It's
given by:

\beqa
B_8 &=&
\left[
\begin{array}{cccc}
U_{00} &\;\;\;&\;\;\;&\;\;\;\\
& U_{01} &&\\
&& U_{10} &\\
&&& U_{11}
\end{array}
\right]
\left[
\begin{array}{cc|cc|cc|cc}
\;\;\; & \;\;\;  &\;\;\; & \;\;\;
& 0 & 0 & 0 & 0 \\
&&&
& \rho_{00} & 0 & 0 & 0 \\ \hline
&&&
& 0 & 0 & 0 & 0 \\
&&&
& 0 & \rho_{01} & 0 & 0 \\ \hline
&&&
& 0 & 0 & 0 & 0 \\
&&&
& 0 & 0 & \rho_{10} & 0 \\ \hline
&&&
& 0 & 0 & 0 & 0 \\
&&&
& 0 & 0 & 0 & \rho_{11} \\
\end{array}
\right]
\\
&=&
F_8
\left[
\begin{array}{cc}
0_4 &  \\
& \Delta_4
\end{array}
\right]
\;,
\eeqa
where

\begin{subequations}
\label{eq-b8-f-del}
\beq
F_8 =
(U_{00}\oplus
U_{01}\oplus
U_{10}\oplus
U_{11})
(I\otimes E_4)(E_4\otimes I)
\;,
\eeq
and

\beq
\Delta_4 = diag(
\rho_{00},\rho_{01},\rho_{10},\rho_{11})
\;.
\eeq
\end{subequations}
When expressed in  operator rather than matrix
notation, Eqs.(\ref{eq-b8-f-del}) become

\begin{subequations}
\beq
F_8(2,1,0) =
U_{00}(0)^{\nbar(2)\nbar(1)}
U_{01}(0)^{\nbar(2)n(1)}
U_{10}(0)^{n(2)\nbar(1)}
U_{11}(0)^{n(2)n(1)}
E_4(1,0)
E_4(2,1)
\;,
\eeq
and

\beq
\Delta_4(1,0) = \sum_{\vecb\in Bool^2}
\rho_\vecb P_\vecb(1,0)
\;.
\eeq
\end{subequations}

The above example
generalizes as follows.
For $\lam=1,2,3,\dots$,

\beq
F_\twotolam(\lam-1, \ldots, 2,1,0)=
\prod_{\vecb=Bool^{\lam-1}}
\{
U_{\vecb}^{P_\vecb(\lam-1,\ldots,2,1)}(0)
\}
E_4(0,1)E_4(1,2)\ldots E_4(\lam-2,\lam-1)
\;,
\eeq
and

\beq
\Delta_{2^{\lam-1}}(\lam-2, \ldots, 2,1,0)=
\sum_{\vecb\in Bool^{\lam-1}} \rho_\vecb
P_\vecb(\lam-2, \ldots, 2,1,0)
\;.
\eeq

In this paper, we are interested
mainly in the case of
balanced binary NAND trees. For
such trees,
$\rho_\vecb$ and $U_\vecb$
are both independent of $\vecb$.
In particular,

\beq
U_{\vecb}=
\frac{1}{\sqrt{2}}
\left[
\begin{array}{cc}
1&1\\
-1&1
\end{array}
\right]
=
H\sigx=e^{i\frac{\pi}{4}\sigy}
\;
\eeq
for all $\vecb$.
Therefore, for balanced binary trees,
$\Delta_\twotolam$ is proportional to
the identity matrix and

\beq
F_\twotolam(\lam-1, \ldots, 2,1,0)=
(H\sigx)(0)
E_4(0,1)E_4(1,2)\ldots E_4(\lam-2,\lam-1)
\;,
\label{eq-f-bb-nand}
\eeq
for all $\lam=1,2,3\dots$.

\section{Compiling Balanced Binary NAND Trees}

In Section \ref{sec-comp-tree-gen},
we found an SVD for each
matrix $B_{\twotolam}$.
Half of its singular values
were zero and the other half
were stored in a diagonal matrix
called
$\Delta_{2^{\lam-1}}$.
In this section,
which contains several subsections, we will
give an approximate compilation of $\exp(iH_{tr})$
for a special type of binary tree
that includes
balanced binary NAND trees.
The special type of
binary trees that we will consider
in this section satisfies

\beq
\Delta_\twotolam = d_\lam I_\twotolam
\;,
\label{eq-const-delta-assump}
\eeq
for all possible $\lam$,
where $d_\lam\in\RR$.
In other words, the $\Delta_\twotolam$
of $H_{tr}$ are all proportional
to the identity matrix.
Balance binary
NAND trees satisfy Eq.(\ref{eq-const-delta-assump})
because  their
coupling constants $\coco_{p,q}$
are the same for all $p,q$.

\subsection{Diagonalizing the Tree Levels}

For $\lam=1,2,3, \dots$,
define

\beq
\calb_\twotolam =
\left[
\begin{array}{cc}
0_{2^{\lam-1}} & \\
& \Delta_{2^{\lam-1}}
\end{array}
\right]
\;.
\eeq
A family of matrices
$\{\cala_\twotolam\}_{\forall \lam}$
can be defined in terms of
the family of matrices
$\{\calb_\twotolam\}_{\forall \lam}$
as follows. Let
$\cala_2=0$, and

\beq
\cala_{2^{\lam+1}}=
\left[
\begin{array}{cc}
\cala_\twotolam & \calb^\dagger_\twotolam\\
\calb_\twotolam & 0
\end{array}
\right]
\;
\label{eq-def-cala}
\eeq
for $\lam=1,2,3, \dots$.
The subscripts of $\cala_\twotolam$
and $\calb_\twotolam$
indicate the dimension of these matrices.

Next, we will explain how
the matrices
$A_{\twotolam}$ are related
to their namesakes
$\cala_{\twotolam}$.
For illustrative purposes,
consider $A_{16}$ and $\cala_{16}$ first.
We claim that if we
define

\begin{eqnarray}
\calf_{16} &=&
diag(I_8,F_8)diag(I_4,F_4,I_4,F_4)
diag(I_2,F_2,I_2,F_2,I_2,F_2,I_2,F_2)
\nonumber\\
&=&
(I_8\oplus F_8)(I_4\oplus F_4)^{\oplus 2}
(I_2\oplus F_2)^{\oplus 4}
\;,
\label{eq-calf16-def-mats}
\end{eqnarray}
then

\beq
A_{16}=
\calf_{16}
\cala_{16}
\calf^\dagger_{16}
\;.
\label{eq-a-is-calf-cala-calf}
\eeq
This claim follows from
the following observations:

\beqa
A_{16} &=&
\left[
\begin{array}{c|c|c|c|cccc}
0_2 & B_2^\dagger
&\multicolumn{2}{c|}{}
&\multicolumn{4}{c}{}
\\ \cline{1-2}
B_2&0_2
&\multicolumn{2}{c|}{\rb{1.5ex}{$B_4^\dagger$}}
&\multicolumn{4}{c}{}
\\ \cline{1-4}
\multicolumn{2}{c|}{}
&\multicolumn{2}{c|}{}
&\multicolumn{4}{c}{}
\\
\multicolumn{2}{c|}{\rb{1.5ex}{$B_4$}}
&\multicolumn{2}{c|}{\rb{1.5ex}{$0_4$}}
&\multicolumn{4}{c}{\rb{4ex}{$B_8^\dagger$}}
\\ \hline
\multicolumn{4}{c|}{}
&\multicolumn{4}{c}{}
\\
\multicolumn{4}{c|}{}
&\multicolumn{4}{c}{}
\\
\multicolumn{4}{c|}{}
&\multicolumn{4}{c}{}
\\
\multicolumn{4}{c|}{\rb{4ex}{$B_8$}}
&\multicolumn{4}{c}{\rb{4ex}{$0_8$}}
\end{array}
\right] \label{eq-pre-super-table}
\\
&=&
\begin{array}{c|c|c||c|c|c|c|c|c|c|c||}
\multicolumn{3}{c||}{}
&
\multicolumn{4}{c|}{I_8}
&
\multicolumn{4}{c||}{F_8^\dagger}
\\ \cline{4-11}
\multicolumn{3}{c||}{}
&
\multicolumn{2}{c|}{I_4}
&
\multicolumn{2}{c|}{F_4^\dagger}
&
\multicolumn{2}{c|}{I_4}
&
\multicolumn{2}{c||}{F_4^\dagger}
\\ \cline{4-11}
\multicolumn{3}{c||}{}
&
I_2
&
F_2^\dagger
&
\;I_2\;
&
\;F_2^\dagger\;
&
\;I_2\;
&
\;F_2^\dagger\;
&
\;I_2\;
&
\;F_2^\dagger\;
\\ \hline\hline

&

&
I_2
&

&
\scriptscriptstyle{{0\atop }}
\scriptscriptstyle{{\atop\Delta_1}}
&

&

&
\multicolumn{2}{c|}{}
&
\multicolumn{2}{c||}{}
\\ \cline{3-7}

&
\rb{1.5ex}{$I_4$}
&
F_2
&
\scriptscriptstyle{{0\atop }}
\scriptscriptstyle{{\atop\Delta_1}}
&

&

&
\Delta_2
&
\multicolumn{2}{c|}{}
&
\multicolumn{2}{c||}{}
\\ \cline{2-11}

&

&
I_2
&

&

&
\multicolumn{2}{c|}{}
&
\multicolumn{2}{c|}{}
&
\multicolumn{2}{c||}{}
\\ \cline{3-5}
\rb{4ex}{$I_8$}
&
\rb{1.5ex}{$F_4$}
&
F_2
&

&
\Delta_2
&
\multicolumn{2}{c|}{}
&
\multicolumn{2}{c|}{}
&
\multicolumn{2}{c||}{\rb{1.5ex}{$\Delta_4$}}
\\ \hline

&

&
I_2
&
\multicolumn{2}{c|}{}
&
\multicolumn{2}{c|}{}
&
\multicolumn{4}{c||}{}
\\ \cline{3-3}

&
\rb{1.5ex}{$I_4$}
&
F_2
&
\multicolumn{2}{c|}{}
&
\multicolumn{2}{c|}{}
&
\multicolumn{4}{c||}{}
\\ \cline{2-7}

&

&
I_2
&
\multicolumn{2}{c|}{}
&
\multicolumn{2}{c|}{}
&
\multicolumn{4}{c||}{}
\\ \cline{3-3}
\rb{4ex}{$F_8$}
&
\rb{1.5ex}{$F_4$}
&
F_2
&
\multicolumn{2}{c|}{}
&
\multicolumn{2}{c|}{\rb{1.5ex}{$\Delta_4$}}
&
\multicolumn{4}{c||}{}
\\
\hline\hline
\end{array}\label{eq-super-table}
\\
&=&
\calf_{16}
\left[
\begin{array}{c|c|c|c|cccc}
0_2 & \calb_2^\dagger
&\multicolumn{2}{c|}{}
&\multicolumn{4}{c}{}
\\ \cline{1-2}
\calb_2&0_2
&\multicolumn{2}{c|}{\rb{1.5ex}{$\calb_4^\dagger$}}
&\multicolumn{4}{c}{}
\\ \cline{1-4}
\multicolumn{2}{c|}{}
&\multicolumn{2}{c|}{}
&\multicolumn{4}{c}{}
\\
\multicolumn{2}{c|}{\rb{1.5ex}{$\calb_4$}}
&\multicolumn{2}{c|}{\rb{1.5ex}{$0_4$}}
&\multicolumn{4}{c}{\rb{4ex}{$\calb_8^\dagger$}}
\\ \hline
\multicolumn{4}{c|}{}
&\multicolumn{4}{c}{}
\\
\multicolumn{4}{c|}{}
&\multicolumn{4}{c}{}
\\
\multicolumn{4}{c|}{}
&\multicolumn{4}{c}{}
\\
\multicolumn{4}{c|}{\rb{4ex}{$\calb_8$}}
&\multicolumn{4}{c}{\rb{4ex}{$0_8$}}
\end{array}
\right]\calf^\dagger_{16}
\\
&=&
\calf_{16}
\cala_{16}
\calf^\dagger_{16}
\;.
\eeqa
Going from Eq.(\ref{eq-pre-super-table})
to Eq.(\ref{eq-super-table})
is the crucial step that requires
assumption Eq.(\ref{eq-const-delta-assump}).
Eq.(\ref{eq-super-table})
is a non-conventional way
of expressing the product of 3 matrices:
the matrix that is totally enclosed by double lines
is being pre-multiplied by matrix
$\calf_{16}$ and post-multiplied by matrix
$\calf^\dagger_{16}$.
$\calf_{16}$ is spread out over the left margin
and $\calf^\dagger_{16}$ over the top margin of
the matrix enclosed by the double-line border.
Writing Eq.(\ref{eq-super-table})
in this nonconventional way
makes it clear why assumption
Eq.(\ref{eq-const-delta-assump})
is needed. For example,
because $\Delta_4$ acts like a scalar,
the matrix $F^\dagger_2$ in the last column
of the top margin commutes with
$\Delta_4$ and can cancel the
$F_2$ in the fourth row of the left margin.

Note that $\calf_{16}$ given
by Eq.(\ref{eq-calf16-def-mats})
can be expressed in operator
rather than matrix
notation as follows:

\beq
\calf_{16}(3,2,1,0)=
F_8(2,1,0)^{n(3)}
F_4(1,0)^{n(2)}
F_2(0)^{n(1)}
\;
\eeq

The above example
generalizes as follows.
For $\lam=1,2,3\ldots$,

\beq
\calf_{2^{\lam+1}}(\lam,\ldots,2,1,0)=
F_{2^\lam}(\lam-1,\ldots,2,1,0)^{n(\lam)}
\ldots
F_4(1,0)^{n(2)}
F_2(0)^{n(1)}
\;,
\label{eq-calf-is-f-prod}
\eeq
and

\beq
A_\twotolam=
\calf_\twotolam
\cala_\twotolam
\calf^\dagger_\twotolam
\;.
\eeq
Thus, for $\Lam=1,2,3, \ldots$,

\beq
\exp(iA_{2^{\Lam+1}})=
\calf_{2^{\Lam+1}}
\exp(i\cala_{2^{\Lam+1}})
\calf^\dagger_{2^{\Lam+1}}
\;.
\label{eq-exp-a-eq-f-exp-cala-f}
\eeq

\subsection{Separating into Strands}

A consequence of diagonalizing the
levels of the binary tree graph is that
the tree graph
is replaced by a collection
of subgraphs shaped like strands.
In this section, we will factor
$\exp(i\cala_{2^{\Lam+1}})$
into separate contributions from each
of these strands.

Let us consider an
example first, before dealing with
the general case. Suppose $\Lam=5$
and consider $\exp(i\cala_{64})$.
By studying carefully the definition
 Eq.(\ref{eq-def-cala}) of
$\cala_{64}$, one can see that

\beq
\exp(i\cala_{64})
=
\prod_{s=0}^{15}
\exp(i\cala^{\sdep}_{64})
\;,
\eeq
where the matrices
$\cala^{\sdep}_{64}$
are defined
in terms of matrices $\cala^\sdep_\twotolam$
and $\calb^\sdep_\twotolam$ as follows.
The index $s$ labels the
16 ``strands" that contribute
to $\exp(i\cala_{64})$.
Suppose $s\in Z_{0,15}$.
Let

\beq
\calb^\sdep_\twotolam =
\left[
\begin{array}{cc}
0_{2^{\lam-1}} & \\
& \Delta^\sdep_{2^{\lam-1}}
\end{array}
\right]
\;
\eeq
for $\lam\in Z_{1,5}$.
Let
$\cala^\sdep_2=0$
and

\beq
\cala_{2^{\lam+1}}^\sdep=
\left[
\begin{array}{cc}
\cala_\twotolam^\sdep &
\calb^{\sdep\dagger}_\twotolam\\
\calb^\sdep_\twotolam & 0
\end{array}
\right]
\;
\eeq
for $\lam\in Z_{1,5}$.

For $s=0$, one finds

\beq
\begin{array}{ll}
\cala_{32}^{(0)}=
(d_1\ket{1}\bra{3} +
d_2\ket{3}\bra{7} +
d_3\ket{7}\bra{15} +
d_4\ket{15}\bra{31}
) + h.c.
&
\calb_{32}^{(0)}= d_5\ket{31}\bra{31}
\\
\cala_{16}^{(0)}=
(d_1\ket{1}\bra{3} +
d_2\ket{3}\bra{7} +
d_3\ket{7}\bra{15}) + h.c.
&
\calb_{16}^{(0)}= d_4\ket{15}\bra{15}
\\
\cala_{8}^{(0)}=
(d_1\ket{1}\bra{3} +
d_2\ket{3}\bra{7}) + h.c.
&
\calb_{8}^{(0)}= d_3\ket{7}\bra{7}
\\
\cala_{4}^{(0)}=
d_1\ket{1}\bra{3} + h.c.
&
\calb_{4}^{(0)}= d_2\ket{3}\bra{3}
\\
\cala_{2}^{(0)}=0
&
\calb_{2}^{(0)}= d_1\ket{1}\bra{1}
\end{array}
\;.
\eeq
For $s=1$, one finds

\beq
\begin{array}{ll}
\cala_{32}^{(1)}=
(d_2\ket{2}\bra{6} +
d_3\ket{6}\bra{14} +
d_4\ket{14}\bra{30}
) + h.c.
&
\calb_{32}^{(1)}= d_5\ket{30}\bra{30}
\\
\cala_{16}^{(1)}=
(d_2\ket{2}\bra{6} +
d_3\ket{6}\bra{14}) + h.c.
&
\calb_{16}^{(1)}= d_4\ket{14}\bra{14}
\\
\cala_{8}^{(1)}=
(
d_2\ket{2}\bra{6}) + h.c.
&
\calb_{8}^{(1)}= d_3\ket{6}\bra{6}
\\
\cala_{4}^{(1)}=0
&
\calb_{4}^{(1)}= d_2\ket{2}\bra{2}
\\
\cala_{2}^{(1)}=0
&
\calb_{2}^{(1)}= 0
\end{array}
\;.
\eeq
For $k\in Z_{0,1}$ (and
$s=2+k\in Z_{2,3})$, one finds

\beq
\begin{array}{ll}
\cala_{32}^{(2+k)}=
(d_3\ket{5-k}\bra{13-k} +
d_4\ket{13-k}\bra{29-k}
) + h.c.
&
\calb_{32}^{(2+k)}= d_5\ket{29-k}\bra{29-k}
\\
\cala_{16}^{(2+k)}=
(
d_3\ket{5-k}\bra{13-k}) + h.c.
&
\calb_{16}^{(2+k)}= d_4\ket{13-k}\bra{13-k}
\\
\cala_{8}^{(2+k)}=0
&
\calb_{8}^{(2+k)}= d_3\ket{5-k}\bra{5-k}
\\
\cala_{4}^{(2+k)}=
\cala_{2}^{(2+k)}=0
&
\calb_{4}^{(2+k)}=
\calb_{2}^{(2+k)}=0
\end{array}
\;.
\eeq
For $k\in Z_{0,3}$ (and
$s=4+k\in Z_{4,7})$, one finds

\beq
\begin{array}{ll}
\cala_{32}^{(4+k)}=
d_4\ket{11-k}\bra{27-k} + h.c.
&
\calb_{32}^{(4+k)}= d_5\ket{27-k}\bra{27-k}
\\
\cala_{16}^{(4+k)}=0
&
\calb_{16}^{(4+k)}= d_4\ket{11-k}\bra{11-k}
\\
\cala_{8}^{(4+k)}=
\cala_{4}^{(4+k)}=
\cala_{2}^{(4+k)}=0
&
\calb_{8}^{(4+k)}=
\calb_{4}^{(4+k)}=
\calb_{2}^{(4+k)}=0
\end{array}
\;.
\eeq
For $k\in Z_{0,7}$ (and
$s=8+k\in Z_{8,15})$, one finds

\beq
\begin{array}{ll}
\cala_{32}^{(8+k)}=0
&
\calb_{32}^{(8+k)}= d_5\ket{23-k}\bra{23-k}
\\
\cala_{16}^{(8+k)}=
\cala_{8}^{(8+k)}=
\cala_{4}^{(8+k)}=
\cala_{2}^{(8+k)}=0
&
\calb_{16}^{(8+k)}=
\calb_{8}^{(8+k)}=
\calb_{4}^{(8+k)}=
\calb_{2}^{(8+k)}=0
\end{array}
\;.
\eeq

The above example
generalizes as follows.
For $\Lam=1,2,3,\ldots$,

\beq
\exp(i\cala_{2^{\Lam+1}})
=
\prod_{s=0}^{2^{\Lam-1}-1}
\exp(i\cala^{\sdep}_{2^{\Lam+1}})
\;,
\label{eq-cala-into-strands}
\eeq
where
$\cala^\sdep_{2^{\Lam+1}}$ is defined
in terms of matrices $\cala^\sdep_\twotolam$
and $\calb^\sdep_\twotolam$ as follows.
Suppose $s\in Z_{0, 2^{\Lam-1}-1}$.
Let

\beq
\calb^\sdep_\twotolam =
\left[
\begin{array}{cc}
0_{2^{\lam-1}} & \\
& \Delta^\sdep_{2^{\lam-1}}
\end{array}
\right]
\;
\eeq
for $\lam\in Z_{1,\Lam}$.
Let
$\cala^\sdep_2=0$
and

\beq
\cala_{2^{\lam+1}}^\sdep=
\left[
\begin{array}{cc}
\cala_\twotolam^\sdep &
\calb^{\sdep\dagger}_\twotolam\\
\calb^\sdep_\twotolam & 0
\end{array}
\right]
\;
\eeq
for $\lam\in Z_{1,\Lam}$.

\begin{figure}[h]
    \begin{center}
    \epsfig{file=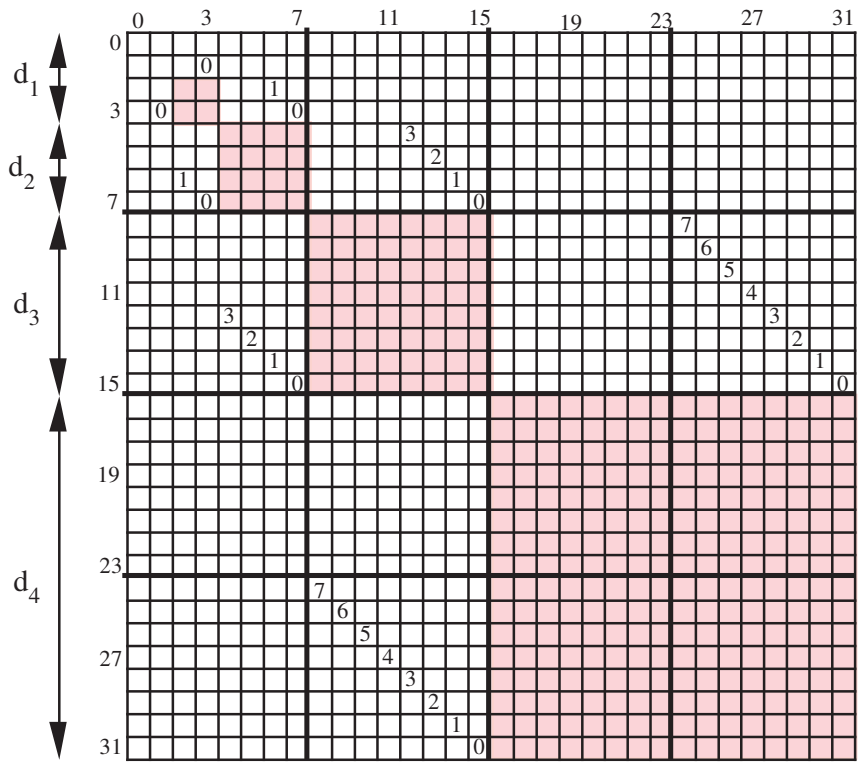, height=4.0in}
    \caption{Illustration of strands as entries of
    matrix $\cala_{32}$. An empty cell
    represents a zero
    matrix entry. The number inside a non-empty cell
    is the number of the strand that owns that cell.
    }
    \label{fig-strands-square}
    \end{center}
\end{figure}

\begin{figure}[h]
    \begin{center}
    \epsfig{file=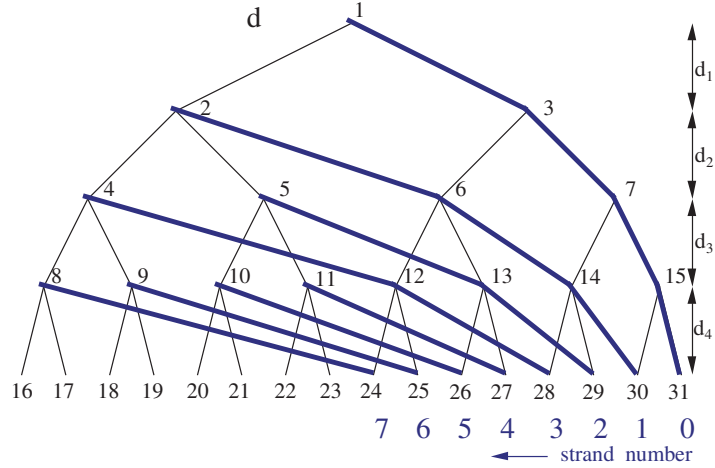, height=2.5in}
    \caption{
    A binary tree with 32 nodes, including a dud node $d$.
    Heavy blue lines
    mark the node strands of the tree.
    }
    \label{fig-strands-tree}
    \end{center}
\end{figure}

\begin{figure}[h]
    \begin{center}
    \epsfig{file=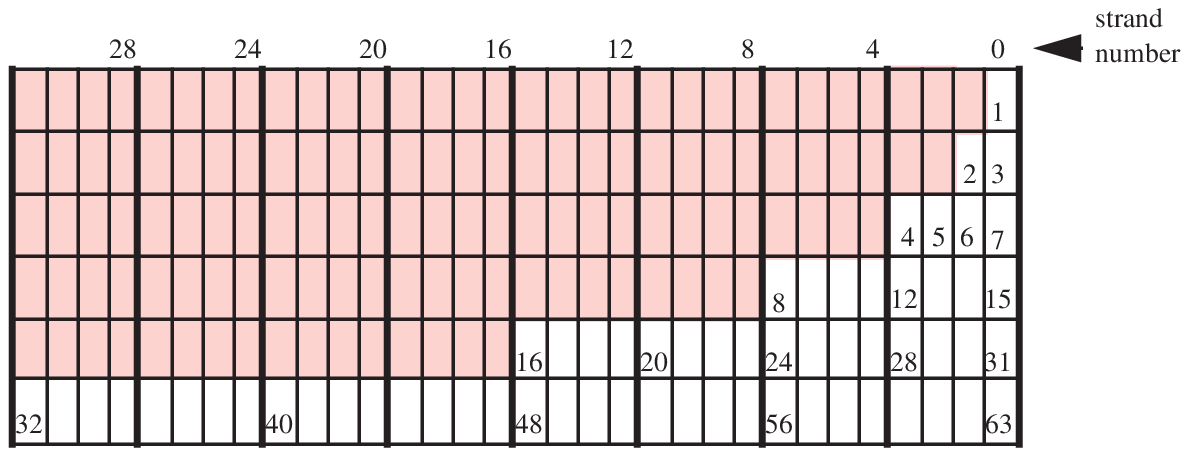, height=2.0in}
    \caption{
    Each column of this
    ``log2 staircase"
    gives
    a different strand of the binary tree
    in Fig.\ref{fig-strands-tree}.
    Strand numbers are given at the top edge.}
    \label{fig-strands-log-staircase}
    \end{center}
\end{figure}

Figures \ref{fig-strands-tree},
\ref{fig-strands-log-staircase}
and \ref{fig-strands-square}
are three alternative ways of
visualizing strands.

It is convenient to
define infinitely many
strands, and to define each strand as an
infinite
sequence of integers.
The
integers
in a strand label some nodes of
an infinite binary tree.
When we consider a finite
tree, the strands are truncated
to a finite length.

For $s=0,1,2,\ldots$, let

\beq
\minlam(s) = \ceil{\log_2(s+1)}+1
\;.
\eeq
Thus,

\beq
\begin{array}{|c|c|}
s & \minlam(s) \\ \hline\hline
0 & 1 \\ \hline
1& 2 \\ \hline
2,3&3 \\ \hline
4-7&4\\ \hline
8-15&5\\ \hline
\end{array}
\;.
\eeq
By the {\bf strand} $s$,
we mean the  infinite sequence

\beq
2^{\minlam(s)}-s-1,\;\;
2^{\minlam(s)+1}-s-1,\;\;
2^{\minlam(s)+2}-s-1,\;\;\ldots
\;.
\eeq
Note
from our $\Lam=5$ example that,
in the definition of
$\cala_\twotolam^\sdep$,
strand $s$ is truncated when it reaches
the term:

\beq
\nu_f(s,\lam) = \twotolam - s -1
\;.
\eeq
Thus, for $\lam\in Z_{1,\Lam}$
and $s\in Z_{0, 2^{\Lam-1}-1}$,

\beq
\cala_\twotolam^\sdep =
\sum_{\lam' = \minlam(s)}^{\lam-1}
d_{\lam'}
\ket{2^{\lam'}-s-1}
\bra{2^{\lam'+1}-s-1}
\;,
\eeq
and

\beq
\calb^\sdep_\twotolam = d_\lam
\ket{\twotolam-s-1}
\bra{\twotolam-s-1}
\;.
\eeq

Note that $2^{\Lam-1}$ strands (
$2^{\Lam-1}=$
a quarter of the number of nodes and
half the number of leaves of the tree)
are active (come into play) in the definition
of $\cala_{2^{\Lam+1}}$. For example,
16 strands are active in the definition
of $\cala_{64}$, and $\cala_{64}$
is associated with a tree with
64 nodes and 32 leaves.

\subsection{Factoring into CSD-ready Evolutions}

In this section, we will
factor
$e^{i \cala^\sdep_{2^{\Lam+1}}}$
for each $s$
into CSD-ready evolutions.

Let us consider an
example first, before dealing with
the general case. Suppose $\Lam=5$
and consider $\exp(i\cala^\sdep_{64})$.
By applying
Eq.(\ref{eq-simplest-recursion})
repeatedly, we can factor
 $e^{i \cala^\sdep_{64}}$
 into CSD-ready evolutions,
as follows:

\beq
e^{i \cala^\sdep_{64}}=
\prod_{\lam=1,2,\ldots,5}
\Gamma^\sdep_\lam
+ \calo(\coco^3)
\;,
\eeq
where

\begin{subequations}
\label{eq-gammas-for-cala}
\beq
\Gamma^\sdep_5 =
\exp(i
\left[
\begin{array}{cc}
0 & \calbbar^{\dagger\sdep}_{32} \\
\calbbar^\sdep_{32} & 0
\end{array}
\right]
)
\;,
\label{eq-gammas5-for-cala}
\eeq

\beq
\Gamma^\sdep_4 =
\exp(i
\left[
\begin{array}{cc|c}
0 & \calbbar^{\dagger\sdep}_{16} & \\
\calbbar^\sdep_{16} & 0 & \\ \hline
& & 0_{32}
\end{array}
\right]
)
=
\exp(i
P_0\otimes\left[
\begin{array}{cc}
0 & \calbbar^{\dagger\sdep}_{16}\\
\calbbar^\sdep_{16} & 0
\end{array}
\right]
)
\;,
\label{eq-gammas4-for-cala}
\eeq

\beq
\Gamma^\sdep_3 =
\exp(i
\left[
\begin{array}{cc|c}
0 & \calbbar^{\dagger\sdep}_{8} & \\
\calbbar^\sdep_{8} & 0 & \\ \hline
& & 0_{48}
\end{array}
\right]
)
=
\exp(i
P_0^{\otimes 2}\otimes
\left[
\begin{array}{cc}
0 & \calbbar^{\dagger\sdep}_{8}\\
\calbbar^\sdep_{8} & 0
\end{array}
\right]
)
\;,
\label{eq-gammas3-for-cala}
\eeq

\beq
\Gamma^\sdep_2 =
\exp(i
\left[
\begin{array}{cc|c}
0 & \calbbar^{\dagger\sdep}_{4} & \\
\calbbar^\sdep_{4} & 0 & \\ \hline
& & 0_{56}
\end{array}
\right]
)
=
\exp(i
P_0^{\otimes 3}\otimes
\left[
\begin{array}{cc}
0 & \calbbar^{\dagger\sdep}_{4}\\
\calbbar^\sdep_{4} & 0
\end{array}
\right]
)
\;,
\label{eq-gammas2-for-cala}
\eeq
and

\beq
\Gamma^\sdep_1 =
\exp(i
\left[
\begin{array}{cc|c}
0 & \calbbar^{\dagger\sdep}_2 & \\
\calbbar^\sdep_2 & 0 & \\ \hline
& & 0_{60}
\end{array}
\right]
)
=
\exp(i
P_0^{\otimes 4}\otimes
\left[
\begin{array}{cc}
0 & \calbbar^{\dagger\sdep}_2\\
\calbbar^\sdep_2 & 0
\end{array}
\right]
)
\;.
\label{eq-gammas1-for-cala}
\eeq
\end{subequations}
$\calbbar^\sdep_\twotolam$ is defined by

\beq
\calbbar^\sdep_\twotolam =
\calb^\sdep_\twotolam
\sinc(\frac{\cala^\sdep_\twotolam}{2})
e^{i\frac{\cala^\sdep_\twotolam}{2}}
\label{eq-caltb-def-5}
\;
\eeq
for $\lam\in Z_{1,5}$.

The above example
generalizes as follows.
For $\Lam=1,2,3,\ldots$,
$s\in Z_{0,2^{\Lam-1}-1}$ and
$\lam\in Z_{1, \Lam}$,
we have

\beq
e^{i \cala^\sdep_{2^{\Lam+1}}}=
\prod_{\lam=1,2,\ldots,\Lam}
\Gamma^\sdep_\lam
+ \calo(\coco^3)
\;,
\label{eq-exp-cala-is-gamma-prod}
\eeq

\beq
\Gamma^\sdep_\lam =
\exp(i P_0^{\otimes \Lam-\lam}\otimes
\left[
\begin{array}{cc}
0 & \calbbar^{\sdep\dagger}_\twotolam\\
\calbbar^\sdep_\twotolam & 0
\end{array}
\right]
)
\;,
\eeq
and

\beq
\calbbar^\sdep_\twotolam=
\calb^\sdep_\twotolam
\sinc(\frac{\cala^\sdep_\twotolam}{2})
e^{i\frac{\cala^\sdep_\twotolam}{2}}
\;.
\label{eq-caltb-def-gen}
\eeq

\subsection{Expressing the
$\calbbar^\sdep_\twotolam$'s
in Bra-Ket Notation}
\label{sec-calbbar-in-braket}

The CSD-ready matrices $\Gamma^\sdep_\lam$
depend on matrices $\calbbar^\sdep_\twotolam$.
In this section, we will express
the $\calbbar^\sdep_\twotolam$
in bra-ket notation.

Let us consider an
example first, before dealing with
the general case. Suppose $\Lam=5$
and consider the matrices
$\calbbar^\sdep_\twotolam$
that enter into the calculation
of $\exp(i\cala^\sdep_{64})$.
To calculate $\calbbar^{(0)}_{32}$
as defined by Eq.(\ref{eq-caltb-def-5}),
we need to calculate
$\sinc(\frac{\cala^{(0)}_{32}}{2})
e^{i\frac{\cala^{(0)}_{32}}{2}}$.
We can calculate
$\sinc(\frac{\cala^{(0)}_{32}}{2})
e^{i\frac{\cala^{(0)}_{32}}{2}}$
using its Taylor expansion  at $x=0$,
provided that we can calculate
all powers of
the matrix $\cala^{(0)}_{32}$
with reasonable efficiency. This can
indeed be done because $\cala^{(0)}_{32}$
is a sparse matrix (it's effectively
a $5\times 5$ ``tridiagonal band matrix").
Indeed, only rows and columns corresponding
to nodes 1,3,7,15 and 31 are nonzero
in $\cala^{(0)}_{32}$. It
can be represented by:

\beq
\cala^{(0)}_{32}=
\begin{array}{c|c|c|c|c|c|}
&\p{1}&\p{3}&\p{7}&\p{15}&\p{31} \\ \hline
\p{1}& & d_1& & &\\ \hline
\p{3}& d_1& &d_2 & &\\ \hline
\p{7}& &d_2& &d_3 & \\ \hline
\p{15}& & &d_3 & &d_4 \\ \hline
\p{31}& & & &d_4&\\ \hline
\end{array}
\;.
\eeq
Hence,

\beq
(\cala^{(0)}_{32})^2=
\begin{array}{c|c|c|c|c|c|}
&\p{1}&\p{3}&\p{7}&\p{15}&\p{31} \\ \hline
\p{1}&d_1^2 & 0& d_1d_2& &\\ \hline
\p{3}& &d_1^2+d_2^2 & &d_2d_3 &\\ \hline
\p{7}&d_1d_2 &&d_2^2+d_3^2 & &d_3d_4 \\ \hline
\p{15}& &d_2d_3 &  &d_3^2+d_4^2& \\ \hline
\p{31}& & & d_3d_4&&d_4^2\\ \hline
\end{array}
\;.
\eeq
Other powers of
$\cala^{(0)}_{32}$ can be calculated
just as easily.
The pattern of entries
that are zero
in
$(\cala^{(0)}_{32})^{n}$
(for some integer $n\geq 1$) can
be easily understood
by appealing to Dirac bra-ket
notation. In that notation, it is
clear that
if a row number of $(\cala^{(0)}_{32})^{n}$
is the starting
point of a staggering drunk,
then the column numbers of the
non-vanishing entries
of that row are the
possible final positions of
the drunk. This,
assuming that the drunk takes
exactly $n$ steps,
either backwards or forwards, on the
5 state runway $1,3,7,15,31$.
These rules imply that, at all times,
the drunk will be
($n$ minus an even number) steps away from his
starting point.

Note that to calculate $\calbbar^{(0)}_{32}$,
 we don't even
need to calculate all the entries of
$(\cala^{(0)}_{32})^{n}$.
In $\calbbar^{(0)}_{32}$,
$(\cala^{(0)}_{32})^{n}$
appears pre-multiplied by
$\calb^{(0)}_{32} = d_5\ket{31}\bra{31}$.
This means that we only need
to calculate the last row of
$(\cala^{(0)}_{32})^{n}$.
For example,

\beq
\calb^{(0)}_{32}(\cala^{(0)}_{32})^2=
d_5 \ket{31}\bra{31}\{\;\;
\ket{31}d_4\av{15|15}d_3\bra{7}
+
\ket{31}d_4\av{15|15}d_4\bra{31}\;\;\}
\;.
\eeq
Thus, the drunk starts at
31. After 2 steps, he must be either:
back at 31, or two steps away, at 7.

The strand illustrations
Figs.\ref{fig-strands-tree},
\ref{fig-strands-log-staircase},
and
\ref{fig-strands-square}
can be used, in conjunction
with our analogy to a drunken walker,
to predict the general
form of all
$\calbbar^\sdep_{\twotolam}$
that contribute to
$\exp(i\cala^\sdep_{64})$.
One concludes that there
must exist\footnote{
An important special case is when
the $d_\lam$ (defined by
Eq.(\ref{eq-const-delta-assump}))
are all equal to $\sqrt{2}\coco$.
Appendix \ref{app-all-d-same}
gives an explicit expression for
the  $b^k_j$ in this case.
}  some real numbers $b^k_j$
such that the following is true.
For $s=0$,

\beq
\begin{array}{l}
\calbbar^{(0)}_{32}=
\ket{31}(
b^{32}_{31}\bra{31} +
ib^{32}_{15}\bra{15} +
b^{32}_{7}\bra{7} +
ib^{32}_{3}\bra{3} +
b^{32}_{1}\bra{1})
\\
\calbbar^{(0)}_{16}=
\ket{15}(
b^{16}_{15}\bra{15} +
ib^{16}_{7}\bra{7} +
b^{16}_{3}\bra{3} +
ib^{16}_{1}\bra{1})
\\
\calbbar^{(0)}_{8}=
\ket{7}(
b^8_{7}\bra{7} +
ib^8_{3}\bra{3} +
b^8_{1}\bra{1})
\\
\calbbar^{(0)}_{4}=
\ket{3}(
b^4_3\bra{3} +
ib^4_1\bra{1}
)
\\
\calbbar^{(0)}_{2}=
\ket{1}(b^2_1\bra{1})
\end{array}
\;.
\eeq
For $s=1$,

\beq
\begin{array}{l}
\calbbar^{(1)}_{32}=
\ket{30}(
b^{32}_{30}\bra{30} +
ib^{32}_{14}\bra{14} +
b^{32}_{6}\bra{6} +
ib^{32}_{2}\bra{2})
\\
\calbbar^{(1)}_{16}=
\ket{14}(
b^{16}_{14}\bra{14} +
ib^{16}_{6}\bra{6} +
b^{16}_{2}\bra{2})
\\
\calbbar^{(1)}_{8}=
\ket{6}(
b^8_{6}\bra{6}+
ib^8_{2}\bra{2}
)
\\
\calbbar^{(1)}_{4}=
\ket{2}(b^4_2\bra{2})
\\
\calbbar^{(1)}_{2}=0
\end{array}
\;.
\eeq
For $k\in Z_{0,1}$ (and
$s=2+k\in Z_{2,3})$,

\beq
\begin{array}{l}
\calbbar^{(2+k)}_{32}=
\ket{29-k}(
b^{32}_{29-k}\bra{29-k} +
ib^{32}_{13-k}\bra{13-k} +
b^{32}_{5-k}\bra{5-k})
\\
\calbbar^{(2+k)}_{16}=
\ket{13-k}(
b^{16}_{13-k}\bra{13-k} +
ib^{16}_{5-k}\bra{5-k}
)
\\
\calbbar^{(2+k)}_{8}=
\ket{5-k}(
b^8_{5-k}\bra{5-k})
\\
\calbbar^{(2+k)}_{4}=
\calbbar^{(2+k)}_{2}=0
\end{array}
\;.
\eeq
For $k\in Z_{0,3}$ (and
$s=4+k\in Z_{4,7})$,

\beq
\begin{array}{l}
\calbbar^{(4+k)}_{32}=
\ket{27-k}(
b^{32}_{27-k}\bra{27-k}+
b^{32}_{11-k}\bra{11-k})
\\
\calbbar^{(4+k)}_{16}=
\ket{11-k}(
b^{16}_{11-k}\bra{11-k})
\\
\calbbar^{(4+k)}_{8}=
\calbbar^{(4+k)}_{4}=
\calbbar^{(4+k)}_{2}=0
\end{array}
\;.
\eeq
For $k\in Z_{0,7}$ (and
$s=8+k\in Z_{8,15})$,

\beq
\begin{array}{l}
\calbbar^{(8+k)}_{32}=
\ket{23-k}(
b^{32}_{23-k}\bra{23-k})
\\
\calbbar^{(8+k)}_{16}=
\calbbar^{(8+k)}_{8}=
\calbbar^{(8+k)}_{4}=
\calbbar^{(8+k)}_{2}=0
\end{array}
\;.
\eeq

The above example
generalizes as follows.
For $\Lam=1,2,3,\ldots$,
$s\in Z_{0,2^{\Lam-1}-1}$ and
$\lam\in Z_{1, \Lam}$,
we have

\beq
\calbbar^\sdep_\twotolam=
\ket{\twotolam-s-1}
\sum_{\lam'\in Z_{\minlam(s),\lam}}
i^{\theta(|\lam-\lam'|\;\;\mbox{\tiny is odd})}
b^\twotolam_{2^{\lam'}-s-1}
\bra{2^{\lam'}-s-1}
\;.
\eeq
Note that
$\calbbar^\sdep_\twotolam=0$ if $\lam< \minlam(s)$.

\subsection{Finding
SVD of $\calbbar^\sdep_{\twotolam}$'s}

In this section, we find an
SVD for each  $\calbbar^\sdep_{\twotolam}$.
Luckily,
such SVD's
are one sided, and can be
found in closed form.

As an example, assume $\Lam=5$,
and consider all $\calbbar^\sdep_\twotolam$
that enter into the
calculation of $\exp(i\cala_{64})$.

\begin{Lemma}
\label{lem-svd-calbbar}
The following SVD holds
for some
 $\rho^{32}_{31},
 \theta^{32}_{31},
 \theta^{32}_{15},
 \theta^{32}_{7},
 \theta^{32}_{3}\in\RR$:

\beq
\calbbar^{(0)}_{32}=
\rho^{32}_{31}
\ket{31}\bra{31}
e^{i\theta^{32}_{31}\sigx(4)}
e^{i\nbar(4)\theta^{32}_{15}\sigx(3)}
e^{i\nbar(4)\nbar(3)\theta^{32}_{7}\sigx(2)}
e^{i\nbar(4)\nbar(3)\nbar(2)\theta^{32}_{3}\sigx(1)}
\;.
\eeq
\end{Lemma}
\proof

In this proof, the symbols $b,\rho,\theta$
should all have a superscript of 32. We
will omit this superscript to simplify the
notation.

In the previous section, we showed
that

\beq
\calbbar^{(0)}_{32}=
\ket{31}(
b_{31}\bra{31}+
ib_{15}\bra{15}+
b_{7}\bra{7}+
ib_{3}\bra{3}+
b_{1}\bra{1})
\;.
\eeq
Note that

\beq
31=(11111)
\;,\;\;
15=(01111)
\;,\;\;
7=(00111)
\;,\;\;
3=(00011)
\;,\;\;
1=(00001)
\;.
\eeq

If we define $\rho_3$ and $\theta_3$ by

\beq
\rho_3=\sqrt{b_3^2+b_1^2}
\;,\;\;
\cos\theta_3=\frac{b_3}{\rho_3}
\;,\;\;
\sin\theta_3=\frac{-b_1}{\rho_3}
\;,
\eeq
then

\beq
(ib_3,b_1)=
(i\rho_3,0)
\left[
\begin{array}{cc}
b_3&-ib_1\\
-ib_1&b_3
\end{array}
\right]
\frac{1}{\rho_3}=
(i\rho_3,0)
e^{i\theta_3\sigx}
\;.
\eeq
Thus,

\beq
\left(
\begin{array}{l}
b_{31}\bra{31}+
ib_{15}\bra{15}+
b_{7}\bra{7}+\\
+ib_{3}\bra{3}
+b_{1}\bra{1}
\end{array}
\right)=
\left(
\begin{array}{l}
b_{31}\bra{31}+
ib_{15}\bra{15}+
b_{7}\bra{7}+\\
+i\rho_{3}\bra{3}
\end{array}
\right)
e^{i\nbar(4)\nbar(3)\nbar(2)\theta_3\sigx(1)}
\;.
\eeq

If we define $\rho_7$ and $\theta_7$ by

\beq
\rho_7=\sqrt{b_7^2+\rho_3^2}
\;,\;\;
\cos\theta_7=\frac{b_7}{\rho_7}
\;,\;\;
\sin\theta_7=\frac{\rho_3}{\rho_7}
\;,
\eeq
then

\beq
(b_7,i\rho_3)=
(\rho_7,0)
\left[
\begin{array}{cc}
b_7&i\rho_3\\
i\rho_3&b_7
\end{array}
\right]
\frac{1}{\rho_7}=
(\rho_7,0)
e^{i\theta_7\sigx}
\;.
\eeq
Thus,

\beq
\left(
\begin{array}{l}
b_{31}\bra{31}+
ib_{15}\bra{15}+
\\
+b_{7}\bra{7}
+i\rho_{3}\bra{3}
\end{array}
\right)=
\left(
\begin{array}{l}
b_{31}\bra{31}+
ib_{15}\bra{15}+
\\
+\rho_{7}\bra{7}
\end{array}
\right)
e^{i\nbar(4)\nbar(3)\theta_7\sigx(2)}
\;.
\eeq

If we define $\rho_{15}$ and $\theta_{15}$ by

\beq
\rho_{15}=\sqrt{b_{15}^2+\rho_7^2}
\;,\;\;
\cos\theta_{15}=\frac{b_{15}}{\rho_{15}}
\;,\;\;
\sin\theta_{15}=\frac{-\rho_7}{\rho_{15}}
\;,
\eeq
then

\beq
(ib_{15},\rho_7)=
(i\rho_{15},0)
\left[
\begin{array}{cc}
b_{15}&-i\rho_7\\
-i\rho_7&b_{15}
\end{array}
\right]
\frac{1}{\rho_{15}}=
(i\rho_{15},0)
e^{i\theta_{15}\sigx}
\;.
\eeq
Thus,

\beq
\left(
\begin{array}{l}
b_{31}\bra{31}+
\\
+ib_{15}\bra{15}+
\rho_{7}\bra{7}
\end{array}
\right)=
\left(
\begin{array}{l}
b_{31}\bra{31}+
\\
+i\rho_{15}\bra{15}
\end{array}
\right)
e^{i\nbar(4)\theta_{15}\sigx(3)}
\;.
\eeq

If we define $\rho_{31}$ and $\theta_{31}$ by

\beq
\rho_{31}=\sqrt{b_{31}^2+\rho_{15}^2}
\;,\;\;
\cos\theta_{31}=\frac{b_{31}}{\rho_{31}}
\;,\;\;
\sin\theta_{31}=\frac{\rho_{15}}{\rho_{31}}
\;,
\eeq
then

\beq
(b_{31},i\rho_{15})=
(\rho_{31},0)
\left[
\begin{array}{cc}
b_{31}&i\rho_{15}\\
i\rho_{15}&b_{31}
\end{array}
\right]
\frac{1}{\rho_{31}}=
(\rho_{31},0)
e^{i\theta_{31}\sigx}
\;.
\eeq
Thus,

\beq
\left(
\begin{array}{l}
b_{31}\bra{31}+
i\rho_{15}\bra{15}
\end{array}
\right)=
\rho_{31}\bra{31}
e^{i\theta_{31}\sigx(4)}
\;.
\eeq
\qed


The techniques
used in the proof of the
above claim can also
be used to find an SVD
of
$\calbbar^\sdep_{\twotolam}$
for all $s\in Z_{0,15}$
and $\lam\in Z_{1,5}$
(These are all the $\calbbar^\sdep_\twotolam$
that enter into the
calculation of $\exp(i\cala_{64})$).
One finds the following.

For $s=0$,

\beq
\begin{array}{l}
\calbbar^{(0)}_{32}=
\rho^{32}_{31}
\ket{31}\bra{31}
e^{i\theta^{32}_{31}\sigx(4)}
e^{i\nbar(4)\theta^{32}_{15}\sigx(3)}
e^{i\nbar(4)\nbar(3)\theta^{32}_{7}\sigx(2)}
e^{i\nbar(4)\nbar(3)\nbar(2)\theta^{32}_{3}\sigx(1)}
\\
\calbbar^{(0)}_{16}=
\rho^{16}_{15}
\ket{15}\bra{15}
e^{i\theta^{16}_{15}\sigx(3)}
e^{i\nbar(3)\theta^{16}_{7}\sigx(2)}
e^{i\nbar(3)\nbar(2)\theta^{16}_{3}\sigx(1)}
\\
\calbbar^{(0)}_{8}=
\rho^{8}_{7}
\ket{7}\bra{7}
e^{i\theta^{8}_{7}\sigx(2)}
e^{i\nbar(2)\theta^{8}_{3}\sigx(1)}
\\
\calbbar^{(0)}_{4}=
\rho^{4}_{3}
\ket{3}\bra{3}
e^{i\theta^{4}_{3}\sigx(1)}
\\
\calbbar^{(0)}_{2}=
\rho^{2}_{1}
\ket{1}\bra{1}
\end{array}
\;.
\label{eq-svd-caltb-s0}
\eeq

For $s=1$,

\beq
\begin{array}{l}
\calbbar^{(1)}_{32}=
\rho^{32}_{30}
\ket{30}\bra{30}
e^{i\theta^{32}_{30}\sigx(4)}
e^{i\nbar(4)\theta^{32}_{14}\sigx(3)}
e^{i\nbar(4)\nbar(3)\theta^{32}_{6}\sigx(2)}
\\
\calbbar^{(1)}_{16}=
\rho^{16}_{14}
\ket{14}\bra{14}
e^{i\theta^{16}_{14}\sigx(3)}
e^{i\nbar(3)\theta^{16}_{6}\sigx(2)}
\\
\calbbar^{(1)}_{8}=
\rho^{8}_{6}
\ket{6}\bra{6}
e^{i\theta^{8}_{6}\sigx(2)}
\\
\calbbar^{(1)}_{4}=
\rho^{4}_{2}\ket{2}\bra{2}
\\
\calbbar^{(1)}_{2}=0
\end{array}
\;.
\label{eq-svd-caltb-s1}
\eeq

For $k\in Z_{0,1}$ (and $s=2+k \in Z_{2,3}$),

\beq
\begin{array}{l}
\calbbar^{(2+k)}_{32}=
\rho^{32}_{29-k}
\ket{29-k}\bra{29-k}
e^{i\theta^{32}_{29-k}\sigx(4)}
e^{i\nbar(4)\theta^{32}_{13-k}\sigx(3)}
\\
\calbbar^{(2+k)}_{16}=
\rho^{16}_{13-k}
\ket{13-k}\bra{13-k}
e^{i\theta^{16}_{13-k}\sigx(3)}
\\
\calbbar^{(2+k)}_{8}=
\rho^{8}_{5-k}
\ket{5-k}\bra{5-k}
\\
\calbbar^{(2+k)}_{4}=
\calbbar^{(2+k)}_{2}=0
\end{array}
\;.
\label{eq-svd-caltb-s2}
\eeq

For $k\in Z_{0,3}$ (and
$s=4+k\in Z_{4,7})$,

\beq
\begin{array}{l}
\calbbar^{(4+k)}_{32}=
\rho^{32}_{27-k}
\ket{27-k}\bra{27-k}
e^{i\theta^{32}_{27-k}\sigx(4)}
\\
\calbbar^{(4+k)}_{16}=
\rho^{16}_{11-k}
\ket{11-k}\bra{11-k}
\\
\calbbar^{(4+k)}_{8}=
\calbbar^{(4+k)}_{4}=
\calbbar^{(4+k)}_{2}=0
\end{array}
\;.
\label{eq-svd-caltb-s4}
\eeq

For $k\in Z_{0,7}$ (and
$s=8+k\in Z_{8,15})$,

\beq
\begin{array}{l}
\calbbar^{(8+k)}_{32}=
\rho^{32}_{23-k}
\ket{23-k}\bra{23-k}
\\
\calbbar^{(8+k)}_{16}=
\calbbar^{(8+k)}_{8}=
\calbbar^{(8+k)}_{4}=
\calbbar^{(8+k)}_{2}=0
\end{array}
\;.
\label{eq-svd-caltb-s8}
\eeq

\subsection{Expressing the
$\Gamma^\sdep_\lam$'s
as EG Circuits}

In this section, we
express each $\Gamma^\sdep_\lam$ as
an EG (elementary gates) circuit.

Let us consider an
example first, before dealing with
the general case. Suppose $\Lam=5$.

Consider $s=0$.

Plugging the SVD of
$\calbbar^{(0)}_{32}$,
given by Eq.(\ref{eq-svd-caltb-s0}),
into the definition
of $\Gamma^{(0)}_5$,
given by Eq.(\ref{eq-gammas5-for-cala}),
one finds

\beqa
\Gamma^{(0)}_5
&=&
\left[
\begin{array}{cc}
U^\dagger_{32}&\\
&I_{32}
\end{array}
\right]
e^{i\rho^{32}_{31}(\ket{63}\bra{31}+h.c.)}
\left[
\begin{array}{cc}
U_{32}&\\
&I_{32}
\end{array}
\right]
\\&=&
U_{32}(4,3,2,1,0)^{\dagger\nbar(5)}
e^{i\rho^{32}_{31}\sigx(5)n(4)n(3)n(2)n(1)n(0)}
U_{32}(4,3,2,1,0)^{\nbar(5)}
\;,
\label{eq-gamma-pre-ckt-s0-5}
\eeqa
where $U_{32}$ is defined by

\beq
U_{32}(4,3,2,1,0)=
e^{i\theta^{32}_{31}\sigx(4)}
e^{i\nbar(4)\theta^{32}_{15}\sigx(3)}
e^{i\nbar(4)\nbar(3)\theta^{32}_{7}\sigx(2)}
e^{i\nbar(4)\nbar(3)\nbar(2)\theta^{32}_{3}\sigx(1)}
\;.
\eeq
Eq.(\ref{eq-gamma-pre-ckt-s0-5})
simplifies to

\beq
\begin{array}{r}
\Gamma_5^{(0)}
=
e^{-i\nbar(5)\nbar(4)\nbar(3)\nbar(2)\theta^{32}_{3}\sigx(1)}
e^{-i\nbar(5)\nbar(4)\nbar(3)\theta^{32}_{7}\sigx(2)}
e^{-i\nbar(5)\nbar(4)\theta^{32}_{15}\sigx(3)}
e^{-i\nbar(5)\theta^{32}_{31}\sigx(4)}
\\
\odot\{
e^{i\rho^{32}_{31}\sigx(5)n(4)n(3)n(2)n(1)n(0)}
\}
\end{array}
\;.
\eeq

Plugging the SVD of
$\calbbar^{(0)}_{16}$,
given by Eq.(\ref{eq-svd-caltb-s0}),
into the definition
of $\Gamma^{(0)}_4$,
given by Eq.(\ref{eq-gammas4-for-cala}),
one finds

\beqa
\Gamma^{(0)}_4
&=&
\left[
\begin{array}{cc}
U^\dagger_{16}&\\
&I_{16}
\end{array}
\right]^{\oplus 2}
e^{iP_0\otimes(\rho^{16}_{15}\ket{31}\bra{15}+h.c.)}
\left[
\begin{array}{cc}
U_{16}&\\
&I_{16}
\end{array}
\right]^{\oplus 2}
\\&=&
U_{16}(3,2,1,0)^{\dagger\nbar(4)}
e^{i\nbar(5)\rho^{16}_{15}\sigx(4)n(3)n(2)n(1)n(0)}
U_{16}(3,2,1,0)^{\nbar(4)}
\;,
\label{eq-gamma-pre-ckt-s0-4}
\eeqa
where $U_{16}$ is defined by

\beq
U_{16}(3,2,1,0)=
e^{i\theta^{16}_{15}\sigx(3)}
e^{i\nbar(3)\theta^{16}_{7}\sigx(2)}
e^{i\nbar(3)\nbar(2)\theta^{16}_{3}\sigx(1)}
\;.
\eeq
Eq.(\ref{eq-gamma-pre-ckt-s0-4})
simplifies to

\beq
\Gamma_4^{(0)}=
e^{-i\nbar(4)\nbar(3)\nbar(2)\theta^{16}_{3}\sigx(1)}
e^{-i\nbar(4)\nbar(3)\theta^{16}_{7}\sigx(2)}
e^{-i\nbar(4)\theta^{16}_{15}\sigx(3)}
\odot\{
e^{i\nbar(5)\rho^{16}_{15}\sigx(4)n(3)n(2)n(1)n(0)}
\}
\;.
\eeq

Similarly, one finds

\beqa
\Gamma^{(0)}_3
&=&
\left[
\begin{array}{cc}
U^\dagger_{8}&\\
&I_{8}
\end{array}
\right]^{\oplus 4}
e^{iP_0^{\otimes 2}\otimes(\rho^8_7\ket{15}\bra{7}+h.c.)}
\left[
\begin{array}{cc}
U_{8}&\\
&I_{8}
\end{array}
\right]^{\oplus 4}
\\&=&
U_{8}(2,1,0)^{\dagger\nbar(3)}
e^{i\nbar(5)\nbar(4)\rho^{8}_7\sigx(3)n(2)n(1)n(0)}
U_{8}(2,1,0)^{\nbar(3)}
\;,
\label{eq-gamma-pre-ckt-s0-3}
\eeqa
where

\beq
U_{8}(2,1,0)=
e^{i\theta^{8}_{7}\sigx(2)}
e^{i\nbar(2)\theta^{8}_{3}\sigx(1)}
\;.
\eeq
Eq.(\ref{eq-gamma-pre-ckt-s0-3})
simplifies to

\beq
\Gamma^{(0)}_3=
e^{-i\nbar(3)\nbar(2)\theta^{8}_{3}\sigx(1)}
e^{-i\nbar(3)\theta^{8}_{7}\sigx(2)}
\odot\{
e^{i\nbar(5)\nbar(4)\rho^8_7\sigx(3)n(2)n(1)n(0)}
\}
\;.
\eeq

Similarly, one finds

\beqa
\Gamma^{(0)}_2
&=&
\left[
\begin{array}{cc}
U^\dagger_{4}&\\
&I_{4}
\end{array}
\right]^{\oplus 8}
e^{iP_0^{\otimes 2}\otimes(\rho^8_7\ket{15}\bra{7}+h.c.)}
\left[
\begin{array}{cc}
U_{4}&\\
&I_{4}
\end{array}
\right]^{\oplus 8}
\\&=&
U_{4}(1,0)^{\dagger\nbar(2)}
e^{i\nbar(5)\nbar(4)\nbar(3)\rho^{4}_3\sigx(2)n(1)n(0)}
U_{4}(1,0)^{\nbar(2)}
\;,
\label{eq-gamma-pre-ckt-s0-2}
\eeqa
where

\beq
U_{4}(1,0)=
e^{i\theta^{4}_{3}\sigx(1)}
\;.
\eeq
Eq.(\ref{eq-gamma-pre-ckt-s0-2})
simplifies to

\beq
\Gamma^{(0)}_2=
e^{-i\nbar(2)\theta^{4}_{3}\sigx(1)}\odot\{
e^{i\nbar(5)\nbar(4)\nbar(3)\rho^4_3\sigx(2)n(1)n(0)}
\}
\;.
\eeq

Similarly, one finds

\beqa
\Gamma^{(0)}_1
&=&
e^{iP_0^{\otimes 4}\otimes(\rho^2_1\ket{3}\bra{1}+h.c.)}
\\&=&
e^{i\nbar(5)\nbar(4)\nbar(3)\nbar(2)\rho^2_1\sigx(1)n(0)}
\;.
\eeqa

\begin{figure}[h]
    \begin{center}
    \epsfig{file=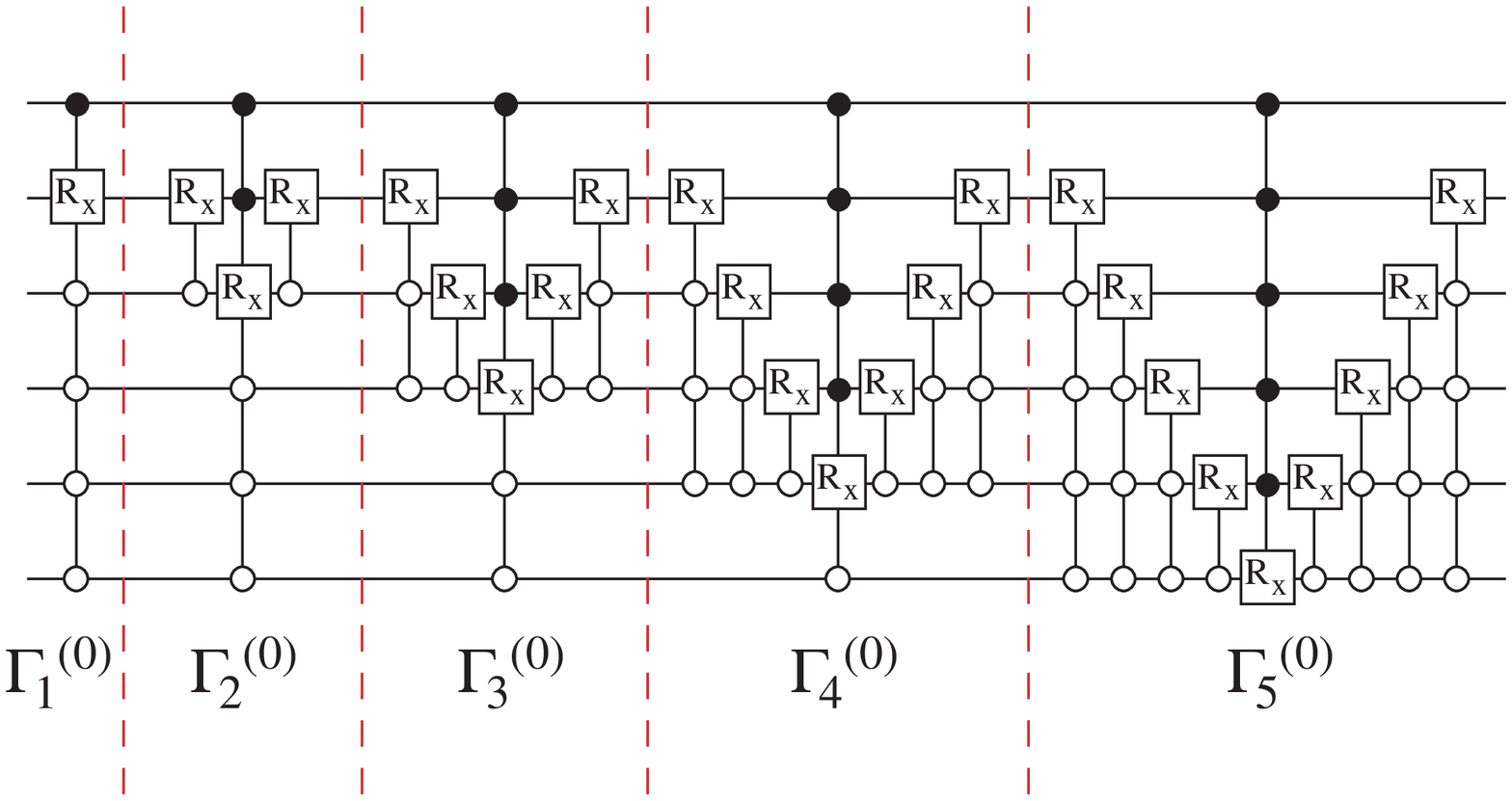, height=2.5in}
    \caption{
    $\Gamma^\sdep_\lam$'s for
    $s=0$.
    }
    \label{fig-Gammas0}
    \end{center}
\end{figure}

To summarize, we have shown that:
\beq
\begin{array}{l}
\begin{array}{r}
\Gamma_5^{(0)}
=
e^{-i\nbar(5)\nbar(4)\nbar(3)\nbar(2)\theta^{32}_{3}\sigx(1)}
e^{-i\nbar(5)\nbar(4)\nbar(3)\theta^{32}_{7}\sigx(2)}
e^{-i\nbar(5)\nbar(4)\theta^{32}_{15}\sigx(3)}
e^{-i\nbar(5)\theta^{32}_{31}\sigx(4)}
\\
\odot\{
e^{i\rho^{32}_{31}\sigx(5)n(4)n(3)n(2)n(1)n(0)}
\}
\end{array}
\\
\Gamma_4^{(0)}=
e^{-i\nbar(4)\nbar(3)\nbar(2)\theta^{16}_{3}\sigx(1)}
e^{-i\nbar(4)\nbar(3)\theta^{16}_{7}\sigx(2)}
e^{-i\nbar(4)\theta^{16}_{15}\sigx(3)}
\odot\{
e^{i\nbar(5)\rho^{16}_{15}\sigx(4)n(3)n(2)n(1)n(0)}
\}
\\
\Gamma^{(0)}_3=
e^{-i\nbar(3)\nbar(2)\theta^{8}_{3}\sigx(1)}
e^{-i\nbar(3)\theta^{8}_{7}\sigx(2)}
\odot\{
e^{i\nbar(5)\nbar(4)\rho^8_7\sigx(3)n(2)n(1)n(0)}
\}
\\
\Gamma^{(0)}_2=
e^{-i\nbar(2)\theta^{4}_{3}\sigx(1)}\odot\{
e^{i\nbar(5)\nbar(4)\nbar(3)\rho^4_3\sigx(2)n(1)n(0)}
\}
\\
\Gamma^{(0)}_1=
e^{i\nbar(5)\nbar(4)\nbar(3)\nbar(2)\rho^2_1\sigx(1)n(0)}
\end{array}
\;.
\label{eq-seos-gamma0}
\eeq
Fig.\ref{fig-Gammas0}
represents
the SEOs of Eq.(\ref{eq-seos-gamma0})
as circuits.

\begin{figure}[h]
    \begin{center}
    \epsfig{file=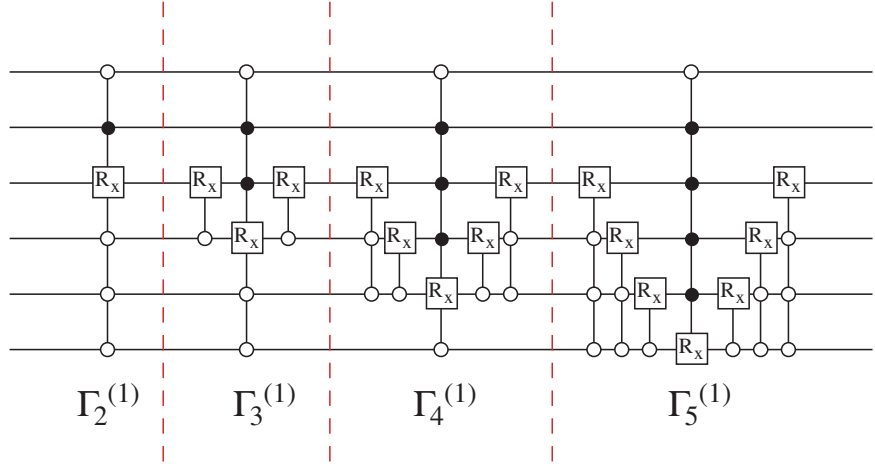, height=2.5in}
    \caption{
    $\Gamma^\sdep_\lam$'s for
    $s=1$. $\Gamma^{(1)}_1$
    is not shown because it equals 1.
    }
    \label{fig-Gammas1}
    \end{center}
\end{figure}

Using the same method that we used
for $s=0$, we find
for $s=1$:

\beq
\begin{array}{l}
\Gamma_5^{(1)}=
e^{-i\nbar(5)\nbar(4)\nbar(3)\theta^{32}_{6}\sigx(2)}
e^{-i\nbar(5)\nbar(4)\theta^{32}_{14}\sigx(3)}
e^{-i\nbar(5)\theta^{32}_{30}\sigx(4)}
\odot\{
e^{i\rho^{32}_{30}\sigx(5)n(4)n(3)n(2)n(1)\nbar(0)}
\}
\\
\Gamma_4^{(1)}=
e^{-i\nbar(4)\nbar(3)\theta^{16}_{6}\sigx(2)}
e^{-i\nbar(4)\theta^{16}_{14}\sigx(3)}
\odot\{
e^{i\nbar(5)\rho^{16}_{14}\sigx(4)n(3)n(2)n(1)\nbar(0)}
\}
\\
\Gamma_3^{(1)}=
e^{-i\nbar(3)\theta^{8}_{6}\sigx(2)}
\odot\{
e^{i\nbar(5)\nbar(4)\rho^8_6\sigx(3)n(2)n(1)\nbar(0)}
\}
\\
\Gamma_2^{(1)}=
e^{i\nbar(5)\nbar(4)\nbar(3)\rho^4_2\sigx(2)n(1)\nbar(0)}
\\
\Gamma_1^{(1)}=1
\end{array}
\;.
\label{eq-seos-gamma1}
\eeq
Fig.\ref{fig-Gammas1}
represents the SEOs
of Eq.(\ref{eq-seos-gamma1})
as circuits.

\begin{figure}[h]
    \begin{center}
    \epsfig{file=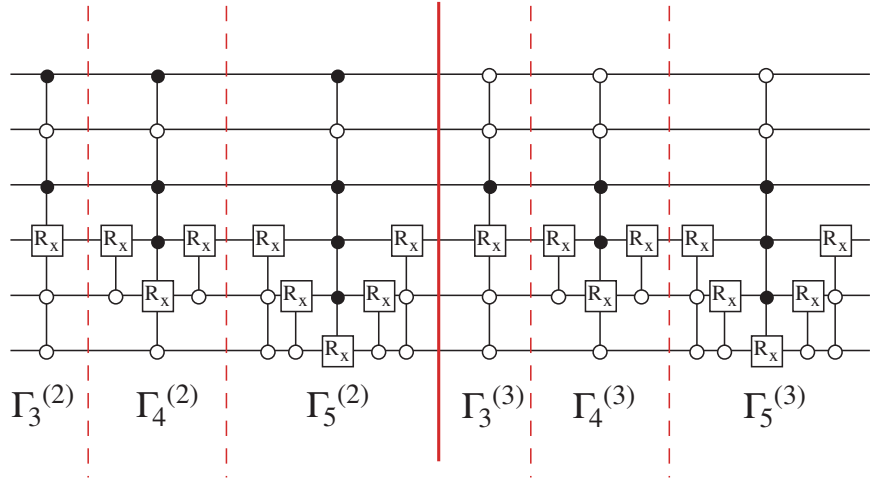, height=2.5in}
    \caption{
    $\Gamma^\sdep_\lam$'s for
    $s=2,3$.
    $\Gamma^\sdep_1$ and
    $\Gamma^\sdep_2$ are not shown
    because they equal 1.
    }
    \label{fig-Gammas2-3}
    \end{center}
\end{figure}

For $s=2$, we find
\beq
\begin{array}{l}
\Gamma_5^{(2)}=
e^{-i\nbar(5)\nbar(4)\theta^{32}_{13}\sigx(3)}
e^{-i\nbar(5)\theta^{32}_{29}\sigx(4)}
\odot\{
e^{i\rho^{32}_{29}\sigx(5)n(4)n(3)n(2)\nbar(1)n(0)}
\}
\\
\Gamma^{(2)}_4 =
e^{-i\nbar(4)\theta^{16}_{13}\sigx(3)}
\odot\{
e^{i\nbar(5)\rho^{16}_{13}
\sigx(4)n(3)n(2)\nbar(1)n(0)}
\}
\\
\Gamma^{(2)}_3 = e^{i\nbar(5)\nbar(4)
\rho^8_5\sigx(3)n(2)\nbar(1)n(0)}
\\
\Gamma^{(2)}_2=\Gamma^{(2)}_1=1
\end{array}
\;.
\label{eq-seos-gamma2}
\eeq
The $\Gamma^\sdep_\lam$ for
$s=3$ are identical to those for
$s=2$,
except that one must
replace $n(0)$ by $\nbar(0)$
in them.
The angle parameters (i.e.,
$\rho^j_k, \theta^j_k, b^j_k$)
also change.
Fig.\ref{fig-Gammas2-3}
represent the SEOs of Eq.(\ref{eq-seos-gamma2})
as circuits.

\begin{figure}[h]
    \begin{center}
    \epsfig{file=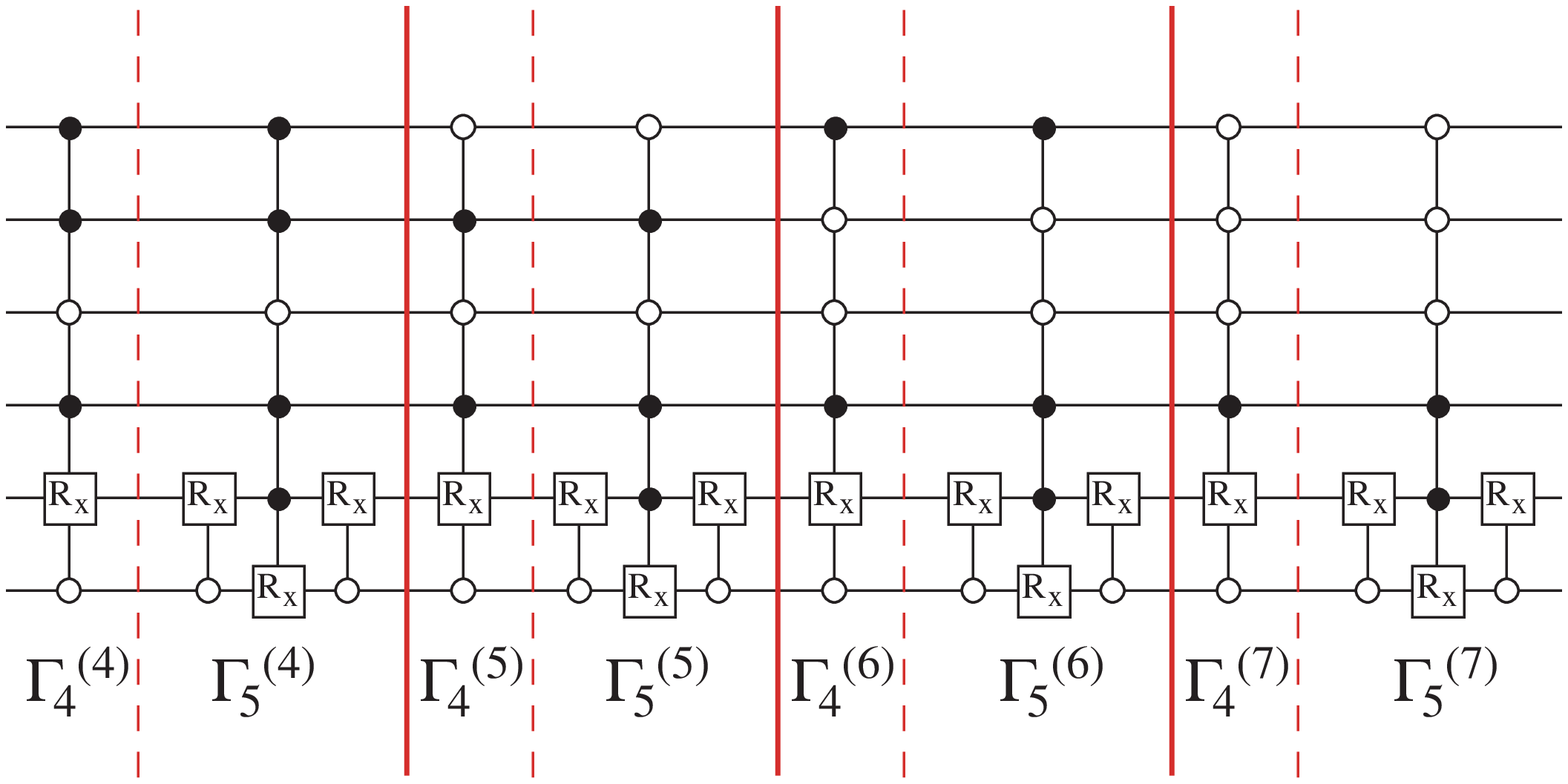, height=2.5in}
    \caption{
    $\Gamma^\sdep_\lam$'s for
    $s=4$ to $7$.
    $\Gamma^\sdep_1$,
    $\Gamma^\sdep_2$ and
    $\Gamma^\sdep_3$ are not shown
    because they equal 1.
    }
    \label{fig-Gammas4-7}
    \end{center}
\end{figure}

For $s=4$, we find

\beq
\begin{array}{l}
\Gamma^{(4)}_5 =
e^{-i\nbar(5)\theta^{32}_{27}\sigx(4)}
\odot\{
e^{i\rho^{32}_{27}\sigx(5)n(4)n(3)\nbar(2)n(1)n(0)}
\}
\\
\Gamma^{(4)}_4 = e^{i\nbar(4)\rho^{16}_{11}
\sigx(4)n(3)\nbar(2)n(1)n(0)}
\\
\Gamma^{(4)}_3=\Gamma^{(4)}_2=\Gamma^{(4)}_1=1
\end{array}
\;.
\label{eq-seos-gamma4}
\eeq
The $\Gamma^\sdep_\lam$ for $s=5,6,7$
are identical to those for
$s=4$, except that we must replace
$n(1)n(0)$ by
$n(1)\nbar(0)$,
$\nbar(1)n(0)$, and
$\nbar(1)\nbar(0)$,
respectively, in them.
The angle parameters also change.
Fig.\ref{fig-Gammas4-7}
represents the SEOs of Eq.(\ref{eq-seos-gamma4})
as circuits.

\begin{figure}[h]
    \begin{center}
    \epsfig{file=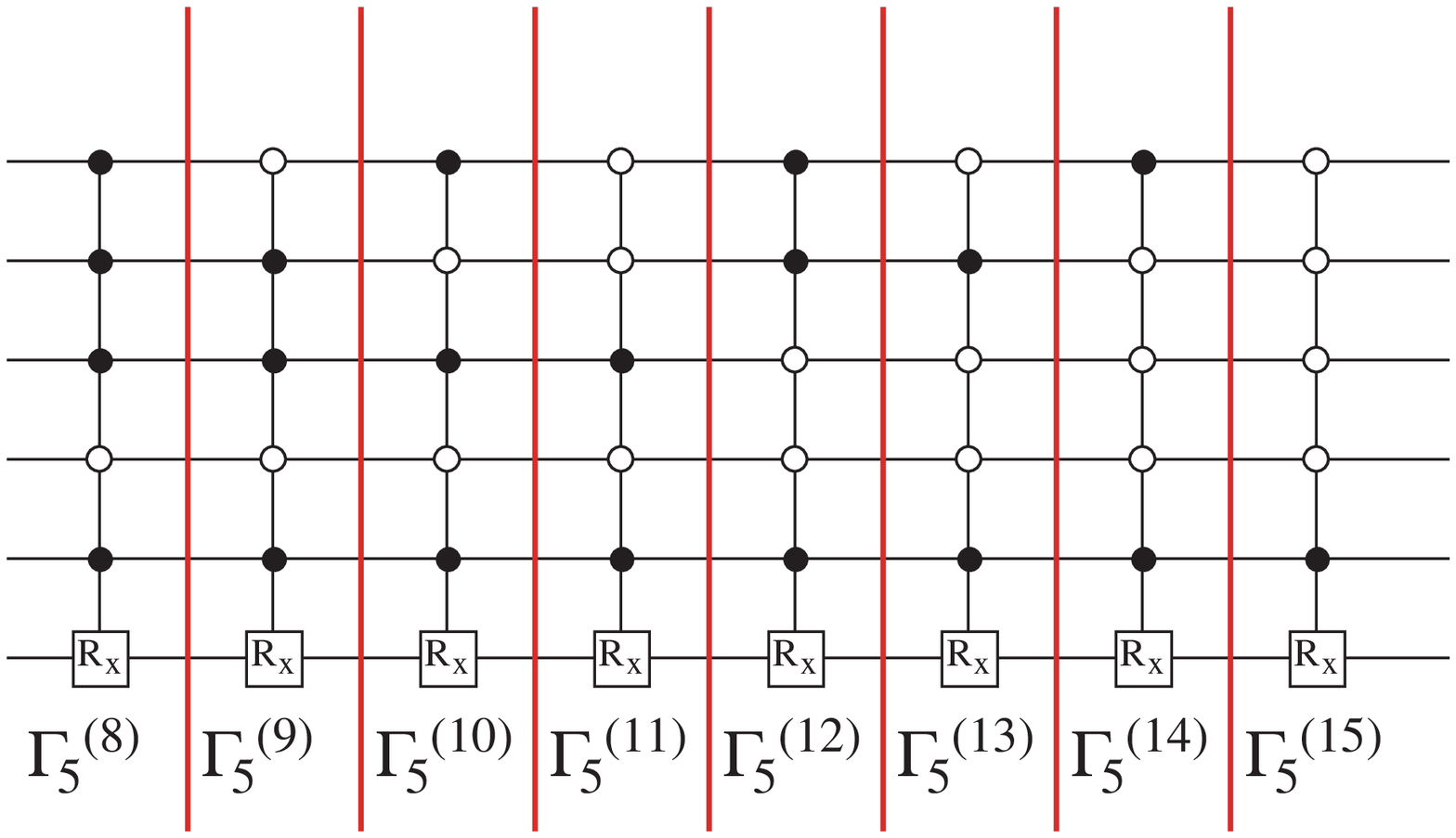, height=2.5in}
    \caption{
    $\Gamma^\sdep_\lam$'s for
    $s=8$ to $15$.
    $\Gamma^\sdep_1$,
    $\Gamma^\sdep_2$,
    $\Gamma^\sdep_3$ and
    $\Gamma^\sdep_4$ are not shown
    because they equal 1.
    }
    \label{fig-Gammas8-15}
    \end{center}
\end{figure}

For $s=8$, we find

\beq
\begin{array}{l}
\Gamma^{(8)}_5 = e^{i\rho^{32}_{23}\sigx(5)n(4)\nbar(3)n(2)n(1)n(0)}
\\
\Gamma^{(8)}_4=\Gamma^{(8)}_3=\Gamma^{(8)}_2=\Gamma^{(8)}_1=1
\end{array}
\;.
\label{eq-seos-gamma8}
\eeq
The $\Gamma^\sdep_\lam$ for $s=9,10,\ldots,15$
are identical to those for
$s=8$, except that we must replace
$n(2)n(1)n(0)$ by the
appropriate projection operator
in them. The angle parameters also change.
Fig.\ref{fig-Gammas8-15}
represents the SEOs of Eq.(\ref{eq-seos-gamma8})
as circuits.

The above example
generalizes as follows.
Suppose $\Lam=1,2,3,\ldots$
and $s\in Z_{0,2^{\Lam-1}-1}$. Then
$\Gamma^\sdep_\lam=0$ if $\lam< \minlam(s)$
since for such $\lam$,
$\calbbar^\sdep_{\twotolam}=0$.
For $\lam\in Z_{\minlam(s),\Lam}$,
there exist $\phi^\sdep_{\lam,j},
\theta^\sdep_\lam\in \RR$ such that

\begin{eqnarray}
\Gamma^\sdep_\lam&=&
\prod_{j=\lam-\minlam(s)-1,\ldots,2,1,0}
\{
e^{i
\nbar(\lam)\nbar(\lam-1)\ldots\nbar(\lam-j)
\phi^\sdep_{\lam,j}\sigx(\lam-j-1)}
\}
\nonumber
\\&&
\odot\{
e^{i\nbar(\Lam)\nbar(\Lam-1)
\ldots\nbar(\lam+1)\theta^\sdep_\lam\sigx(\lam)
P_{\overline{bin(s)}}(\lam-1,\ldots,1,0)}
\}
\;.
\label{eq-gen-gamma-s-lam}
\end{eqnarray}
In Eq.(\ref{eq-gen-gamma-s-lam}), we use
$bin(s)=(s_{\lam-1}s_{\lam-2}\ldots s_1s_0)$
to denote the binary representation of $s$, and
$\overline{bin(s)}=
(\sbar_{\lam-1}\sbar_{\lam-2}
\ldots \sbar_1 \sbar_0)$ to denote its
bitwise negation.

\subsection{Weaving the Strands}

In this section, we finally
achieve our goal of
compiling
$e^{iH_{tr}}$.

Suppose $\vecalp$ and
$\vecbet$ are n-tuples (ordered sets),
not necessarily of the same length,
of distinct qubit positions. Assume that
these two n-tuples are disjoint (if
treated as non-ordered sets).
Let $\pi$ and $\pi'$ be
commuting projection operators
acting on qubits $\vecbet$.
Let $M$ be any operator acting on qubits $\vecalp$.
A simple identity that will
be useful in what follows is

\beq
M(\vecalp)^{\pi(\vecbet)}
M(\vecalp)^{\pi'(\vecbet)}
=
M(\vecalp)^{\pi(\vecbet) + \pi'(\vecbet)}
\;.
\label{eq-id-combine-projs}
\eeq
This identity allows us to combine
operations that are being needlessly
performed separately.

Define $\Gamma^\sdep$
for $s\in Z_{0,2^{\Lam-1}-1}$ by

\beq
\Gamma^\sdep=
e^{i\cala^\sdep_{2^{\Lam+1}}}=
\prod_{\lam=1,2,\ldots,\Lam}
\Gamma^\sdep_\lam
+ \calo(\coco^3)
\;,
\label{eq-gamma-s-def}
\eeq
where we have used Eq.(\ref{eq-exp-cala-is-gamma-prod}).

Note that the $\Gamma^\sdep$
for different strands
commute, because
any two different strands ``live" on
disjoint sets of nodes.
The main point of this section is
that many operations can be combined
if we multiply the
$\Gamma^\sdep$'s  judiciously.

For definiteness, consider our
usual example of $\Lam=5$.

Note from Fig.\ref{fig-strands-tree},
that strands
can be grouped into equivalence classes
$\{0\}, \{1\}$, $Z_{2,3}$, $Z_{4,7}$
and $Z_{8,15}$.
All members of
the same equivalence class start and end
at the same tree levels. They also have
the same strengths $d_i$'s
for transitions between levels.
One could say that they are
identical in every respect except that
the names of their nodes differ.

Now consider Fig.(\ref{fig-Gammas8-15}),
which gives circuits for the $\Gamma^\sdep_\lam$'s,
 for
strand class $Z_{8,15}$.
This figure does not tell us
what are the specific angles for the
qubit rotations of the type $R_x$.
But from our observation that
equivalent strands are ``identical
except for node relabelling",
it is clear that any qubit rotation
of type $R_x$ in a strand $s_1$
is identical to the ``analogous" one
in any strand $s_2$ of the same class.
Therefore,
we can apply identity
Eq.(\ref{eq-id-combine-projs}) to combine
the $\Gamma^\sdep$ of all the strands
of class $Z_{8,15}$:

\begin{subequations}
\label{eq-gamma-class-prods}
\beq
\prod_{s\in Z_{8,15}}
\Gamma^\sdep=
\begin{array}{c}
\Qcircuit @C=1em @R=.25em @!R{
&\qw&\qw
\\
&\qw&\qw
\\
&\qw&\qw
\\
&\ogate\qwx[2]&\qw
\\
&\dotgate&\qw
\\
&\gate{R_x}&\qw
}
\end{array}
\;.
\eeq
Likewise, by
applying identity Eq.(\ref{eq-id-combine-projs})
to Fig.(\ref{fig-Gammas4-7}),
one finds that

\beq
\prod_{s\in Z_{4,7}}
\Gamma^\sdep=
\begin{array}{c}
\Qcircuit @C=1em @R=.25em @!R{
&\qw&\qw&\qw&\qw&\qw
\\
&\qw&\qw&\qw&\qw&\qw
\\
&\ogate&\qw&\ogate&\qw&\qw
\\
&\dotgate&\qw&\dotgate&\qw&\qw
\\
&\gate{R_x}\qwx[-2]&\gate{R_x}&\dotgate\qwx[-2]&\gate{R_x}&\qw
\\
&\ogate\qwx[-1]&\ogate\qwx[-1]&\gate{R_x}\qwx[-1]&\ogate\qwx[-1]&\qw
}
\end{array}
\;.
\eeq
Finally, by applying
identity Eq.(\ref{eq-id-combine-projs})
to Fig.(\ref{fig-Gammas2-3}), one finds

\beq
\prod_{s\in Z_{2,3}}
\Gamma^\sdep=
\begin{array}{c}
\Qcircuit @C=1em @R=.25em @!R{
&\qw&\qw&\qw&\qw&\qw&\qw&\qw&\qw&\qw&\qw
\\
&\ogate&\qw&\ogate&\qw&\qw&\qw&\ogate&\qw&\qw&\qw
\\
&\dotgate&\qw&\dotgate&\qw&\qw&\qw&\dotgate&\qw&\qw&\qw
\\
&\gate{R_x}\qwx[-2]\qwx[1]&\gate{R_x}\qwx[1]&\dotgate&\gate{R_x}\qwx[1]&\gate{R_x}\qwx[1]&\qw&\dotgate&\qw&\gate{R_x}\qwx[1]&\qw
\\
&\ogate&\ogate&\gate{R_x}\qwx[-3]&\ogate&\ogate&\gate{R_x}&\dotgate&\gate{R_x}&\ogate&\qw
\\
&\ogate\qwx[-1]&\qw&\ogate\qwx[-1]&\qw&\ogate\qwx[-1]&\ogate\qwx[-1]&\gate{R_x}\qwx[-4]&\ogate\qwx[-1]&\ogate\qwx[-1]&\qw
}
\end{array}
\;.
\eeq
\end{subequations}

\subsection{Complexity of $e^{iH_{tr}}$
Compilation}

In this section, we will show that
our SEO
for $e^{iH_{tr}}$
contains
$\calo(\Lam^4)$ CNOTs,
in the case of balanced binary NAND trees.

If $S$ represents a SEO or
its associated circuit,
let  $\calc(S)$ denote
its circuit complexity.
We will evaluate the circuit complexity
of our SEO for $e^{iH_{tr}}$ by counting
the number of control vertices in its
quantum circuit.

Assume a tree with
$2^{\Lam+1}$ nodes. According
to Eq.(\ref{eq-exp-a-eq-f-exp-cala-f}),

\beq
\calc(e^{iH_{tr}})=
2\calc(\calf_{2^{\Lam+1}})
+\calc(\exp(i\cala_{2^{\Lam+1}}))
\;.
\label{eq-nk-h-tr}
\eeq

According to Eq.(\ref{eq-calf-is-f-prod}),

\beq
\calc(\calf_{2^{\Lam+1}})=
\sum_{\lam=1}^\Lam
\calc(F_\twotolam(\lam-1,\dots,2,1,0)^{n(\lam)})
\;.
\label{eq-nk-calf}
\eeq
An exchange operator $E_4(\alpha,\beta)$
can be expressed as
product of 3 CNOTs. So a
singly controlled exchange operator
has six control vertices. This
observation and
Eq.(\ref{eq-f-bb-nand}) imply that

\beq
\calc(F^{n(\lam)}_\twotolam)=
6\lam-5
\;.
\label{eq-nk-f}
\eeq
Combining Eqs.(\ref{eq-nk-calf})
and (\ref{eq-nk-f})
yields

\beq
\calc(\calf_{2^{\Lam+1}})=\calo(\Lam^2)
\;.
\label{eq-nk-calf-fin}
\eeq

According to
Eqs.(\ref{eq-cala-into-strands}) and
(\ref{eq-gamma-s-def}),

\beq
\calc(\exp(i\cala_{2^{\Lam+1}}))
=
\calc(\prod_{s=0}^{2^{\Lam-1}-1}\Gamma^\sdep)
=
\calc(\Gamma^{(0)})+
\sum_{\sigma\in Z_{0,\Lam-2}}
\calc(\prod_{s=2^\sigma, 2^\sigma+1,
\dots 2^{\sigma+1}-1}\Gamma^\sdep)
\;.
\label{eq-nk-exp-cala}
\eeq
Here $\sigma$ labels the
different strand equivalence classes.
According to Eqs.(\ref{eq-gamma-class-prods}),
for $\sigma\in Z_{0,\Lam-2}$,

\beq
\calc(\prod_{s=2^\sigma, 2^\sigma+1,
\dots 2^{\sigma+1}-1}\Gamma^\sdep) =
\sum_{\lam=1}^{\Lam}
\{
\calc(\Gamma^{(2^\sigma)}_\lam) -\sigma
\}
\;.
\label{eq-nk-gamma-class}
\eeq
According to Eq.(\ref{eq-gen-gamma-s-lam})
and Figs.\ref{fig-Gammas0}
to \ref{fig-Gammas8-15},

\beq
\calc(\Gamma^\sdep_\lam)=
\Lam + 2 \sum_{j=0}^{\lam-\minlam(s)-1}
\{j+1\}
\;.
\label{eq-nk-gamma-s-lam}
\eeq
Combining Eqs.
(\ref{eq-nk-exp-cala}),
(\ref{eq-nk-gamma-class}),
(\ref{eq-nk-gamma-s-lam}),
and doing some algebra,
yields

\beq
\calc(\exp(i\cala_{2^{\Lam+1}}))=
\calo(\Lam^4)
\;.
\label{eq-exp-cala-fin}
\eeq

Now we can use Eq.(\ref{eq-nk-h-tr}),
(\ref{eq-nk-calf-fin})
and (\ref{eq-exp-cala-fin}),
to conclude that

\beq
\calc(e^{iH_{tr}})=
\calo(\Lam^4)
\;.
\eeq

\section{Compiling Input Graph (``Oracle")}
In this section, we will show how
to compile the evolution operator
$e^{iH_{in}}$ for an
input graph (``oracle").

Suppose $\Lam=1,2,3\ldots$. Consider a tree with
$\nlvs = 2^\Lam$ leaves,
with leaf inputs $x_k \in Bool$
for $k\in Z_{0, 2^\Lam-1}$.
The input Hamiltonian for such a tree is:

\beqa
H_{in}&=&
\coco
\left[
\begin{array}{cccc|cccc}
&&&&x_0&&&\\
&&&&&x_1&&\\
&&&&&&\ddots&\\
&&&&&&&x_{\nlvs-1}\\ \hline
x_0&&&&&&&\\
&x_1&&&&&&\\
&&\ddots&&&&&\\
&&&x_{\nlvs-1}&&&&
\end{array}
\right]
\\
&=&
\coco\sigx\otimes\sum_{\vecb\in Bool^\Lam}
x_\vecb P_\vecb
\;,
\label{eq-h-in-disordered-xk}
\eeqa
where $\coco\in\RR$, and
$x_\vecb$ is defined to equal $x_{dec(\vecb)}$.

Define operators $U$
and $\pi$ by

\beq
U= e^{i\coco\sigx}
\;,\;\;
\pi = \sum_{\vecb\in Bool^\Lam} x_\vecb P_\vecb
\;.
\label{eq-u-pi-defs}
\eeq
Let

\beq
\vecxi=(\Lam-1, \dots, 1,0)
\;.
\eeq
Then

\beqa
e^{iH_{in}}&=&
[U(\Lam)]^{\pi(\vecxi)}
\\
&=&
\sigx(\alpha)^{\pi(\vecxi)}
U(\Lam)^{n(\alpha)}
\sigx(\alpha)^{\pi(\vecxi)}
\ket{0}_\alpha
\;.
\label{eq-oracle-to-2-cnots}
\eeqa
Here $\alpha$ is an ancilla qubit.
To get Eq.(\ref{eq-oracle-to-2-cnots}),
we did the same as we did in
Eq.(\ref{eq-contr-u}): we replaced the
multiply controlled $U(2)$
operator by a singly controlled one,
and two MCNOTs.
So, to compile $e^{iH_{in}}$,
all we need to do is to compile
$\sigx(\alpha)^{\pi(\vecxi)}$.

The operator $\sigx(\alpha)^{\pi(\vecxi)}$
appears frequently
in the quantum computation literature,
but usually in a different guise.
It's more common to
denote the function
$x_.$ in Eq.(\ref{eq-u-pi-defs})
by a more standard name
for a function, such as
$f_\Lam:Bool^\Lam\rarrow Bool$.
In this notation,
the operator $\pi$ of Eq.(\ref{eq-u-pi-defs})
becomes

\beq
\pi=\sum_{\vec{x'}\in Bool^\Lam}
f_\Lam(\vec{x'})P_{\vec{x'}}
\;.
\label{eq-f-lam-def}
\eeq
Thus,

\beq
\sigx(\alpha)^{\pi(\vecxi)}
\ket{y}_\alpha
\ket{\vec{x}}_{\vecxi}
=
\ket{y\oplus f_\Lam(\vec{x})}_\alpha
\ket{\vec{x}}_{\vecxi}
\;.
\eeq
The best general way of
compiling $\sigx(\alpha)^{\pi(\vecxi)}$
 is not known.
It is known that there
are some $f_\Lam()$ for which
$\sigx(\alpha)^{\pi(\vecxi)}$
has complexity $\calo(2^\Lam)=\calo(\nlvs)$.
In Appendix \ref{app-banded-oracle},
we give a
compilation
for a special case, that
of a ``banded oracle"
(an oracle for which  the ordered set
of NAND formula inputs
$(x_k)_{\forall k}$
has a fixed number of bands of ones).
Our compilation of the banded oracle
evolution operator has complexity $\calo(\Lam^2)$.

\section{Overall Circuit Complexity}

The following table summarizes
the circuit complexity of the various
compilations that
have been presented in this paper:

\begin{tabular}{|l|l|l|l|}
\hline
evolution&
description &
number of CNOTs in SEO & exact?
\\
operator&&&
\\\hline\hline
$e^{iH_{tr}}$ & balanced binary NAND tree &
$\calo(\Lam^4)$ & no
\\
&with $2^\Lam=\nlvs$ leaves & &
\\
\hline
$e^{iH_{lp}}$&
loop of length
$\ns$ &$\calo(\ns)$ & yes
\\
&&
$\calo(\nlvs^{\frac{1}{2}})$
for $\ns\approx\nlvs^{\frac{1}{2}}$&
\\
\hline
$e^{iH_{gl}}$&
glue connecting tree and loop&
$\calo(\Lam)$ & yes
\\
\hline
$e^{iH_{cut}}$&
cuts a loop open&
$\calo(\Lam)$& yes
\\
\hline
$e^{iH_{line}}$&
line of length $\ns$&
$\calo(\nb^2)$ & no
\\\hline
$e^{iH_{in}}$&
oracle, with $2^\Lam=\nlvs$ inputs&
Could reach $\calo(\nlvs)$.&
\\
&&$\calo(\Lam^2)$
for banded oracle.&\\
\hline
\end{tabular}

Henceforth, we will say that
a SEO is tractable if
it has $\calo(\Lam^k)$
CNOTs, for some $k\geq 0$.

Our plan is to combine
$e^{iH_{bulk}}$ and
$e^{iH_{corr}}$ via Trotterized Suzuki.
Let $N_{exp}$ be the number
of times $e^{iH_{bulk}}$ or
$e^{iH_{corr}}$ appear in the
Trotterized Suzuki product that approximates
$e^{i(H_{bulk}+ H_{corr})}$.
According to Eq.(\ref{eq-best-nexp}), if we
want to evaluate a NAND formula
in time
$\approx \calo((\nlvs)^{\frac{1}{2}+\epsilon})$,
then

\beq
N_{exp}
\approx \calo((\nlvs)^{\frac{1}{2}+\epsilon})
\;.
\eeq
Thus

\beqa
\calc(e^{i(H_{bulk}+H_{corr})})
)&=&N_{exp}\calc(e^{iH_{bulk}}e^{iH_{corr}})\\
&\approx&
\calo((\nlvs)^{\frac{1}{2}+\epsilon})
\calc(e^{iH_{bulk}}e^{iH_{corr}})
\;.
\eeqa
If $e^{iH_{bulk}}e^{iH_{corr}}$
is tractable, then
$\calc(e^{i(H_{bulk}+H_{corr})})
=\calo((\nlvs)^{\frac{1}{2}+\epsilon})$.

\appendix

\section{Appendix: Exponential of
Anti-block Diagonal Matrix}\label{app-exp-anti-block}

\begin{Lemma}\label{lem-anti-d-sans-svd}
Suppose $F\in \CC^{n\times n}$. Then

\beq
\exp(i
\left[
\begin{array}{cc}
0 & F^\dagger \\
F & 0
\end{array}
\right]
)
=
\left[
\begin{array}{cc}
\cos(\sqrt{F^\dagger F}) &
i\frac{\sin(\sqrt{F^\dagger F  })}
{\sqrt{F^\dagger F  }}
F^\dagger \\
i\frac{\sin(\sqrt{F F^\dagger })}
{\sqrt{F F^\dagger }}
F &
\cos(\sqrt{F F^\dagger })
\end{array}
\right]
\;.
\eeq
\end{Lemma}
\proof

Note that
\beq
\left[
\begin{array}{cc}
0 & F^\dagger \\
F & 0
\end{array}
\right]
=
\sigm\otimes F +
\sigp\otimes F^\dagger
\;.
\label{eq-sigp-fh-sigm-f}
\eeq
Since $\sigp = \ket{0}\bra{1}$
and $\sigm = \ket{1}\bra{0}$,
$\sigp^2=\sigm^2=0$. Thus,
when we raise the right hand
side of Eq.(\ref{eq-sigp-fh-sigm-f})
 to a power,
only terms proportional to an
alternating sequence of
$\sigp$ and $\sigm$ survive.
For $n=1,2,3,\ldots$, one gets

\beqa
(\sigm\otimes F +
\sigp\otimes F^\dagger
)^{2n}
&=&
(\sigp\sigm)^n\otimes(F^\dagger F)^n
+
(\sigm\sigp)^n\otimes(F F^\dagger)^n
\\
&=&
P_0\otimes(F^\dagger F)^n
+
P_1\otimes(F F^\dagger)^n
\;,
\eeqa
and

\beqa
(\sigm\otimes F +
\sigp\otimes F^\dagger
)^{2n+1}
&=&
P_0\sigp\otimes(F^\dagger F)^n F^\dagger
+
P_1\sigm\otimes(F F^\dagger)^n F
\\
&=&
\sigp\otimes(F^\dagger F)^n F^\dagger
+
\sigm\otimes(F F^\dagger)^n F
\;.
\eeqa
Thus

\begin{eqnarray}
\lefteqn{\sum_{n=0}^{\infty}
\frac{(i)^{2n}}{(2n)!}
(\sigm\otimes F +
\sigp\otimes F^\dagger
)^{2n}=} \nonumber\\
&=& P_0\otimes \cos(\sqrt{F^\dagger F})
+
P_1\otimes \cos(\sqrt{F F^\dagger })
\;,
\end{eqnarray}
and

\begin{eqnarray}
\lefteqn{\sum_{n=0}^{\infty}
\frac{(i)^{2n+1}}{(2n+1)!}
(\sigm\otimes F +
\sigp\otimes F^\dagger
)^{2n+1}=} \nonumber\\
&=& \sigp\otimes i
\frac{\sin(\sqrt{F^\dagger F})}
{\sqrt{F^\dagger F}}
F^\dagger
+
\sigm\otimes i
\frac{\sin(\sqrt{F F^\dagger })}
{\sqrt{F F^\dagger }}
F.
\;
\end{eqnarray}
\qed

\section{Appendix: Special Case $d_\lam=
\sqrt{2}g$ for all $\lam$}
\label{app-all-d-same}

An important special case is when
$d_\lam = \sqrt{2}\coco$ for all $\lam$,
where
$d_\lam$ is defined by
Eq.(\ref{eq-const-delta-assump}).
This appendix will discuss some
idiosyncrasies of this special case.

When the $d_\lam$ are
all equal, it is possible
to make a slight
redefinition of $\Gamma^\lam$ in
Eq.(\ref{eq-gamma-defs})
so that the decomposition
of $e^{i\cala_{2^{\Lam+1}}}$
into $\Gamma_{2^\lam}$ given by
Eq.(\ref{eq-strati}) is
good to order $\calo(\coco^4)$
instead of $\calo(\coco^3)$.
We show this next for $\Lam=3$.
Generalization to higher $\Lam$ will
be obvious.

\begin{Lemma}
Let

\beq
f = 1 + \frac{\coco^2}{6}
\;.
\eeq
Then

\beq
e^{i \cala_{16}}=
\Gamma_1^{<3>} \Gamma_2^{<3>}
\Gamma_3^{<3>} + \calo(\coco^4)
\;,
\eeq
where

\begin{subequations}
\beq
\Gamma_3^{<3>} =
\exp(i
\left[
\begin{array}{cc}
0 & \calbbar_8 \\
\calbbar_8 & 0
\end{array}
\right]
)
\;,
\eeq

\beq
\Gamma_2^{<3>} =
\exp(i f
\left[
\begin{array}{cc|c}
0 & \calbbar_4 & \\
\calbbar_4 & 0 & \\ \hline
& & 0_{8}
\end{array}
\right]
)
\;,
\eeq

\beq
\Gamma_1^{<3>} =
\exp(i f
\left[
\begin{array}{cc|c}
0 & \calbbar_2 & \\
\calbbar_2 & 0 & \\ \hline
& & 0_{12}
\end{array}
\right]
)
\;,
\eeq
\end{subequations}
and

\beq
\calbbar_\twotolam=
\calb_\twotolam
\sinc(\frac{\cala_\twotolam}{2})
e^{i\frac{\cala_\twotolam}{2}}
\;,\;\;{\rm for}\;\; \lam \in Z_{1,3}
\;.
\eeq
\end{Lemma}
\proof
According to Eq.(\ref{eq-approx-with-t1-t2}),

\begin{eqnarray}
e^{i\cala_{16}} &=&
\exp(i
\left[
\begin{array}{cc}
\cala_8 & \calb_8 \\
\calb_8 & 0
\end{array}
\right]
)\\
&=&
\exp(i
\left[
\begin{array}{cc}
\cala_8 &  \\
 & 0_8
\end{array}
\right]
)
\exp(
\left[
\begin{array}{cc}
T_1 + T_2 & i\calbbar_8\\
i\calbbar_8 & -i \frac{\calb_8\cala_8\calb_8}{6}
\end{array}
\right]
)
+ \calo(\coco^5)
\;.
\end{eqnarray}
One finds that

\beq
T_1 =i\frac{\coco^2}{6}
\left[
\begin{array}{cc}
0 & \calb_4\\
\calb_4 & 0
\end{array}
\right]
\;,\;\;
T_2 = \calo(\coco^4)
\;,
\eeq
and

\beq
\calb_8 \cala_8 \calb_8 = 0
\;.
\eeq
Therefore,

\begin{eqnarray}
e^{i\cala_{16}} &=&
\exp(i
\left[
\begin{array}{cc}
\cala_8 &  \\
 & 0_8
\end{array}
\right]
)
\exp(i
\left[
\begin{array}{cc}
\frac{\coco^2}{6}\left[
\begin{array}{cc}
0 & \calb_4\\
\calb_4 & 0
\end{array}
\right]
& \calbbar_8\\
\calbbar_8 & 0_8
\end{array}
\right]
)
+ \calo(\coco^4)
\label{eq-const-dlam-proof-1}
\\
&=&
\exp(i
\left[
\begin{array}{cc}
\left[
\begin{array}{cc}
\cala_4 & f\calb_4\\
f\calb_4 & 0
\end{array}
\right] &  \\
 & 0_8
\end{array}
\right]
)
\Gamma_3^{<3>}
+ \calo(\coco^4)
\label{eq-const-dlam-proof-2}
\;.
\end{eqnarray}
To go from Eq.(\ref{eq-const-dlam-proof-1})
to Eq.(\ref{eq-const-dlam-proof-2}),
we used
$\cala_8 =
\left[
\begin{array}{cc}
\cala_4 & \calb_4\\
\calb_4 & 0
\end{array}
\right]$.
Furthermore,

\begin{eqnarray}
\lefteqn{
\exp(i
\left[
\begin{array}{cc}
\left[
\begin{array}{cc}
\cala_4 & f\calb_4\\
f\calb_4 & 0
\end{array}
\right] &  \\
 & 0_8
\end{array}
\right]
)
=} \nonumber\\
&=&
\exp(i
\left[
\begin{array}{cc}
\left[
\begin{array}{cc}
\cala_4 & \\
& 0_4
\end{array}
\right] &  \\
 & 0_8
\end{array}
\right]
)
\exp(i
\left[
\begin{array}{cc}
    \left[
    \begin{array}{cc}
        \frac{\coco^2}{6}
        \left[
        \begin{array}{cc}
        0 & f\calbbar_2\\
        f\calbbar_2& 0
        \end{array}
        \right]
    &f\calbbar_4\\
    f\calbbar_4& 0_4
    \end{array}
    \right]
&\\
&0_8
\end{array}
\right]
)
\\
&=&
\Gamma_1^{<3>} \Gamma_2^{<3>}
\;,
\end{eqnarray}
where we used
$\cala_4 =
\left[
\begin{array}{cc}
\cala_2 & \calb_2\\
\calb_2 & 0
\end{array}
\right]$ and $\coco^2 f \approx \coco^2$.
\qed

Another useful consequence
of $d_\lam = \sqrt{2}\coco$
for all $\lam$ is that
in this case it is easy to
calculate the $b^k_j$
coefficients used in Section
\ref{sec-calbbar-in-braket}.
Indeed, observe that in this
case all non-zero entries of
$\cala_{2^\lam}$ equal $\sqrt{2}\coco$.
Define

\beq
b = \coco \sqrt{2}
\sinc( \frac{\coco\sqrt{2}}{2} )
e^{i \frac{\coco \sqrt{2}}{2}}
\;.
\eeq
Let $b_r$ and $b_i$
denote the real and imaginary
parts of $b$, respectively.
Now we can use the following rule.
In Section \ref{sec-calbbar-in-braket},
in the bra-ket expansions
of $\calbbar^\sdep_\twotolam$,
replace, all $b^k_j$ that are
preceded by an imaginary $i$, by $b_i$ ,
and replace, all
$b^k_j$ that are NOT preceded
by an imaginary $i$, by $b_r$.
For example, in the
example of Lemma \ref{lem-svd-calbbar},
one has

\beq
\rho_3 = \sqrt{b_r^2 + b_i^2} = |b|\;,\;\;
\cos\theta_3 = \frac{b_i}{|b|}\;,\;\;
\sin\theta_3 = \frac{-b_r}{|b|}
\;,
\eeq

\beq
\rho_7 = \sqrt{b_r^2 + |b|^2}\;,\;\;
\cos\theta_7 = \frac{b_r}{\rho_7}\;,\;\;
\sin\theta_7 = \frac{\rho_3}{\rho_7}
\;,
\eeq

\beq
\rho_{15} = \sqrt{b_i^2 + \rho_7^2}\;,\;\;
\cos\theta_{15} = \frac{b_i}{\rho_{15}}\;,\;\;
\sin\theta_{15} = \frac{-\rho_7}{\rho_{15}}
\;,
\eeq

\beq
\rho_{31} = \sqrt{b_r^2 + \rho_{15}^2}\;,\;\;
\cos\theta_{31} = \frac{b_r}{\rho_{31}}\;,\;\;
\sin\theta_{31} = \frac{\rho_{15}}{\rho_{31}}
\;.
\eeq

\section{Appendix: Compiling Banded
Oracle}\label{app-banded-oracle}

Suppose
$x_\vecb\in Bool$ for $\vecb\in Bool^\Lam$,
and
$\nlvs=2^\Lam$.
Suppose $\alpha$ is a qubit,
and $\vecxi=(\Lam-1, \dots, 1, 0)$
is a vector of $\Lam$ qubits,
and all these $\Lam+1$ qubits are distinct.
In this Appendix, we
will show how to compile
the oracle evolution operator

\beq
U_O = \sigx(\alpha)^{
\sum_{\vecb\in Bool^\Lam}
x_\vecb P_\vecb(\vecxi) }
\;,
\eeq
when $\vec{x}=[x_0,x_1,\ldots,x_{\nlvs-1}]^T$
is banded.
If we envision $\vec{x}$ as a sequence
of ones and zeros, then
we will call a band of $\vec{x}$,
any subsequence of $\vec{x}$
consisting of adjacent terms,
all of which are one.
It $\vec{x}$ contains a fixed
($\Lam$ independent) number of
bands, we will say it is banded.

\begin{figure}[h]
    \begin{center}
    \epsfig{file=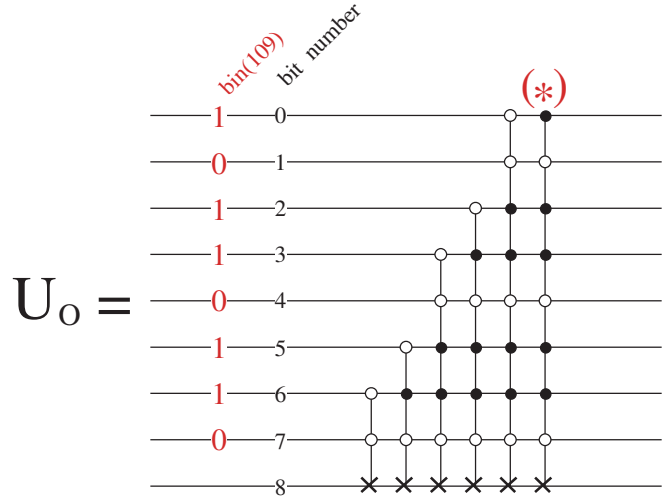, height=2.7in}
    \caption{Circuit diagram for
    Eq.(\ref{eq-def-uo}),
    when
    the first 109 components of $\vec{x}$
     equal one,
    and the final $2^8-109=
    256-109=147$ components equal zero.
    }
    \label{fig-H-in}
    \end{center}
\end{figure}

We will consider first the case
when $\vec{x}$ has a single
front band. By this we mean that the first
$N_1$ terms of $\vec{x}$ are one, and the
rest are zero. Thus

\beq
\vec{x}=
[\underbrace{1,1,\dots,1}_{N_1\;\;{\rm times}}
,\underbrace{0,0,\dots,0}_{\nlvs-N_1\;\;{\rm times}}]^T
\;.
\eeq
Define a binary vector $\vecbmax$
and its corresponding set of
binary vectors $S(\vecbmax)$ by

\beq
\vecbmax = bin(N_1-1)
\;,
\;\;
S(\vecbmax)= \{\vecb\in Bool^\Lam:
0\leq dec(\vecb)
\leq dec(\vecbmax)\}
\;.
\eeq
In the
single front band case,
$U_O$ reduces to:

\beq
U_O =
\sigx(\alpha)^{\sum_{\vecb\in S(\vecbmax)}
P_\vecb(\vecxi)}
\;.
\label{eq-def-uo}
\eeq
$U_O$, in the form given by
Eq.(\ref{eq-def-uo}), is already compiled.
But the length of this compilation can be reduced
significantly by
reducing $\sum_{\vecb\in S(\vecbmax)}P_\vecb$.
Let us consider an
example first, before dealing with
the general case. Suppose
$\Lam=8$ and
$\vecbmax=(0110,1101)=bin(109)$.
Although $\sum_{\vecb\in S(bin(109))} P_\vecb$
is a sum of 109 projection operators
of the type $P_\vecb$
where $\vec{b}\in Bool^8$,
it can be expressed as a sum of
just six simpler projection operators:

\beq
\sum_{\vecb\in S(bin(109))} P_\vecb
=
\left\{
\begin{array}{ll}
P_{00\cdot\cdot,\cdot\cdot\cdot\cdot}&
\mbox{(sum from 0000,0000 to 0011,1111)}
\\
+P_{010\cdot,\cdot\cdot\cdot\cdot}&
\mbox{(sum from 0100,0000 to 0101,1111)}
\\
+P_{0110,0\cdot\cdot\cdot}&
\mbox{(sum from 0110,0000 to 0110,0111)}
\\
+P_{0110,10\cdot\cdot}&
\mbox{(sum from 0110,1000 to 0110,1011)}
\\
+P_{0110,1100}
\\
+P_{0110,1101}&(*)
\end{array}
\right.
\;.
\label{eq-proj-sum-109}
\eeq
Fig.\ref{fig-H-in}
gives a circuit diagram
for Eq.(\ref{eq-def-uo}),
assuming $dec(\vecbmax) = 109$, and
with $\sum_{\vecb\in S(bin(109))} P_\vecb$
expressed in the simplified
form given by the right hand side of
Eq.(\ref{eq-proj-sum-109}).

It's also interesting to
consider an example
in which $dec(\vecbmax)$ is an even number
instead of an odd one. When
$dec(\vecbmax)=108$ instead of
109, we must remove the
operator marked by an asterisk in
Eq.(\ref{eq-proj-sum-109}),
and the operator marked by an
asterisk in Fig.\ref{fig-H-in}.

The pattern of the control vertices in
Eq.(\ref{eq-proj-sum-109}) and
Fig.\ref{fig-H-in} is not hard to uncover.
There is exactly one MCNOT
for each nonzero bit in $bin(109)$.
In addition, there is one ``final"
MCNOT with controls equal to $P_{bin(109)}$.
Call $\calk(\beta)$ the
``non-final" MCNOTs. There is precisely one of
these for each
bit $\beta\in Z_{0,\Lam}$
such that the $\beta$ component of $\vecbmax$
(i.e., $(\vecbmax)_\beta$)
equals one,
and none when $(\vecbmax)_\beta=0$.
$\calk(\beta)$ always has control
$P_0$ at bit $\beta$.
At bits $\alpha\in Z_{\beta+1,\Lam-1}$,
$\calk(\beta)$ has a control
$P_{(\vecbmax)_\alpha}$.
Hence, in general,

\beq
\sum_{\vecb\in S(\vecbmax)}
P_\vecb(\vecxi)=
\left\{
\begin{array}{l}
\sum_{\beta\in Z_{0,\Lam-1}}
\delta((\vecbmax)_\beta,1)
P_0(\beta)
\prod_{\alpha\in Z_{\beta+1, \Lam-1}}
P_{(\vecbmax)_\alpha}(\alpha)

\\
+ P_\vecbmax(\vecxi)
\end{array}
\right.
\;.
\eeq

Note that the number of MCNOTs
in $U_O$ is $\calo(\Lam)$.
The number of CNOTs
in each of these MCNOTs
is $\calo(\Lam)$. Thus,
the number of CNOTs in $U_O$
is $\calo(\Lam^2)$.

Now suppose $\vec{x}$ has a single band
which is, however, not at the front.
Suppose it ranges from $dec(\vecb_{min})\neq 0$
to $dec(\vecb_{max})$.
Then just multiply $U_O$ for
a single front band up to $dec(\vecb_{min})-1$
times
$U_O$ for
a single front band up to $dec(\vecb_{max})$.
Multiple bands can be handled similarly.

Sometimes it is possible to apply
$\sigx$ on individual qubits and/or apply
qubit
permutations
 to $U_O$ so as to get a new
$U_O$ with  fewer bands. Fewer bands will lead
to a shorter SEO of the type proposed in this
appendix. $\sigx$ on individual qubits and/or qubit
permutations\footnote{Don't confuse qubit
permutations with state permutations.}
 do not increase the length of a
SEO if they are applied
to the SEO  on both sides,
via a $\odot$ product.

\section{Appendix: Compiling Evolution Operators in\\
Grover's Algorithm}\label{app-grovers}

The goal of this appendix is not
to say something new about
Grover's algorithm\cite{grovers}.
After all, Grover's algorithm has been
studied so extensively in the literature
that it's almost impossible to say
anything new about it.
The goal of this appendix is, rather, to
review how one compiles
the oracle and non-oracle evolution operators
associated with Grover's algorithm.
This will allow the reader
to compare the compilation of
Grover's algorithm with the compilation of
FGG07
presented in this paper.

Let $\nb$ be the number of qubits
and $\ns=2^\nb$ the number of states
used in Grover's algorithm.
Let $\vec{x}_o \in Bool^\nb$
be the target state of the algorithm.
Define

\beq
\ket{\mu} = \frac{1}{\sqrt{\ns}}
\sum_{\vecx\in Bool^\nb}\ket{\vecx}=
\frac{1}{\sqrt{\ns}}
[1,1,\dots,1]^T
\;,
\eeq
and

\beq
\mu = \ket{\mu}\bra{\mu}
=\frac{1}{\ns}
\left[
\begin{array}{ccccc}
1&1&\dots&&1\\
1&1&\dots&&1\\
\vdots&\vdots&\ddots&&\vdots\\
1&1&\dots&&1\\
1&1&\dots&&1
\end{array}
\right]
\;.
\eeq
Thus, in matrix notation, $\ket{\mu}$ is
an $\ns$ dimensional column vector,
and $\mu$ is an $\ns\times\ns$ matrix.

In Grover's algorithm, one alternates
between
an oracle evolution operator,
call it $U_{corr}$, and
a non-oracle evolution operator,
call it $U_{bulk}$.
The oracle evolution operator
depends on the target state
whereas the non-oracle evolution
operator does
not. More precisely,
these two evolution operators are
defined by:

\beq
U_{corr}=(-1)^{\ket{\vecx_o}\bra{\vecx_o}}=
(-1)^{P_{\vec{x}_o}}=e^{i\pi P_{\vec{x}_o}}
\;,
\eeq
and

\beq
U_{bulk}=
2\ket{\mu}\bra{\mu}-1=
-(-1)^{\ket{\mu}\bra{\mu}}=
e^{i\pi(\mu+1)}
\;.
\eeq

In Grover's algorithm, one
applies $\sqrt{\ns}$ times the
product $U_{corr}U_{bulk}$.
A variant of this would be to apply
$\nt=\ns^{\frac{1}{2}+\delta}$ times the product
$U_{corr}^\frac{1}{\ns^\delta}
U_{bulk}^\frac{1}{\ns^\delta}$,
for some $\delta>0$.
One can think of this variant of
Grover's algorithm as
a Trotterized Lie approximation:

\beq
e^{i(H_{corr}+H_{bulk})}
\approx
\left(
e^{i\frac{H_{corr}}{\ns^{\frac{1}{2}+\delta}}
}
e^{i\frac{H_{bulk}}{\ns^{\frac{1}{2}+\delta}}
}
\right)^{\ns^{\frac{1}{2}+\delta}}
\;,
\eeq
for some $t\in\RR$,
where the Hamiltonians
$H_{corr}$ and $H_{bulk}$ are
given by

\beq
\begin{array}{l}
e^{i
\frac{H_{corr}}{\ns^{\frac{1}{2}+\delta}}
}
=
U_{corr}^\frac{1}{\ns^{\delta}}
\Longrightarrow
\frac{H_{corr}}{\ns^{\frac{1}{2}+\delta}}
=
\frac{\pi P_{\vec{x}_o}}{\ns^\delta}
\\
e^{i
\frac{H_{bulk}}{\ns^{\frac{1}{2}+\delta}}
}=
U_{bulk}^\frac{1}{\ns^\delta}
\Longrightarrow
\frac{H_{bulk}}{\ns^{\frac{1}{2}+\delta}}
=
\frac{\pi (1+\mu)}{\ns^\delta}
\end{array}
\;.
\label{eq-suggested-hams}
\eeq
This variant of
Grover's algorithm is a
quantum walk over a fully connected
graph with $\ns$ nodes.
Transitions occur on this graph
along edges connecting distinct nodes
and also from a node back to
itself.
Self transitions occur
with strength proportional to $1+\frac{1}{\ns}$
for all nodes except the
one representing the target state.
Self transitions for the target
node occur with strength proportional to
$2+\frac{1}{\ns}$, larger than
the strength for the other self transitions.
Transitions along the edges connecting distinct
nodes occur with vanishing strength
proportional to
$\frac{1}{\ns}$.

Compiling the oracle evolution
operator for Grover's algorithm is trivial.
Let $\vecxi$ denote the $\nb$-dimensional
vector of qubit positions for the
$\nb$ primary qubits used in Grover's algorithm,
and let $\alpha$ denote the qubit
position of an additional
(not in $\vecxi$) ancilla qubit.
Then

\beqa
U_{corr}=(-1)^{P_{\vec{x}_o}(\vecxi)}
&=&
(-1)^{n(\alpha)P_{\vec{x}_o}(\vecxi)}
\ket{1}_\alpha
\\
&=&
\sigz(\alpha)^{P_{\vec{x}_o}(\vecxi)}
\ket{1}_\alpha
\\
&=&
H(\alpha)\sigx(\alpha)^{P_{\vec{x}_o}(\vecxi)}
(H\sigx)(\alpha)
\ket{0}_\alpha
\;.
\eeqa
We used $\sigz=(-1)^n$, $\sigx\ket{0}=\ket{1}$
and $H\sigx H = \sigz$.

Compiling the non-oracle evolution
operator for Grover's algorithm is also
trivial. One notes that

\beq
\ket{\mu}=H^{\otimes \nb}\ket{0}
\;,
\eeq
since the first column of
$H^{\otimes \nb}$ is all ones.
Therefore,

\beq
\mu=H^{\otimes \nb}\;\ket{0}\bra{0}\;H^{\otimes \nb}
\;.
\label{eq-mu-svd}
\eeq
Eq.(\ref{eq-mu-svd})
is merely the eigenvalue decomposition
(and SVD)
of $\mu$. Since $\mu$ is a circulant
matrix, we could have
obtained
Eq.(\ref{eq-mu-svd})
from the eigenvalue decomposition
of circulant matrices presented in Section
\ref{sec-circulant}.
An immediate consequence of
Eq.(\ref{eq-mu-svd}) is

\beq
U_{bulk}=-
H^{\otimes \nb}\;
(-1)^{\ket{0}\bra{0}}
\;H^{\otimes \nb}
= -
H^{\otimes \nb}\;
(-1)^{P_{0,0,\dots,0}}
\;H^{\otimes \nb}
\;.
\eeq
One can compile
$(-1)^{P_{0,0,\dots,0}}$ in the
same way that
$(-1)^{P_{\vec{x}_o}}$
was compiled above.

Note that we have expressed both
$U_{bulk}$ and
$U_{corr}$ as
SEOs containing a
single MCNOT.
This MCNOT can
be expressed as a SEO with $\calo(\nb)$ CNOTs.


\begin{thebibliography}{99}

\bibitem{FGG07}
E. Farhi, J. Goldstone, S. Gutmann,
``A Quantum Algorithm for the
Hamiltonian NAND Tree",
arXiv:quant-ph/0702144

\bibitem{Cleve-quick}
A. Childs, R. Cleve,
S. Jordan, D. Yeung,
``Discrete-query quantum
algorithm for NAND trees",
arXiv:quant-ph/0702160

\bibitem{caltech}
A. Childs, B. Reichardt,
R. Spalek, S. Zhang,
``Every NAND formula on
N variables can be evaluated
in time $\calo(N^{1/2+\epsilon})$",
arXiv:quant-ph/0703015


\bibitem{Amb}
A. Ambainis,
``A nearly optimal discrete query quantum algorithm
for evaluating NAND formulas",
arXiv:0704.3628

\bibitem{Cleve-two-trees}
A. Childs, R. Cleve, E. Deotto,
E. Farhi, S. Gutmann, D. Spielman,
``Exponential algorithmic
speedup by quantum walk",
arXiv:quant-ph/0209131


\bibitem{Cleve-using-Suz}
D. Berry, G. Ahokas, R. Cleve,
B. Sanders,
``Efficient quantum algorithms
for simulating sparse Hamiltonians",
arXiv:quant-ph/0508139

\bibitem{Childs-thesis}
A. Childs,
``Quantum information processing
in continuous time"
(MIT 2004 Ph.D. thesis)

\bibitem{Hines}Andrew P. Hines, P.C.E. Stamp,
``Quantum Walks, Quantum Gates and Quantum Computers",
 arXiv:quant-ph/0701088

\bibitem{Golub}
G.H. Golub and C.F. Van Loan,
{\it Matrix Computations, Third Edition}
(John Hopkins Univ. Press, 1996).

\bibitem{Paulinesia}
R.R.Tucci,
``QC Paulinesia",
quant-ph/0407215

\bibitem{circulant}R. M. Gray,
 ``Toeplitz and Circulant
 Matrices: A Review",
 available at
 http://www-ee.stanford.edu/\~{}gray/toeplitz.pdf

\bibitem{HatSuz}
N.Hatano, M.Suzuki,
``Finding Exponential Product
Formulas of Higher Orders",
arXiv:math-ph/0506007

\bibitem{Zachos}
C. Zachos, ``Crib Notes on
Campbell-Baker-Hausdorff expansions",
http://www.hep.anl.gov/czachos/CBH.pdf

\bibitem{Reinsch}M. Reinsch,
``A Simple Expression for
the Terms in the Baker-Campbell-Hausdorff
Series", math-ph/9906007


\bibitem{Bar}
Barenco et al.,
``Elementary Gates for Quantum Computation",
arXiv:quant-ph/9503016


\bibitem{Tuc-dressed-nots}
R. R. Tucci,
``Simplifying Quantum Circuits via
Circuit Invariants and Dressed CNOTs",
arXiv:quant-ph/0606061


\bibitem{Tuc99}
R.R. Tucci,
``A Rudimentary Quantum
Compiler(2cnd Ed.)",
arXiv:quant-ph/9902062

\bibitem{Hels04b}
 V. Bergholm, J. Vartiainen,
 M.Mottonen, M. Salomaa,
 ``Quantum circuit for
 a direct sum of two-dimensional
 unitary operators",
 quant-ph/0410066


\bibitem{qbtr}
www.ar-tiste.com/qubiter.htm

\bibitem{Mich04}
V. Shende,
S. Bullock, I. Markov,
``A Practical Top-down
Approach to Quantum Circuit
Synthesis", quant-ph/0406176




\bibitem{Tuc-qbtr-mod}
R.R. Tucci,
``Qubiter Algorithm Modification,
Expressing Unstructured Unitary
Matrices with Fewer CNOTs",
arXiv:quant-ph/0411027

\bibitem{VD}
G. Vidal, C.M. Dawson,
``A Universal Quantum Circuit
for Two-qubit
Transformations with 3 CNOT Gates",
quant-ph/0307177

\bibitem{bound}
V.V.Shende, I.L. Markov, S.S. Bullock,
``On Universal Gate Libraries and Generic
Minimal Two-qubit Quantum Circuits", quant-ph/0308033


\bibitem{Tuc-fft}
R.R. Tucci,
``Quantum Fast Fourier
Transform Viewed as a
Special Case of Recursive Application
of Cosine-Sine Decomposition",
arXiv:quant-ph/0411097



\bibitem{Yumi}
Y. Nakajima, Y. Kawano, H. Sekigawa,
``A new algorithm for
producing quantum circuits
using KAK decompositions",
arXiv:quant-ph/0509196

\bibitem{Copper}Don Coppersmith,
``An approximate Fourier
transform useful in quantum factoring",
(1994 IBM Internal Report),
quant-ph/0201067

\bibitem{grovers}
See
http://en.wikipedia.org/wiki/Grover's\_algorithm

\end{thebibliography}
\end{document}